# UCSB South Pole 1994 CMB anisotropy measurement constraints on open and flat-Λ CDM cosmogonies

Ken Ganga[1], Bharat Ratra[2], Joshua O. Gundersen[3], and Naoshi Sugiyama[4]

## ABSTRACT

We develop methods to account for experimental and observational uncertainties in likelihood analyses of data from cosmic microwave background (CMB) anisotropy experiments and apply them to an analysis of the UCSB South Pole 1994 (SP94) experiment. Observationally motivated open and spatially-flat $\Lambda$, cold dark matter cosmogonies are considered. Among the models we consider, the full SP94 data set is most consistent with $\Omega_0 \sim 0.1-0.2$ open models and less so with old ($t_0 \gtrsim 15-16$ Gyr), high baryon density ($\Omega_B \gtrsim 0.0175h^{-2}$), low density ($\Omega_0 \sim 0.2-0.4$), flat-$\Lambda$ models. The SP94 data do not rule out any of the models we consider at the $2\sigma$ level. The SP94 experiment is most sensitive to anisotropies on a somewhat larger, model-dependent, angular scale than the scale at which the window function peaks.

For establishing the significance of a detection of CMB anisotropy we derive limits using the highest posterior density (HPD) prescription, since it yields smaller lower limits. Since HPD limits lead to tighter constraints on the CMB amplitude, they also provide for greater discrimination between models. Model normalizations deduced from the SP94 data subsets are mostly consistent with those deduced from the two-year *COBE*-DMR data, although the Ka-band data prefer a normalization $\sim 1\sigma$ lower than do the Q-band data, the Q and Ka

---

[1]Division of Physics, Mathematics and Astronomy, MS 59–33, California Institute of Technology, Pasadena, CA 91125. Current address: IPAC, MS 100–22, California Institute of Technology, Pasadena, CA 91125.

[2]Center for Theoretical Physics, Massachusetts Institute of Technology, Cambridge, MA 02139. Current address: Department of Physics, Kansas State University, Manhattan, KS 66506.

[3]Department of Physics, University of California, Santa Barbara, CA 93106. Current address: Department of Physics, Brown University, Box 1843, Providence, RI 02912.

[4]Department of Physics and Research Center for the Early Universe, University of Tokyo, Tokyo 113. Current address: Department of Physics, Kyoto University, Kitashirakawa-Oiwakecho, Sakyo-ku, Kyoto 606.

+ Q data favour a slightly higher normalization for the $\Omega_0 = 0.1$ open model than does the DMR, and the Ka and Ka + Q data prefer a somewhat lower normalization for the older, higher $\Omega_B$, low-density $\Lambda$ models than does the DMR.

## 1. Introduction

Low-density cold dark matter (CDM) cosmogonies, either with flat spatial hypersurfaces and a cosmological constant, $\Lambda$, or with open spatial hypersurfaces and no $\Lambda$, are consistent with a large fraction of present observational data. For flat-$\Lambda$ models, see Peebles (1993), Stompor, Górski, & Banday (1995), Scott, Silk, & White (1995), Ostriker & Steinhardt (1995), and Ratra et al. (1997, hereafter RSBG). For open models, see Ratra & Peebles (1994), Kamionkowski et al. (1994), Górski et al. (1995), Liddle et al. (1996), and RSBG.

These observational data include:

- measurements of the Hubble parameter $H_0$ ($= 100h\,\mathrm{km\ s^{-1}\,Mpc^{-1}}$), which suggest $h > 0.55$ (e.g., Pierce & Jacoby 1995; Mould et al. 1995; Tanvir et al. 1995; Höflich & Khokhlov 1996; Baum et al. 1995; Whitmore et al. 1995; but also see, e.g., Nugent et al. 1995), and globular cluster age estimates which suggest that the age of the universe $t_0 > 11$ Gyr (e.g., Bolte & Hogan 1995; Jimenez et al. 1996; Chaboyer et al. 1996);
- dynamical estimates of the mass clustered on scales $\lesssim 10h^{-1}$Mpc, which suggest $\Omega_0 \sim 0.05 - 0.35$ (Peebles 1993) [here the present value of the clustered-mass density parameter $\Omega_0 = 8\pi G\rho_b(t_0)/(3{H_0}^2)$, where $G$ is the gravitational constant and $\rho_b(t_0)$ is the mean clustered-mass density now];
- estimates of $\Omega_0$ from measurements of the plasma mass fraction of the rich clusters, standard nucleosynthesis theory, and the observed light element abundances (e.g., White & Fabian 1995; David, Jones, & Forman 1995);
- dynamical estimates of the mass clustered on scales $\gtrsim 10h^{-1}$Mpc (e.g., Cole, Fisher, & Weinberg 1995; Loveday et al. 1996; Baugh 1996; there is, however, some scatter; see, e.g., Dekel et al. 1993; Shaya, Peebles, & Tully 1995);

- the shape of the observed galaxy fluctuation power spectrum (e.g., Peacock & Dodds 1994; Tadros & Efstathiou 1995);

- the observed cluster mass and correlation functions (e.g., Liddle et al. 1996; Viana & Liddle 1996);

- the presence of high redshift ($z \sim 0.5 - 1$) clusters (e.g., Mellier et al. 1994; Luppino & Gioia 1995; Smail & Dickinson 1995), high redshift ($z > 2$) galaxy groups (e.g., Francis et al. 1996; Pascarelle et al. 1996), high redshift ($z \gtrsim 3 - 4$) damped Lyman-$\alpha$ systems (e.g., Giallongo et al. 1994; Lu et al. 1996; Wampler et al. 1996), and the observed similarity of the giant elliptical luminosity functions at $z = 0$ and $z \sim 1$ (e.g., Djorgovski et al. 1995; Glazebrook et al. 1995; Lilly et al. 1995);

- the large angular scale cosmic microwave background (CMB) anisotropy measured by the *COBE*-DMR experiment and analyzed in the context of these cosmogonical models (e.g., Bunn & Sugiyama 1995; Górski et al. 1995; Stompor et al. 1995); and

- the smaller-scale CMB spatial anisotropy detected by a number of experiments (RSBG; Ganga, Ratra, & Sugiyama 1996, hereafter GRS).

We emphasize that a number of the above observational estimates are still tentative, and that some have large error bars. In particular, any individual set of measurements does not have the power to seriously constrain the range of cosmological parameters such as $\Omega_0$ and $h$. It is only the combination of all these different measurements that indicates a low-density cosmogony may be a more appropriate model of the universe.

Following the DMR discovery of large-scale CMB anisotropy and the confirmation by FIRS (Smoot et al. 1992; Bennett et al. 1992; Wright et al. 1992; Ganga et al. 1994), attention has been focussed on attempts to use the DMR data to constrain cosmogonies which are consistent with other cosmological data (e.g., Bunn & Sugiyama 1995, Górski et al. 1995, Stompor et al. 1995, Yamamoto & Bunn 1996). While the DMR data does allow for a much more accurate determination of the normalization of cosmological models than has previously been possible, it does not have the discriminative power to significantly constrain other cosmological parameters.

Besides the DMR and FIRS measurements of large-scale CMB anisotropy, there are now a number of other detections of anisotropy on smaller angular scales (Hancock et al. 1996; Piccirillo et al. 1997; Netterfield et al. 1997; Gundersen et al. 1995, hereafter G95; Ruhl et al. 1995; de Bernardis et al. 1994; Tanaka et al. 1996; Cheng et al. 1996; Griffin et al. 1997). It is of some interest to determine whether these smaller-scale anisotropy measurements can help to constrain cosmological model parameters. Given the error bars associated with these measurements, it is not surprising that, even if one makes use of the DMR data to fix one free parameter and normalize cosmological models, all the smaller-scale anisotropy experiments combined do not very significantly constrain other cosmological parameters (RSBG; GRS). In particular, a quantitative goodness-of-fit comparison of the CMB anisotropy predictions from two-year DMR-normalized, open and flat-$\Lambda$ CDM models to all available CMB detections by GRS found that in most cases the DMR error bars precluded robust conclusions about model viability.[5]

The comparisons between the theoretical predictions and the data done in RSBG and GRS did not make use of the complete data from each experiment. Rather, they used a single number with error bars related to the rms anisotropy detected by the experiment. There are two significant issues related to such a sparse representation of the observational data from an experiment. Since the procedure used to extract this single number from the observational data set typically assumes a flat bandpower spectrum or a gaussian autocorrelation function for the anisotropy, it is of interest to examine how sensitively the final result depends on this assumption and to determine the result using more realistic forms for the spectrum. Secondly, although the experiments are sensitive to a range of spatial scales, representing the observational data by a single number effectively discards additional spatial correlation information. It is therefore also of interest to examine whether this additional spatial correlation information aids in discriminating between model spectra.

In this paper we address these issues for the SP94 data set of G95, using anisotropy spectra derived in gaussian, adiabatic, open and flat-$\Lambda$, CDM cosmogonies (RSBG). Bond & Jaffe (1996) have also analyzed the SP94 data, mostly making use of CMB anisotropy spectra in spatially-flat models, but considering a wider variety of these models than we have here. The SP94 windows are sensitive to anisotropy on angular scales where the open and flat-$\Lambda$ model CMB spatial anisotropy spectra are not very scale dependent, so we expect only a weak dependence of the deduced signal amplitude on the assumed CMB anisotropy shape.

SP94 is the most recent of the SP CMB anisotropy experiments at the South Pole. The observational results of SP89 and the ACME telescope are discussed by Meinhold &

---

[5]The new four-year DMR results (e.g., Bennett et al. 1996; Górski et al. 1996) indicate a normalization slightly lower than that found from the two-year data. This is mostly a consequence of more detailed modelling of foreground Galactic emission (Banday et al. 1997; Kogut et al. 1996). From Figs. 5 and 6 of GRS, which correspond to the models normalized at the $-1\sigma$ value of the two-year normalization (or, $\sim \sigma/3$ below the nominal value of the four-year DMR normalization), one sees that most models considered by GRS and here are consistent, in the goodness-of-fit sense, with the small-scale CMB anisotropy detection data.

Lubin (1991) and Meinhold et al. (1993). Analyses of the SP89 data, in the context of fiducial scale-invariant, spatially-flat CDM models and some isocurvature models is given in Vittorio et al. (1991) and Bond et al. (1991). The SP91 observational results are in Gaier et al. (1992) and Schuster et al. (1993). The Gaier et al. data have been analyzed in the context of fiducial CDM by Górski, Stompor, & Juszkiewicz (1993), Muciaccia et al. (1993) (who also consider tilted CDM models), Dodelson & Jubas (1993), Bunn et al. (1994), and Stompor & Górski (1994). Stompor & Górski also consider the now more fashionable flat-$\Lambda$ CDM models and analyze the Schuster et al. (1993) data. These analyses only used the data from the highest-frequency channel of the SP91 scans to draw conclusions about the CMB anisotropy. Off-diagonal channel-channel noise correlations were accounted for in the multifrequency reanalysis of SP91 given in G95, who made use of the flat bandpower approximation. The inclusion of these off-diagonal correlations has little effect on the most probable sky signal amplitude. However, they increase the $\pm 1\sigma$ error bars by $\sim 20\%$.

The SP94 experiment extends the SP91 frequency range by taking data with a Q-band receiver, as well as with a Ka-band receiver. Descriptions of the SP94 experiment may be found in Gundersen et al. (1994), G95, and Gundersen (1995). The far-field beam patterns of the Ka-band and Q-band radiometers, coupled to the ACME telescope, were measured in both azimuth and elevation, at 27.7 GHz for Ka and 41.5 GHz for Q. These measurements indicate that in both azimuth and elevation the beams are well-approximated by gaussians down to 30 dB (Gundersen 1995). The beamwidths measured in azimuth and elevation differ from the average beamwidth by $\sim 1.5 - 2\%$ ($1\sigma$). In our analysis, we do not explicitly account for this ellipticity but assume circular beams with beamwidth errors bars that do account for this difference. Multifrequency beam patterns estimated from observations of the Moon with the Q-band system and observations of the Eta Carina region with both systems are consistent with the far-field beam measurements. Far-field multifrequency two-dimensional beam maps of the far side lobes have not been made. Based on the one-dimensional far-field measurements of the beam patterns, the main beam efficiency is estimated to be $> 98\%$. The $< 2\%$ loss to the beam side lobes is not corrected for in this analysis. A more detailed discussion of these issues may be found in Gundersen (1995).

The Ka-band ($26 - 36$ GHz) is multiplexed into four channels centered at $\nu = 27.25$, 29.75, 32.25, and 34.75 GHz, with 3 dB bandwidths of 2.5 GHz. The measurements described above indicate that the individual Ka channels have a frequency-dependent gaussian beamwidth $\sigma_G^{(\mathrm{Ka})} = (0.70 \pm 0.04)^\circ \times [27.7\ \mathrm{GHz}/\nu]$ ($1\sigma$ error). The Q-band ($38 - 45$ GHz) is multiplexed into three equal-width channels centered at $\nu = 39.15$, 41.45, and 43.75 GHz with 3 dB bandwidths of 2.3 GHz and a frequency-dependent gaussian beamwidth $\sigma_G^{(\mathrm{Q})} = (0.47 \pm 0.04)^\circ \times [41.5\ \mathrm{GHz}/\nu]$ ($1\sigma$ error). The measured passband central frequency errors are $\sim 1\%$, and are ignored in our analysis. Far-field measurements of the Ka- and Q-band beamwidths were performed at 27.7 and 41.5 GHz respectively. We therefore assume that all four Ka-band beamwidth uncertainties and all three Q-band beamwidth uncertainties shift together. However, this is not necessarily the case, since the Ka and Q measurements were performed consecutively, with different horns, but with the same telescope optics.

The SP94 Q and Ka observations were performed consecutively. Since the multiplexing was done after amplification and initial filtering, the HEMT amplifiers and the atmosphere introduce significant intraband channel-channel correlations, which are accounted for in the analysis.

Unlike the stepped scans of SP89 and SP91, SP94 data were taken during smooth, azimuthal, constant declination, constant velocity scans extending $20^\circ$ on the sky. While observing, the beam was sinusoidally chopped with a half peak-to-peak chop amplitude of $1.5^\circ$ on the sky. The rms fractional uncertainty in the peak-to-peak chop angle was measured to be 0.04%, and is ignored in our analysis. The two-beam/single-difference data were binned into 43 bins, with a bin size of $(20/43)^\circ$ on the sky. As discussed in G95, the relative-pointing uncertainty "on the sky" is $\pm 0.02^\circ$ in azimuth and $\pm 0.05^\circ$ in elevation, and is ignored in our analysis here. The absolute-pointing uncertainty on the sky is $\pm 0.12^\circ$ in azimuth and in elevation (G95). The absolute-pointing uncertainty is irrelevant for either the Ka- or Q-scan data analyses, but could be a significant issue when the Ka and Q data are combined for the Ka + Q analysis, or when comparing the results of the Ka and Q analyses. However, for reasons discussed below, this is ignored in the analysis here.

The G95 data were taken in smooth scans of $\pm 10^\circ$ on the sky centered at $\alpha = 45^\circ$ and $\delta = -61.8^\circ$ (1994), with a total sky coverage $\sim 20$ deg$^2$. The data editing, as described in G95, resulted in the removal of $\sim 25\%$ of the data — less for Ka, more for Q — and left 88 h and 108 h of usable Ka- and Q-band data respectively. After editing, an offset and linear gradient were removed from the data for each channel and each scan (of angular extent $20^\circ$ on the sky; we note that the SP group sometimes calls this a half-scan), and for each channel the binned data from the individual scans were coadded (G95). The data were also corrected for atmospheric absorption (G95). This is the reduced data we analyze in this paper.

While in this analysis we assume that the SP94 data is purely CMB anisotropy, it is prudent to bear in mind that one cannot yet conclusively rule out a small amount of undetected, non-CMB contamination in the data. The G95 time subset check, which shows that when the data is divided into 4 contiguous time series, all 4 subsets and all combinations of 3 of the 4 subsets are mutually consistent, indicates that transient atmospheric phenomena are not a significant issue. G95 also show that the contamination expected from Galactic diffuse synchrotron emission, from the Sunyaev-Zel'dovich effect, and from Galactic 20 K dust emission are small compared to the signal they see. See Gundersen (1995) for a more detailed discussion. G95 could not, however, rule out non-CMB foreground discrete radio source contamination (also see Gundersen et al. 1996).

The Ka-band and Q-band radiometers were calibrated using a combination of different methods. Without the ACME telescope, the radiometer V/K were determined for an ambient temperature Eccosorb load and for a liquid nitrogen cold load. These were used in combination with sky zenith scans and an atmosphere model to determine the atmospheric contribution to the antenna temperature and to derive a sky zenith temperature. The sky zenith scans were performed with the radiometers coupled to the ACME telescope. The calibration of the radiometers, determined from the three different combinations of ambient load, cold load, and sky load, were found to have a channel to channel uncertainty of 3% and an absolute uncertainty of 10% ($1\sigma$). This determination of the calibration was consistent with that determined from a cryogenic termination that was used to produce a load similar to that of the sum of the atmosphere and the CMB. Measured microwave emission from the Moon was also consistent with a model of the Moon's emission (Keihm 1983) to within the $\sim 20\%$ uncertainty. Some CMB anisotropy experiments calibrate on a point source to determine V/Jy. SP94 does not, so for SP94 the beamwidth uncertainty does not influence the calibration uncertainty. In our analysis, the SP94 beamwidth and radiometer-calibration uncertainties are treated as independent sources of uncertainty. A more detailed discussion of these points may be found in Gundersen (1995). In what follows, we treat the absolute-radiometer-calibration uncertainty as an additional, purely statistical, uncertainty to be included in the likelihood analyses.

Besides accounting for beamwidth and calibration uncertainties, and using anisotropy spectra in observationally motivated cosmogonies, our analysis differs from that of G95 in two other respects. Unlike G95, we do not make use of data-weighted windows to quote limits, and we also use a different statistical prescription to determine limits from the probability density distribution functions. A more detailed discussion of these points is given in §2 below.

In §2 we outline the computational techniques used in our analysis. Our results and a discussion are in §3, and we conclude in §4.

## 2. Computation

The CMB spatial anisotropy temperature variations, $\delta T/T$, can be decomposed into spherical harmonics,

$$\frac{\delta T(\theta,\phi)}{T} = \sum_{l=2}^{\infty}\sum_{m=-l}^{l} a_{lm} Y_{lm}(\theta,\phi). \tag{1}$$

For gaussian models we may characterize the anisotropy by its power spectrum, $C_l$, which is defined in terms of the ensemble average

$$\langle a_{lm} a^{*}_{l'm'}\rangle = C_l\, \delta_{ll'}\delta_{mm'}. \tag{2}$$

One would like to use anisotropy observations to measure the power spectrum over the entire range of $l$ to which the experiment is sensitive. However, the broad width of the SP94 windows and the limited sky coverage prevents one from extracting each individual $C_l$. It is therefore necessary to assume a functional form for the $C_l$s over the range of $l$ which SP94 is sensitive to, with the overall normalization allowed to be a free parameter and with the shape of the $C_l$ allowed to depend on a small number of other parameters. One may then compare these $C_l$s to the spectrum of the SP94 data and determine the value of the normalization and of the shape parameters which best reproduce the data spectrum.

On large angular scales, when the effects of the pressure of the photon-baryon fluid and matter velocity perturbations at photon decoupling can be ignored, an approximation to the CMB anisotropy spectrum in the fiducial CDM model is provided by the flat CMB angular spectrum (Peebles 1982),

$$C_l = \frac{6C_2}{l(l+1)} = \frac{24\pi}{5}\frac{(Q_{\rm rms-PS}/T_0)^2}{l(l+1)}, \tag{3}$$

where $Q_{\rm rms-PS}$ is the corresponding quadrupole-moment amplitude of the model CMB anisotropy and $T_0$ is the CMB temperature now. This spectrum, normalized to best reproduce the two-year DMR sky maps is shown in Figure 1 (line labelled "Flat"; Górski et al. 1994). The DMR range for $Q_{\rm rms-PS}$ in this model is given in the last line of the third column of Table 12.

The fiducial CDM model is an Einstein-de Sitter model with gaussian, adiabatic, scale-invariant energy density perturbations (Harrison 1970; Peebles & Yu 1970; Zel'dovich 1972), standard recombination (Peebles 1993, §6), Hubble parameter $h = 0.5$, and baryon density parameter $\Omega_B = 0.0125h^{-2}$. Scale-invariant energy density perturbations are generated by quantum-mechanical fluctuations during an early epoch of inflation in a variety of spatially-flat inflation models (e.g., Fischler, Ratra, & Susskind 1985). In inflation models the small observed CMB anisotropy could be the consequence of the small ratio of the inflation epoch mass scale to the Planck mass (Ratra 1991, and references therein; also see Banks et al. 1995).

The spectrum for this model (O14 in Table 2), normalized to best reproduce the two-year DMR sky maps, is shown in Figure 1 (Górski et al. 1995). The range for $Q_{\rm rms-PS}$ allowed by the DMR and a subset of the SP94 data is given in the line labelled O14 of Table 12. Note that Models O14 and Λ12 have identical CMB anisotropy spectral shape, but are normalized using the slightly different DMR galactic- and ecliptic-coordinate maps. The fiducial CDM model is no longer thought to provide an adequate representation of the observed universe.

On the other hand, low-density open and flat-Λ CDM cosmogonies are consistent with present observational data. For historical reasons (Guth 1981, also see Kazanas 1980; Sato 1981a,b), no longer valid (Gott 1982; Guth & Weinberg 1983), the low-density Λ CDM models considered here are taken to have flat spatial sections, so the simplest power spectrum for gaussian, adiabatic, energy-density perturbations in these models is the scale-invariant one. Such a spectrum is generated by quantum zero-point fluctuations during an early epoch of inflation in a spatially-flat inflation cosmogony. The simplest spectrum for gaussian adiabatic energy-density perturbations consistent with open spatial sections is that generated by zero-point fluctuations during an early epoch of inflation in an open model (Ratra & Peebles 1994, 1995; Bucher, Goldhaber, & Turok 1995; Lyth & Woszczyna 1995; Yamamoto, Sasaki, & Tanaka 1995; Bucher & Turok 1995).

The values of the parameters $\Omega_0$, $h$, and $\Omega_B$ used here are chosen to be roughly consistent with present observational estimates of $\Omega_0$, $h$, the age of the universe, and the constraints on $\Omega_B$ that follow from the observed light element abundances in the standard nucleosynthesis model (RSBG). Two standard-recombination open and two standard-recombination flat-Λ spectra are shown in Figure 1.

In this analysis we wish to draw conclusions about parameters representing effects that must exist; i.e., $\Omega_0$, $h$, and $\Omega_B$. To simplify the comparison to SP94 data, the effects of tilt, primordial gravity waves, and reionization are ignored. Tilt, gravity waves, and early reionization are unlikely to be significant in viable open models, although there are mild indications that some such effect might be required to bring some flat-Λ CDM models into agreement with observational data (Stompor et al. 1995; Scott et al. 1995; Ostriker & Steinhardt 1995; RSBG; Klypin, Primack, & Holtzman 1996; GRS). We emphasize that any effect which significantly modifies the shape of the spectra used in our analysis here, such as isocurvature perturbations or nongaussian anisotropy, will affect the final results.

The general procedure for the computation of the CMB spectra is discussed by Sugiyama (1995). A discussion of the various assumptions and approximations involved in such a computation, as well as a discussion of the resulting accuracy of the computation, may be found in Hu et al. (1995).

Figure 2 shows the zero-lag windows, $W_l$, for the seven individual SP94 channels, at the nominal beamwidths. Figure 2 also shows the uncertainties due to the $1\sigma$ beamwidth uncertainty in the highest and lowest frequency channels.

It has become conventional to summarize anisotropy experiment windows in terms of a few parameters (e.g., Bond 1996). These are the value of $l$ where $W_l$ is greatest, $l_{\rm m}$, the two values of $l$, $l_{e^{-0.5}}$, where $W_{l_{e^{-0.5}}} = e^{-0.5} W_{l_{\rm m}}$, and the effective multipole $l_{\rm e} = I(lW_l)/I(W_l)$, where

$$I(W_l) = \sum_{l=2}^{\infty} \frac{(l+0.5)W_l}{l(l+1)}. \tag{4}$$

The values of these parameters for each of the seven channels are given in Table 1, both for the nominal beamwidths, and for beamwidths $1\sigma$ smaller and larger than the nominal ones.[6]

The range of multipole moments to which an anisotropy experiment is sensitive depends on both the window $W_l$ and the sky signal $C_l$. Depending on the form of the sky signal, the window parameters $l_{\rm e}$, $l_{\rm m}$, and $l_{e^{-0.5}}$ do not necessarily give a good indication of the multipoles to which an experiment is sensitive.

CMB anisotropy experiments are sensitive to $\delta T_{\rm rms}$, the rms of the anisotropy seen through their window. Here

$$(\delta T_{\rm rms})^2 = \sum_{l=2}^{\infty} ({\delta T_{\rm rms}}^2)_l, \tag{5}$$

where

$$({\delta T_{\rm rms}}^2)_l = {T_0}^2 \frac{(2l+1)}{4\pi} C_l W_l, \tag{6}$$

and $T_0 = 2.726 \pm 0.010$ K (Mather et al. 1994). Given an anisotropy model, $({\delta T_{\rm rms}}^2)_l$ provides a convenient way of establishing the approximate range of multipoles $l$ an experiment is sensitive to. It is approximate since it ignores spatial correlation information. In Figure 3 we plot $({\delta T_{\rm rms}}^2)_l$ for the models shown in Figure 1, for two SP94 window functions. In Table 2 we list $l_{\rm m}$, the value of $l$ at which $({\delta T_{\rm rms}}^2)_l$ is at a maximum, and $l_{e^{-0.5}}$ the two multipoles where $(\delta T_{\rm rms}{}^2)_{l_{e^{-0.5}}} = e^{-0.5}(\delta T_{\rm rms}{}^2)_{l_{\rm m}}$, for these two windows, and for all the models we consider here.

---

[6]Note that in our internal computations these and other numerical values are not truncated at 2 or 3 significant figures as in the tables; as a consequence, when the truncated numerical values of the tables are used to rederive some of our results there will be small differences from the values listed in the tables.

The range of multipole moments to which each channel is sensitive is quite model dependent, and this range is not adequately summarized by the $W_l$ parameters $l_{\rm e}$ and $l_{e^{-0.5}}$. For the spectra considered here, the SP94 experiment is sensitive to a somewhat larger angular scale than the effective angular scale determined by the $W_l$ parameter $l_{\rm e}$. In this case, however, the $(\delta T_{\rm rms}{}^2)_l$ parameters $l_{\rm m}$ and $l_{e^{-0.5}}$ do provide an adequate, model-dependent, characterization of the range of $l$ to which the SP94 experiment is sensitive. This is not true in general, and only works here because the models considered here have spectra that are fairly smooth in the relevant range of $l$-space.

In what follows, we shall have need for the bandtemperature (e.g., Bond 1996),

$$\delta T_l = \frac{\delta T_{\rm rms}}{\sqrt{I(W_l)}}. \tag{7}$$

Note that despite the notation, $\delta T_l$ does not depend explicitly on $l$. Compared to $\delta T_{\rm rms}$, $\delta T_l$ has the advantage of being insensitive to the normalization of $W_l$. As we shall see below, it is still fairly sensitive to the precise shape of $W_l$.

The reduced SP94 data used in the analysis of G95 are shown in Figure 4. In the line labelled "Sky" in Table 3, we give the estimated rms of the anisotropy, computed from the data of Figure 4 as the square root of the difference between the variance of the mean temperatures and the variance of the error bars. As this ignores off-diagonal noise and spatial correlations, as well as the oversampling of points on the sky[7], which are all accounted for in the likelihood analysis, it could deviate from the true sky rms.

For the purpose of the following discussion we assume that the SP94 sky signal is purely CMB anisotropy without any offset or gradient, which would have been removed while fitting out an experimental drift. Following Bond et al. (1991), the likelihood function at a given beamwidth and nominal calibration (hereafter, the "bare" likelihood function) for a given model $C_l$ is computed according to

$$L(C_l) \propto \frac{1}{\sqrt{\det(\mathbf{f}^T\mathbf{M}^{-1}\mathbf{f})\det(\mathbf{M})}} e^{-\chi^2/2}, \tag{8}$$

where

$$\chi^2 = \mathbf{\Delta}^T(\mathbf{M}^{-1} - \mathbf{M}^{-1}\mathbf{f}(\mathbf{f}^T\mathbf{M}^{-1}\mathbf{f})^{-1}\mathbf{f}^T\mathbf{M}^{-1^T})\mathbf{\Delta}. \tag{9}$$

[7]The SP94 FWHM beamwidths are significantly larger than the data-bin-separation on the sky.

Here $\mathbf{\Delta}$ is the vector of temperature deviations (with $N = 43 \times 7$ elements for the full data set, for example) and $\mathbf{M} = \mathbf{C} + \mathbf{\Sigma}$ is the correlation matrix of the data in the context of the model considered. $\mathbf{C}$ represents correlations arising from the specific model considered and can be computed using eq. (2) of G95, while $\mathbf{\Sigma}$ represents instrumental and atmospheric noise. The unwieldy term in the expression for $\chi^2$ arises from assuming a uniform prior in the amplitude of the drifts and marginalizing over all possible values of offset and gradient. $\mathbf{f}$ is the array of functions that have been fit out of the data, in this case an offset and gradient from each channel; it is thus an $N \times 2$ matrix for the full data set. Here, because the bin weights for each channel are fairly uniform, the difference between using unweighted fitting functions and weighted ones, as recommended by Bunn et al. (1994), is small. This would not be the case, however, for experiments such as FIRS which do not uniformly sample the area covered. As discussed above, we parametrize the model $C_l$ using $Q_{\rm rms-PS}$, $\Omega_0$, $h$, and $\Omega_B$.

With this prescription, for a given data set, all models yield the same likelihood at $Q_{\rm rms-PS} = 0\ \mu$K, i.e., no sky signal. This provides a convenient way to normalize across models.

In Table 3, the lines labelled "FBP" are central $\delta T_{\rm rms}$ values derived from likelihood analyses using the flat bandpower (FBP) spectrum (eq. [3]) at the nominal value of the beamwidths for each channel.

To derive the $Q_{\rm rms-PS}$ central value and limits from the likelihood function for a given $C_l$ and data set, we adopt the following prescription (Berger 1985, p. 140; Myers, Readhead, & Lawrence 1993, §3.1; Górski et al. 1996).

Bayes's theorem states that the posterior probability density distribution is proportional to the likelihood function multiplied by the prior probability. Lacking good information to the contrary, here we assume a uniform prior in $Q_{\rm rms-PS}(\geq 0)$, resulting in a posterior probability density distribution equal to the likelihood function.

The central value is taken to be the value of $Q_{\rm rms-PS}$ at which the probability density distribution peaks. The $2\sigma$ limits, $Q_{-2\sigma}$ and $Q_{+2\sigma}$, are defined as the two values of $Q_{\rm rms-PS}$ for which

$$\int_{Q_{-2\sigma}}^{Q_{+2\sigma}} L\, dQ_{\rm rms-PS} = 0.9545 \int_0^\infty L\, dQ_{\rm rms-PS} \tag{10}$$

and for which $Q_{+2\sigma} - Q_{-2\sigma}$ is minimized. This is the highest posterior density prescription, hereafter HPD. If $Q_{-2\sigma}$ is greater than $0\ \mu$K, we say that there is a ($2\sigma$) detection; we then determine $Q_{+1\sigma}$ and $Q_{-1\sigma}$ analogously. If instead $Q_{-2\sigma}$ is zero, we say that there is no detection and integrate the probability density function over $Q_{\rm rms-PS}$, starting from $0\ \mu$K until we get to the value of $Q_{\rm rms-PS}$ that includes $1 - 0.5(1 - 0.9545) = 97.72\%$ of the total area under the probability density function and call this value of $Q_{\rm rms-PS}$ the $2\sigma$ upper limit. This is the equal tail prescription, hereafter ET. Of course, the choice of how to define limits depends on what one wishes to use the data for, and will hopefully not be a significant issue when the data improves. The definition we choose to adopt yields smaller lower limits for detections and yields larger upper limits for nondetections. It also leads to the tightest constraint on $Q_{\rm rms-PS}$ and therefore provides for greater discrimination between models.

In all cases we computed limits up to $\pm 3\sigma$ (99.73% HPD). We do not quote them here, but emphasize that even accounting for calibration and beamwidth uncertainties, the Ka, Q, and Ka + Q data subsets show evidence for at least $3\sigma$ detections.

G95 quote $\pm 1\sigma$ ET limits for the Ka, Q, and Ka + Q data. Their prescription differs from that adopted here. In Table 4 we give the G95 central values and $\pm 1\sigma$ ET limits along with our central values and $\pm 1\sigma$ limits computed using both the G95 ET prescription and the HPD prescription. These are given in terms of bandtemperature for the combined Ka, Q, and Ka + Q data, extracted using the flat bandpower angular spectrum. They are calculated ignoring calibration and beamwidth uncertainties. The last two columns of Table 4 give the average of the $\pm 1\sigma$ error bars in $\mu$K and as a percentage of the central value.

We now want to account for the additional uncertainties induced by the radiometer-calibration and beamwidth uncertainties.

Most experiments quote a $1\sigma$ uncertainty in the widths of their beams. We assume that the probability distribution for the beamwidth, $b$, is gaussian,

$$P(b) = \frac{1}{\sqrt{2\pi}\,\sigma_b} e^{-(b-b_0)^2/(2{\sigma_b}^2)}, \tag{11}$$

for $b > 0$,where $b_0$ is the nominal beamwidth and $\sigma_b$ is the uncertainty in the beamwidth. The likelihood function marginalized over beamwidth uncertainty is then

$$L(Q_{\rm rms-PS}) = \int_0^\infty db\, P(b)\, L(Q_{\rm rms-PS}, b), \tag{12}$$

where $L(Q_{\rm rms-PS}, b)$ is the bare likelihood function derived if a beamwidth of $b$ is assumed. To evaluate the expression in eq. (12), we determine a select few beamwidths at which to compute the likelihood function and use Gauss-Hermite quadrature summation to approximate the integral. In this case we used three point quadrature (e.g., Press et al. 1992), which will be accurate so long as $\sigma_b \ll b_0$.

As discussed above, we assume that the beamwidth uncertainties for all channels shift together. This is justified for the individual Ka channels and for the individual Q channels.

However, since the Ka-band and Q-band measurements were performed consecutively, using different horns and the same telescope, the Ka and Q beamwidth uncertainties do not necessarily shift together. Since we have analyzed this data set assuming that the Ka- and Q-band beamwidth uncertainties shift together, in this case our limits are somewhat overconservative. Our neglect of the absolute-pointing uncertainty when combining the Ka- and Q-scan data sets (for the Ka + Q analysis) compensates approximately for this effect. We note that a proper accounting of the absolute-pointing uncertainty would require analyses of the data with the Ka- and Q-scan bins offset on the sky. These need to be performed at various values of the offset, and one then "marginalizes" over the offset. Other uncertainties, such as that in the chop amplitude, which is exceedingly small for SP94 and is ignored here, may also be accounted for by a straightforward generalization of our method.

In passing, we note that the technique used by G95 to measure their beamwidths simplifies the issue of accounting for the beamwidth uncertainty. If the beamwidths of the different channels are measured at different frequencies, or were found not to factorize into a frequency-dependent term and a frequency-independent term, it would be a slightly more complex matter to properly account for the beamwidth uncertainty.

Most experiments quote a $1\sigma$ fractional uncertainty in their calibration. We assume that the probability distribution for the calibration, $C$, is gaussian:

$$P(C) = \frac{1}{\sqrt{2\pi}\,\sigma_C} e^{-(C-1)^2/(2{\sigma_C}^2)}, \tag{13}$$

for $C > 0$, where $\sigma_C$ is the $1\sigma$ fractional uncertainty in the calibration, 0.1 in this case, and $C$ is greater than 0.

If $C$ is not equal to 1, then the "real" value of $Q_{\rm rms-PS}$ is related to the "bare" value of $Q_{\rm rms-PS}$, which we will hereafter refer to as $Q'_{\rm rms-PS}$, through

$$Q_{\rm rms-PS} = \frac{Q'_{\rm rms-PS}}{C}. \tag{14}$$

The likelihood function marginalized over calibration uncertainty is then

$$L(Q_{\rm rms-PS}) = \int_0^\infty dC\, P(C) L(Q'_{\rm rms-PS}). \tag{15}$$

With eqs. (13) and (14), this implies

$$\begin{aligned} L(Q_{\rm rms-PS}) \;&=\; \frac{1}{\sqrt{2\pi}\,\sigma_C Q_{\rm rms-PS}} \\ &\times \int_0^\infty dQ'_{\rm rms-PS}\, e^{-(Q'_{\rm rms-PS}-Q_{\rm rms-PS})^2/[2(\sigma_C Q_{\rm rms-PS})^2]} L(Q'_{\rm rms-PS}). \end{aligned} \tag{16}$$

In passing we note that the factor of $Q_{\rm rms-PS}$ in the denominator of the prefactor on the right-hand side of this equation makes the likelihood marginalized over $Q_{\rm rms-PS}$ logarithmically divergent at large $Q_{\rm rms-PS}$. Since this occurs at extremely large values of $Q_{\rm rms-PS}$, this is not an issue that need bother us here, and is most likely due to the invalidity of our assumption (in eq. [13]), that there is a nonzero probability for $C = 0$.

Some intuition about eq. (16) may be gotten by studying how it transforms simple analytical "bare" likelihoods. If there is no noise and sample variance, and the "bare" likelihood indicates a perfect, Dirac delta, detection at $Q_0$,

$$L(Q'_{\rm rms-PS}) = \delta(Q'_{\rm rms-PS} - Q_0), \tag{17}$$

then from eq. (16),

$$L(Q_{\rm rms-PS}) = \frac{1}{\sqrt{2\pi}\,\sigma_C Q_{\rm rms-PS}} e^{-(Q_{\rm rms-PS}-Q_0)^2/[2(\sigma_C Q_{\rm rms-PS})^2]}; \tag{18}$$

i.e., the calibration uncertainty skews, and broadens the "bare" likelihood, and $L(Q_{\rm rms-PS})$ has an amplitude ($Q_{\rm rms-PS}$) dependent width. If on the other hand, one has a good "bare" detection at $Q_0$, for which we assume a gaussian form for the "bare" likelihood,

$$L(Q'_{\rm rms-PS}) = \frac{1}{\sqrt{2\pi}\,\sigma_0} e^{-(Q'_{\rm rms-PS}-Q_0)^2/(2{\sigma_0}^2)}, \tag{19}$$

then in the limit when $\sigma_0 \ll Q_0$ (and assuming that $\sigma_0 \not\gg \sigma_C Q_{\rm rms-PS}$ and that $\sigma_C Q_{\rm rms-PS} \not\gg Q_0$) it may be shown that eq. (16) results in

$$L(Q_{\rm rms-PS}) = \frac{1}{\sqrt{2\pi\{(\sigma_C Q_{\rm rms-PS})^2 + {\sigma_0}^2\}}} e^{-(Q_{\rm rms-PS}-Q_0)^2/[2\{(\sigma_C Q_{\rm rms-PS})^2+{\sigma_0}^2\}]}; \tag{20}$$

i.e., the calibration uncertainty again skews and broadens the "bare" likelihood function, and the calibration-uncertainty-corrected likelihood has an amplitude-dependent width. This is what is intuitively expected, and differs a little from the usual naive prescription for accounting for the fractional calibration uncertainty by adding it in quadrature to the error bars derived from the "bare" likelihood function (e.g., RSBG).

Although the beamwidth and calibration probability distributions are not known for any experiment, the gaussian assumptions (eqs. [11] & [13]) are probably quite reasonable, and we use these in what follows. It is important, however, to bear in mind that this is an assumption which could be checked by direct measurements.

Finally, we estimate the anisotropy spectral index $\beta$ (Wollack et al. 1993) using both the estimated SP94 "sky" rms, and a flat bandpower model spectrum (eq. [3]) with the

SP94 $W_l$. For a frequency spectrum $\delta T_{\rm rms}(\nu) \propto \nu^\beta$, we have

$$\beta = \ln\left[\frac{\delta T_{\rm rms}(\nu_1)}{\delta T_{\rm rms}(\nu_2)}\right]\left[\ln\left(\frac{\nu_1}{\nu_2}\right)\right]^{-1}. \tag{21}$$

For a flat bandpower model spectrum, this may be rewritten as

$$\beta = \ln\left[\sqrt{\frac{I(W_l[\nu_1])}{I(W_l[\nu_2])}}\right]\left[\ln\left(\frac{\nu_1}{\nu_2}\right)\right]^{-1}, \tag{22}$$

where $I(W_l)$ is defined in eq. (4), and we have used eq. (7) and the fact that for a flat bandpower angular spectrum $\delta T_l = \sqrt{12/5}\, Q_{\rm rms-PS}$. Such a computation makes sense for SP94 since the bandpass filters are fairly well modelled as narrow tophats. The central values of $\beta$ are given in Table 5 for the nominal beamwidth SP94 windows (these computations use the values for $\nu$ and $\sqrt{I(W_l)}$ for the nominal $W_l$s given in Table 1, and the values in the row labelled "Sky" of Table 3). The $\beta$ values estimated using eq. (22) do not account for off-diagonal noise and spatial correlations, nor for the oversampling of points.

In Table 6 we give bandtemperature central values and both ET and HPD $\pm 1\sigma$ limits derived from flat bandpower likelihood analyses of the individual-channel SP94 data sets. These ignore beamwidth and calibration uncertainty. The last two columns of Table 6 give the average of the $\pm 1\sigma$ error bars, in $\mu$K and as a percentage of the central values.

In Table 7 we give the corresponding numbers derived from flat bandpower likelihood analyses of the combined Ka, Q, and (full) Ka + Q data subsets, now accounting for calibration and beamwidth uncertainties using two prescriptions: (1) that suggested by one of us (JOG) to account for these uncertainties by adding 15% of the central value in quadrature to the G95 ET error bars, which is the de facto "standard" prescription used in the CMB anisotropy field; and (2) that described above to first correct the likelihood functions for these uncertainties, and then to use these corrected likelihood functions to derive the relevant numbers.

Central values and limits for $Q_{\rm rms-PS}$, both ignoring and accounting for beamwidth and calibration uncertainties, are given in Tables 8 and 9 for the Ka2, Q2, Ka, and Ka + Q data sets. Some of the likelihood functions used in the derivation of these numerical values are shown in Figs. 5. In Fig. 6 we compare the calibration- and beamwidth-uncertainty corrected likelihood functions for the selected models. These data sets were chosen for illustrative purposes only; the Ka2 data does not have a detection, and the Q2 data has one of the best detections.

In Tables 10 and 11 we list the values of the probability density distribution functions at the peak, and the marginalized (over $Q_{\rm rms-PS}$) probability density distribution values[8], both ignoring and accounting for beamwidth and calibration uncertainties, for the Ka + Q, Ka, Ka2, and Q2 data sets. The models shown in the Figures were chosen on the basis of their marginal probability distribution values for the Ka + Q data set, accounting for beamwidth and calibration uncertainties. For all the combined SP94 data sets, and some of the individual-channel ones as well, the likelihood functions are peaked, and well-separated from 0 $\mu$K. See Figs. 5 and 6. For instance, for the Ka + Q data set the amplitude of the likelihood function at the peak is approximately $5 \cdot 10^{19}$ times larger than that at 0 $\mu$K. For such likelihood functions it makes sense to choose between models on the basis of the value of the marginal probability distribution function. Model O1 (an $\Omega_0 = 0.1$ open model) is the most likely model, followed, amongst our selected models, by Flat (flat bandpower), models O11 and $\Lambda$2 (an open $\Omega_0 = 0.5$ model and a spatially-flat $\Omega_0 = 0.2$ model), O14 (fiducial CDM), and, the least likely one, $\Lambda$10 (a flat-$\Lambda$ $\Omega_0 = 0.4$ model). These selected models include the most and least likely open and flat-$\Lambda$ ones amongst the models we consider here.

Since both the open and flat-$\Lambda$ model CMB spatial anisotropy spectra shapes depend on $\Omega_0$, $h$, and $\Omega_B$, the 14 open and 12 flat-$\Lambda$ model spectra we use in the analyses here do not provide enough resolution in model-parameter space for us to be able to marginalize over these parameters to extract most likely values for each separately.

In Tables 12 and 13 we list central values and limits for $Q_{\rm rms-PS}$ for all the SP94 data sets. These account for beamwidth and calibration uncertainties. These tables also give the corresponding two-year DMR results, accounting for both statistical and systematic DMR uncertainties (Stompor et al. 1995, also see RSBG). Table 14 shows $\delta T_l$, computed from the $Q_{\rm rms-PS}$ values of Tables 12 and 13 using eqs. (5) – (7). Except for the flat bandpower model, it is not possible to compute $\delta T_l$ for the Ka, Q, and Ka + Q data sets (since they are combinations of data from different windows). In Tables 15 and 16 we list the values of the probability density distribution functions at the peak and the marginal probability density distribution function values, for all the data sets, accounting for beamwidth- and calibration-uncertainty corrections.

Figures 7 show selected model spectra normalized to the SP94 (Ka, Q, and Ka + Q) and DMR data sets, as well as the corresponding individual-channel SP94 results.

---

[8]The marginalized probability density distribution function is the probability density distribution function for just the cosmological parameters $\Omega_0$, $h$, and $\Omega_B$. In what follows we usually call this the “marginal probability distribution function”.

## 3. Results and Discussion

From Table 3 we see that the values of the rms estimated from likelihood analyses using flat bandpower spectra agree quite well with the value of the "sky" rms estimated from the data. For four channels they agree to better than 5% and for six channels they agree to better than 25%. This gives some indication of the importance of the off-diagonal noise, spatial correlations and oversampling of points, and provides semiquantitative confirmation of the likelihood method results. The Ka4 channel "sky" rms is $\sim 50\%$ larger than that derived from the likelihood analysis. The data for this channel does look somewhat different from that for the other channels [see the lowest panel of Fig. 4(a)]. To examine whether this can significantly affect the results deduced from the likelihood analyses, we have reanalyzed the combined Ka and Ka + Q data subsets with the Ka4 channel excluded from the analyses. When the Ka4 channel data is dropped from the Ka analysis, the central $Q_{\rm rms-PS}$ values rise by $\sim 3-6\%$ (depending on model), and the $1\sigma$ error range broadens by $\sim 5-8\%$, depending on the model. When it is excluded from the Ka + Q analysis, the central values drop a little and the $1\sigma$ error ranges slightly tighten. These changes are not big enough to warrant dropping the Ka4 channel data, so the numbers we quote here are from the full analyses, including the Ka4 data.

In Table 4 we compare the central values and $\pm 1\sigma$ limits on bandtemperature derived from likelihood analyses with the flat bandpower spectrum. The numbers are the 16.0% and 84.0% ET limits derived by G95, and both the 15.87% and 84.13% ET as well as the 68.27% HPD limits found here for the Ka, Q, and Ka + Q data sets ignoring beamwidth and calibration uncertainty. The computations described here use more accurate window functions than those used by G95; they differ by $\sim 0.5-2\%$. Scaling from analyses done at three different beamwidths, we estimate that these window function differences and the slight differences in ET limit definitions are probably responsible for most of the difference between the $\delta T_l$ values computed here and in G95.

Table 4 also gives the 68.27% HPD $\pm 1\sigma$ limits. We adopt this prescription to quote limits for detections, since it results in a more conservative $-1\sigma$ limit than does the corresponding ET prescription. It also leads to slightly tighter average constraints on the bandtemperature, and so provides for slightly greater discrimination between models. Both these effects are evident in the numbers of Table 4: comparing, e.g., the third and fourth rows, we see that the HPD prescription results in somewhat smaller upper error bars and somewhat larger lower error bars than does the ET prescription.

Unlike the Saskatoon (SK) experiments (Wollack et al. 1993; Netterfield et al. 1995, hereafter N95; Netterfield 1995; Netterfield et al. 1997) where the shape of the window functions are frequency-independent, the SP94 window functions are frequency-dependent.

As this couples spatial ($l$) and frequency ($\nu$) information, some care is needed when interpreting the results of a $\beta$ analysis. In the lower left-hand triangle of Table 5 we list the $\beta$s derived by comparing the "sky" rmses of each pair of channels. To estimate combined-channel $\beta$s we ignore the Ka4 channel, since we know that the "sky" rms for this channel is significantly higher than that estimated from the likelihood analyses, and average over all the other $\beta$s. We find $\beta_{\rm Ka} \approx -0.2$, $\beta_{\rm Q} \approx 2.2$, and $\beta_{\rm Ka+Q} \approx 0.9$. These should be compared to those derived in G95 from likelihood analyses using the flat bandpower spectrum: $\beta_{\rm Ka} = 0.2^{+0.9}_{-1.4}$, $\beta_{\rm Q} = 1.7^{+1.5}_{-1.6}$, and $\beta_{\rm Ka+Q} = 0.9^{+0.3}_{-0.6}$. Our approximate estimates of the $\beta$ central values reproduce the more accurate estimates of G95 to within $\sigma/3$, where the $\sigma$ are the G95 error bars on the $\beta$ values. This agreement gives some indication of the importance of off-diagonal noise and spatial correlations, as well as of the effect of oversampling.

The upper right-hand triangle of Table 5 gives the $\beta$s which should be found if the sky were described by a flat bandpower CMB spectrum. Averaging as before, but this time including the Ka4 channel, we find $\beta_{\rm Ka} \approx 0.5$, $\beta_{\rm Q} \approx 0.4$, and $\beta_{\rm Ka+Q} \approx 0.4$. A similar analysis was carried through for all the other model CMB angular spectra used here, with the result that $\beta_{\rm Ka} \sim 0.5 - 0.8$, $\beta_{\rm Q} \sim 0.4 - 0.6$, and $\beta_{\rm Ka+Q} \sim 0.4 - 0.7$, with the steeper slopes corresponding to the spatially-flat $\Lambda$ CDM models. That is, since the SP94 window functions couple spatial and frequency information one does not expect to find $\beta = 0$ even for a pure CMB spectrum. Our approximate $\beta$s agree with the values found by G95 using flat bandpower likelihood analyses to $\sim \sigma/3$ for the Ka data and to $\sim 0.8\sigma$ for the Q and Ka + Q data. We therefore conclude that the G95 data could be even more consistent with a CMB spectrum than was indicated there. Since the SP94 $W_l$ couple $l$- and $\nu$-space, the only way to see how consistent the G95 $\beta$ values are with a CMB anisotropy spectrum is to use a model for both the spatial ($l$) and frequency ($\nu$) dependence of the assumed non-CMB foregrounds. While not statistically significant at even the $1\sigma$ level it is noted that for all CMB spectra we consider $\beta_{\rm Q} < \beta_{\rm Ka}$.

As shown in the second last column of Table 4, the deduced Ka + Q data average absolute uncertainties are larger than those for the Ka data and only slightly smaller than those for the Q data. Provided that there is no significant non-CMB contamination in one of the bands and that the absolute-pointing uncertainty can be ignored, these error bars only account for instrumental and atmospheric noise, which must integrate down, and sample variance due to the limited number of independent pixels of data, which will not integrate down without more, independent, sky coverage.

To investigate the behaviour of the error bars, in Table 6 we list the bandtemperature central values and ET and HPD limits derived for each channel using a flat bandpower spectrum. The Q1 channel data has a detection for flat bandpower, as well as for some of the low-density open model CMB spectra. However, accounting for calibration and beamwidth uncertainties, the flat bandpower model $-2\sigma$ HPD $\delta T_l$ limit is $\sim 3\mu$K, so it is not a robust detection. The Ka2 channel does not have a detection, independent of the assumed CMB anisotropy spectrum.

We may approximately model the total average absolute error bars for the combined Ka and Q data sets of Table 4, and for the individual channel data sets of Table 6, using

$$\begin{aligned} \sigma_{\rm tot.,com.} &= \sqrt{\sigma_{\rm SV}{}^2 + \sigma_{\rm N}{}^2/N_I} \\ \sigma_{\rm tot.,ind.} &= \sqrt{\sigma_{\rm SV}{}^2 + \sigma_{\rm N}{}^2}, \end{aligned} \qquad (23)$$

where $\sigma_{\rm SV}$ is the uncertainty due to sample variance, $\sigma_{\rm N}$ that due to intrinsic noise, and $N_I$ is the number of channels that contribute to the corresponding combined data set. We emphasize that the model of eqs. (23) is approximate, since it neglects correlations between channels.

To determine $\sigma^{\rm A}_{\rm tot.,ind.}$, where A is either Ka or Q, we use $\sigma^{\rm A}_{\rm tot.,ind.} = [\sum_{i=1}^{N_I}(\sigma^{{\rm A}i})^2/N_I]^{1/2}$, where the $\sigma^{{\rm A}i}$ are the individual-channel HPD $\delta T_l$ average absolute error bars of Table 6. With robust detections only, $\sigma^{\rm Ka}_{\rm tot.,ind.} = 14.6\ \mu$K ($N_I = 3$) and $\sigma^{\rm Q}_{\rm tot.,ind.} = 12.1\ \mu$K ($N_I = 2$). With all channels, $\sigma^{\rm Ka}_{\rm tot.,ind.} = 13.8\ \mu$K ($N_I = 4$) and $\sigma^{\rm Q}_{\rm tot.,ind.} = 12.3\ \mu$K ($N_I = 3$).

Using these numbers and the average HPD $\delta T_l$ error bars of Table 4, 10.1 $\mu$K for Ka and 11.7 $\mu$K for Q, to explain the behaviour of the Ka and Q data error bars the sample variance and intrinsic noise contributions to the $\delta T_l$ error bars need to be

$$\begin{aligned} \sigma^{\rm Ka}_{\rm SV} &= 6.8(8.5)\ \mu{\rm K}, \qquad \sigma^{\rm Ka}_{\rm N} = 13(11)\ \mu{\rm K}, \\ \sigma^{\rm Q}_{\rm SV} &= 11(11)\ \mu{\rm K}, \qquad \sigma^{\rm Q}_{\rm N} = 4.4(4.6)\ \mu{\rm K}. \end{aligned} \qquad (24)$$

Here the term in parentheses on the right hand side of each equation is that determined using all channels, and the first terms are those determined using just the robust detection channels. This approximate estimate suggests that the Q data total error bars are dominated by sample variance, the Ka data total error bars are dominated by intrinsic noise, and that the Q data is intrinsically less noisy than the Ka data. The behaviour of the error bars (in Table 4) when the Ka and Q data are combined to form the Ka + Q data set then does not seem very surprising.

It is possible to simply estimate the sample variance error bars, though a more accurate estimate would make use of simulations, and would be somewhat model dependent. We first illustrate the method using the SK93 overlap part of the SK94 experiment (N95), since N95 have performed the requisite simulations and we can use their result to judge how accurate the estimate is.

Most of the weight of the SK93 overlap data comes from the Ka-band observations (L. Page, private communication 1995), which were performed with a beamwidth $\sigma^{(\rm Ka)}_{\rm FWHM} = 1.4^\circ$. The beam centre traces out a circle of radius $4.4^\circ$ centered at the NCP, so the experiment has $N_{\rm pix} = (2\pi \times 4.4^\circ)/1.4^\circ = 20$ pixels (the data is binned into 24 bins, so the SK93 overlap experiment somewhat oversamples). The sample variance error is $\approx 1/\sqrt{2N_{\rm pix}} = 16\%$. From the $\rm K_a$93 entry in Table 1 of N95, $\Delta_{\rm rms} = 37\ \mu$K, so the approximate estimate of the sample variance error bar is 5.9 $\mu$K, which should be compared to the N95 result of 6.5 $\mu$K (last paragraph on p. L71; note that this is the sample variance contribution to the $\delta T_{\rm rms}$ error bars). The simple analytical estimate of sample variance needs to be increased by 10% to reproduce the N95 result.

Although there are a number of differences between the SK93 overlap and SP94 experiments, the analytical method to estimate sample variance should be fairly accurate for SP94. At the central Ka-band and Q-band SP94 frequencies, the beamwidths are $\sigma^{(\rm Ka)}_{\rm FWHM} = 1.47^\circ$ and $\sigma^{(\rm Q)}_{\rm FWHM} = 1.11^\circ$, so for a $20^\circ$ scan $N^{(\rm Ka)}_{\rm pix} = 13.6$ and $N^{(\rm Q)}_{\rm pix} = 18.0$, so SP94 is more oversampled than the SK93 overlap. Accounting for the 10% correction, the SP94 Ka sample variance is 21.1%, and the Q sample variance is 18.4%. Using the Ka and Q $\delta T_l$ HPD central values of Table 4, 30.4 $\mu$K and 41.4 $\mu$K respectively, and eqs. (23), we find for the sample variance and intrinsic noise contributions to the SP94 $\delta T_l$ average error bars,

$$\begin{aligned} \sigma^{\rm Ka}_{\rm SV} &= 6.4(6.4)\ \mu{\rm K}, \qquad & \sigma^{\rm Ka}_{\rm N} &= 13(12)\ \mu{\rm K}, \\ \sigma^{\rm Q}_{\rm SV} &= 7.6(7.6)\ \mu{\rm K}, \qquad & \sigma^{\rm Q}_{\rm N} &= 9.4(9.7)\ \mu{\rm K}, \end{aligned} \tag{25}$$

which should be compared to the estimates of eqs. (24). Again, the first term on the right hand side of each equation is that determined using just the robust detection channels, and the terms in parentheses use data from all channels. This approximate estimate of the Ka sample variance and intrinsic noise is close to what is required (eqs. [24]) to explain the behaviour of the Ka data error bars. However, the approximate estimate of the Q sample variance indicates that it is not as important (relative to the Q noise) as what is required (eqs. [24]) to explain the behaviour of the Q data error bars. Given the small sky coverage of SP94, this discrepancy might not be statistically significant. More data, with greater sky coverage, is needed to resolve this issue. However, it is clear that sample variance is a significant contributor to the SP94 error bars.

In Table 7 we give bandtemperature values for the combined Ka, Q and Ka + Q data subsets corrected for beamwidth and calibration uncertainties in two different ways. The corresponding uncorrected numbers are given in Table 4. As mentioned previously, we have ignored the absolute-pointing uncertainty when combining data from the Ka and Q scans, and for this Ka + Q data set analysis we have also assumed that the Ka and Q beamwidth uncertainties shift together. These effects may approximately compensate, but we emphasize that the results from the Ka + Q data set analysis are less robust than those from either the Ka or Q data subset analyses.

The approximate prescription generally used in the CMB anisotropy field to account for these uncertainties results in total $1\sigma$ error bar ranges that are not very different from the total $1\sigma$ HPD error bar ranges derived here (rows labelled JOG vs. rows labelled HPD of Table 7). Note, however, that this is because the upper HPD error bars derived here are somewhat smaller and the lower HPD error bars are somewhat larger than the error bars previously used to represent the SP94 combined Ka and Q results (RSBG; GRS). Given our understanding of the numerical uncertainties of the computations here, we believe that these differences are numerically significant. If a similar result holds for a number of other small-scale CMB spatial anisotropy experiment data analysis results, then this would raise the lower values of reduced-$\chi^2$ found by GRS and probably also tighten the range in the $\chi^2$ values for a given model.

Tables 8 and 9 show the effects of the beamwidth- and calibration-uncertainty corrections on the deduced $Q_{\rm rms-PS}$ central values and limits for some of the SP94 data subsets. The corresponding likelihood functions for the selected models are shown in Figs. 5 and 6.

The beamwidth-uncertainty correction has a bigger effect for those data sets that probe a region of $l$-space where the model $C_l$ spectra changes rapidly with $l$. This is because in this case the contributions from the $\pm 1\sigma$ beamwidth analyses no longer approximately compensate for each other. This effect is clearly evident in the numbers of Tables 8 and 9.

As discussed in the previous section, the calibration uncertainty correction broadens and skews the probability density distribution functions toward higher values of $Q_{\rm rms-PS}$, resulting in a small increase in the central $Q_{\rm rms-PS}$ values and a larger increase in the $1\sigma$ limit range. These effects are also clearly evident in the numbers of Tables 8 and 9.

Tables 10 and 11 show the effects of the beamwidth- and calibration-uncertainty corrections on the maximum and marginalized values of the probability density distribution functions for some of the SP94 data subsets. Note that the Ka2 data set does not have a detection. From Table 11, we see that these corrections do not significantly affect the ordering of models with respect to the marginal probability distribution function value.

It is interesting that at a given beamwidth- and calibration-uncertainty treatment, there is a greater range in marginal probability values for the Ka data then there is for the Ka + Q data (see Table 11); this larger range is also evident for the Q data (see Table 16). In all cases, among all the models considered here, low-density $\Omega_0 \sim 0.1 - 0.2$ open models (models O1 – O4) are favoured by the SP94 data. If we assume that the marginal probability distribution (the probability density distribution for just the cosmological parameters $\Omega_0$, $h$, and $\Omega_B$) is gaussian, then relative to a model with a renormalized marginal value of unity, a model $1\sigma$ away has a value = 0.61, and one $1.5\sigma$ away has a value = 0.32. Clearly, the SP94 data do distinguish between the model CMB anisotropy shapes, but not at a very significant level.

Tables 12 – 16 summarize our SP94 data set analyses.

Table 14 lists the central values and $1\sigma$ ranges of band temperature for the individual-channel data sets. For a given data set the bandtemperature values vary by $\sim 0 - 10\%$ from model to model, with a typical range of $\lesssim 5\%$. Given the other uncertainties, an analysis based on the flat bandpower CMB spectrum is still quite useful for deriving approximate constraints on models, but it is not ideal. This is because a flat bandpower analysis cannot provide information about the relative probability of the model CMB anisotropy shapes. Since the combined Ka, Q, and (full) Ka + Q data subset analyses are based on data from a combination of windows, it is not possible to accurately compute the corresponding $\delta T_l$ central values and limits, except for the flat bandpower model (and these are given in Table 7).

Note that the variation in $\delta T_l$ with beamwidth is approximately $0 - 4$ percentage points less than the variation in $Q_{\rm rms-PS}$ with beamwidth which, depending on model and data subset, can be $\sim 15\%$ between the $-1\sigma$ and $+1\sigma$ beamwidths. Hence the variation in $\delta T_l$ with beamwidth is still quite significant. This is particularly true for the flat bandpower spectrum, where $\delta T_l$ varies as much as $Q_{\rm rms-PS}$, and so does not take out any of the effect of $W_l$ varying with beamwidth.

Focussing on just the SP94 $Q_{\rm rms-PS}$ central values and limits from the combined data sets (last column of Table 12 and last two columns of Table 13), one notices that for each model the Q-data $Q_{\rm rms-PS}$ values are somewhat higher than those for the Ka data. However, this is only at approximately $1\sigma$ of either the Ka or Q data error bars, and is not significant. Comparing the corresponding $\delta T_l$ values for the flat bandpower model (fourth, seventh, and tenth row of Table 7) one again notices this effect, but we emphasize that given the errors this is not significant. Furthermore, the Ka + Q data set values are consistent with those from the Ka and Q data subsets, although the Ka + Q data values are much closer to the Q data values.

The individual-channel $\delta T_l$ central values and ranges (Tables 14) are approximately model independent for each data set, and so provide a convenient, approximate summary of the SP94 observational results. Focussing on a given model and recalling that the frequency dependent SP94 window function couples different channels to different regions of $l$-space, one sees that, in a given line in Table 14, the SP94 $\delta T_l$s seem to mildly drop or remain constant from $l \sim 40$ to $l \sim 50$. They then seem to mildly rise from $l \sim 50$ to $l \sim 60$, or at least stay approximately constant but at a higher value than for the $40 \lesssim l \lesssim 50$ range. We emphasize that the precise numerical $l$ values are model dependent, and that the quoted numbers correspond approximately to those for model O1. We also note that the higher values at larger $l$ are from the Q data, so it is prudent to bear in mind the earlier discussions about $\beta$ and sample variance. This effect may be seen more clearly in Figs. 7, and is related to the effect mentioned in the previous paragraph; again, this is probably not significant. It is interesting that the mildly falling or flat $\delta T_l$ values at lower $l$ are qualitatively consistent with what is seen in the SK94 and SK95 experiments (Netterfield 1995; Netterfield et al. 1997), and that the "dip" or "break" at $l \sim 50 - 60$ is also qualitatively consistent with what is seen in the SK95 experiment (Netterfield et al. 1997).[9] It would be of interest to determine whether or not the SP94 and SK95 "dips" are quantitatively consistent. We emphasize that here we have assumed that the SP94 data is purely CMB anisotropy. The "dip" however might be a consequence of as yet undiscovered small systematic effects, foreground contamination, statistical noise, neglect of the absolute pointing uncertainty, sample variance, or just a result of the CMB anisotropy not being well-described by the models we have used here. Even if there is a "dip" in the CMB anisotropy, it is not very significant in the SP94 data. More data will be needed to resolve these issues.

The marginal probability distribution values for the combined SP94 data sets (last three columns of Table 16) are similar, although there is a larger range for the Ka and Q data sets than for the Ka + Q data set. CMB anisotropy shapes that are favoured by the Ka data are also favoured by the Q data and hence by the Ka + Q data. Those that are not favoured by the Q data are also not favoured by the Ka data and hence neither by the Ka + Q data.

The Ka + Q data, accounting for beamwidth- and calibration-uncertainty, favour (last column of Table 16) an open $\Omega_0 = 0.1$ model (model O1), among the models we consider here. We do not have the resolution in model-parameter space to construct the complete marginal probability distribution function (the probability density distribution function for just the cosmological parameters), and so cannot properly judge the significance of the values we have computed at isolated points in model-parameter space. To get some idea of the significance of these values we may assume that the marginal distribution is not far from a gaussian. As mentioned above, for a gaussian marginal distribution, a model $1\sigma$ away from the most favoured low-density open model CMB anisotropy shape has a marginal value of 0.61, and a model with marginal value 0.32 is $1.5\sigma$ away from the most favoured low-density open model. If the marginal probability distribution function is narrower than a gaussian the above $\sigma$-values are overconservative, and if it is wider than a gaussian they are overoptimistic for the purpose of ruling out models. Under the gaussian assumption, the difference between the most favoured low-density open model and the least favoured model, which is always a flat-$\Lambda$ one, is $\sim 1.6\sigma$ for Ka + Q and $\sim 1.8\sigma$ for Ka or Q, which are not very significant.

[9]Note that in Figs. 7 – 9 of Netterfield et al. (1997) the SK data are plotted at $l_{\rm e}$ of the $W_l$, and not at $l_{\rm m}$ of the $(\delta {T_{\rm rms}}^2)_l$ as done in Fig. 7 here.

We conclude that the SP94 data are most consistent with the CMB anisotropy shape in low-density open CDM models with $\Omega_0 \sim 0.1 - 0.3$ and 0.4 with larger $h$ and smaller $\Omega_B$, and with the flat bandpower shape, at least among the models we consider here. Since the fiducial CDM and flat-$\Lambda$ models have CMB anisotropy shapes that are always more than $1\sigma$ away, under the gaussian assumption, from the most favoured low-density open model, the SP94 data do not favour these models. This is especially true for old ($t_0 \gtrsim 15$ – 16 Gyr), large baryon density ($\Omega_B \gtrsim 0.0175h^{-2}$), low-density ($\Omega_0 \sim 0.2 - 0.4$), flat-$\Lambda$ ones. These results are mostly consistent with the qualitative ones of RSBG, and the quantitative ones of GRS. We emphasize, however, that under the gaussian marginal assumption the SP94 data do not rule out any of the models we consider here at the $2\sigma$ level, on the basis of their CMB anisotropy shape alone.

Finally, we note that the model normalizations deduced from the two-year DMR data and from the SP94 data sets are mostly consistent although, as mentioned above, the Ka data prefer a normalization $\sim 1\sigma$ below that favoured by the Q data. The SP94 Q and Ka + Q data favour a slightly higher normalization for the $\Omega_0 = 0.1$ open model than do the two-year DMR data, and the SP94 Ka and Ka + Q data favour a slightly lower normalization of the older (smaller $h$), higher $\Omega_B$, low-density flat-$\Lambda$ models than do the two-year DMR data. These results are consistent with those of GRS. It will be of interest to see if the $\Omega_0 = 0.1$ open model can be significantly ruled out on the basis of CMB anisotropy data alone when the new, slightly lower, four-year DMR normalization for this model is combined with our results here.

## 4. Conclusion

We have developed general methods to account for various sources of experimental and observational uncertainty in the likelihood analysis of observational data from a CMB spatial anisotropy experiment.

We have accounted for beamwidth- and calibration-uncertainty in likelihood analyses of the SP94 observational data that make use of theoretical CMB spatial anisotropy spectra in a variety of observationally-motivated, open and spatially-flat $\Lambda$, CDM cosmogonies.

Absolute-radiometer-calibration and beamwidth uncertainties are the largest known sources of uncertainty for the Ka- and Q-scan data from the SP94 experiment that were not previously explicitly accounted for. In our analysis we have not explicitly accounted for a number of other known but very much smaller uncertainties, including: the ellipticity and possible mild nongaussianity of the beams; the relative beamwidth uncertainty of the individual-channel beams; the chop-amplitude uncertainty; the effect of the main beam efficiency; the relative-radiometer-calibration uncertainty of the individual channels; the uncertainty in the measured passband frequencies; and the relative-pointing uncertainty. Collectively, these are likely to contribute to the error bars at the percentage-point level. In addition, when combining the Ka- and Q-scan data we have ignored the absolute-pointing uncertainty. We have also analyzed the Ka + Q data set assuming that the Ka and Q beamwidth uncertainties shift together. These two effects, quite likely, compensate, and our Ka + Q data results are not grossly incorrect; however, the Ka and Q data subset results are more robust.

In our analyses of the SP94 data we have assumed that it is purely CMB spatial anisotropy; justification for this assumption may be found in G95 and Gundersen (1995).

The combined Ka, Q, and (full) Ka + Q data subsets show evidence for at least $3\sigma$ detections of anisotropy. Six of the seven individual-channel "sky" rms values agree with the rms values derived from individual-channel flat bandpower analyses to within 25%. Predicted flat bandpower $\beta$ values agree with those found from flat bandpower likelihood analyses by G95 to within a third (Ka) and 0.8 (Q and Ka + Q) of the G95 error bars. Hence the G95 data could be even more consistent with a CMB $\nu$-spectrum than was indicated there. However, since the SP94 $W_l(\nu)$ couple $l$- and $\nu$-space, the only way of establishing from a $\beta$-analysis how inconsistent the SP94 data is with non-CMB contamination would be to use a model for both the $l$- and $\nu$-dependence of the suspected non-CMB foreground in likelihood analyses of the data. Only if a multiple-frequency CMB anisotropy experiment has individual-channel window shapes (in $l$) that are identical does $\beta \neq 0$ indicate that the data are inconsistent with a CMB $\nu$-spectrum. The behaviour of the Ka data error bars is consistent with what is expected from our approximate estimate of sample variance; the behaviour of the Q data error bars is not as consistent. For all the model CMB spatial anisotropy spectra we consider here, the Q data prefer a somewhat higher normalization than do the Ka data, but this is at only $\sim 1\sigma$ of either the Ka- or Q-data error bars. To see if the differences between the Ka and Q data are significant will require more data. Until this is accomplished it would be prudent to not draw conclusions using only the full Ka + Q analysis, but to also consider the implications of both the Ka and Q analyses.

The HPD bandtemperature estimates of Table 7 provide an approximate summary of the SP94 data, and can be used to approximately constrain other models of CMB spatial anisotropy, although likelihood analyses, like those done here, are needed if one wishes to utilize all the information in the SP94 data. Furthermore, as discussed above, the deduced bandtemperature values typically vary by $\sim 5\%$ from model to model.

While the marginal probability distribution function values show that the SP94 data are clearly sensitive to spatial structure on the sky, they do not significantly distinguish between models, although they do favour low-density open models over older, high $\Omega_B$, low-density, flat-$\Lambda$ models. No model considered here is ruled out at the $2\sigma$ level by the SP94 data alone, at least in the gaussian marginal probability distribution approximation. The SP94 model normalizations are mostly consistent with those deduced from the two-year DMR data. These results are mostly compatible with what one concludes from a goodness-of-fit analysis of all presently available CMB anisotropy detection data (GRS).

The major contributors to the SP94 Ka- and Q-scan data analyses error bars are (alphabetically): (1) beamwidth uncertainty; (2) calibration uncertainty; (3) intrinsic noise uncertainty; and, (4) sample variance uncertainty. More sky coverage and less oversampling will reduce the sample variance contribution, and longer integration will reduce the intrinsic noise contribution. Measured beamwidth and calibration distribution functions should help reduce the beamwidth- and calibration-uncertainty contributions, unless it turns out that our gaussian approximation is actually narrower than the real distribution function. It should be possible, in the next few years, to acquire data with error bars significantly smaller than the present ones ($\sim 30 - 36\%$).

If measured distribution functions for the parameters of CMB experiments, at least for those that are known to have relatively large uncertainty, become available, the methods we have developed here will allow for a more robust estimate of the sky signal detected in CMB anisotropy experiments, even for presently available CMB anisotropy data.

We acknowledge helpful discussions with D. Bond, K. Górski, J. Peebles and the referee, E. Wright, and are especially indebted to L. Page. KG acknowledges support from NASA grants NAGW-4623 and 4870. BR acknowledges support from NSF grant EPS-9550487 and matching support from the state of Kansas.

Table 1: Numerical Values for the Individual-Channel Zero-Lag Window Function Parameters[a]

| | Chan.: | Ka1 | Ka2 | Ka3 | Ka4 | Q1 | Q2 | Q3 |
|---|---|---|---|---|---|---|---|---|
| Beamwidth | $\nu$(GHz): | 27.25 | 29.75 | 32.25 | 34.75 | 39.15 | 41.45 | 43.75 |
| | $l_{e^{-0.5}}$ | 34 | 35 | 37 | 38 | 40 | 41 | 42 |
| | $l_{\rm e}$ | 55.2 | 58.0 | 60.5 | 62.8 | 67.1 | 68.8 | 70.4 |
| $-1\sigma$ | $l_{\rm m}$ | 62 | 65 | 67 | 70 | 73 | 75 | 76 |
| | $l_{e^{-0.5}}$ | 95 | 100 | 104 | 107 | 114 | 116 | 118 |
| | $\sqrt{I(W_l)}$ | 1.04 | 1.09 | 1.14 | 1.17 | 1.24 | 1.27 | 1.29 |
| | $l_{e^{-0.5}}$ | 33 | 34 | 36 | 37 | 39 | 40 | 40 |
| | $l_{\rm e}$ | 53.4 | 56.1 | 58.7 | 61.0 | 64.4 | 66.1 | 67.8 |
| Nom. | $l_{\rm m}$ | 60 | 63 | 65 | 68 | 71 | 73 | 74 |
| | $l_{e^{-0.5}}$ | 92 | 97 | 101 | 105 | 110 | 112 | 115 |
| | $\sqrt{I(W_l)}$ | 1.01 | 1.06 | 1.10 | 1.14 | 1.20 | 1.23 | 1.26 |
| | $l_{e^{-0.5}}$ | 32 | 33 | 35 | 36 | 38 | 38 | 39 |
| | $l_{\rm e}$ | 51.6 | 54.4 | 56.9 | 59.2 | 61.9 | 63.7 | 65.3 |
| $+1\sigma$ | $l_{\rm m}$ | 58 | 61 | 64 | 66 | 69 | 70 | 72 |
| | $l_{e^{-0.5}}$ | 89 | 94 | 98 | 102 | 106 | 109 | 111 |
| | $\sqrt{I(W_l)}$ | 0.997 | 1.03 | 1.07 | 1.11 | 1.16 | 1.19 | 1.22 |

[a]Computed according to the prescriptions above and in eq. (4).

Table 2: Numerical Values for Parameters Characterizing the Shape of $(\delta T_{\rm rms}{}^2)_l$

| Window: | | +1σ Ka1 | | | −1σ Q3 | | |
|---|---|---|---|---|---|---|---|
| # | $(\Omega_0, h, \Omega_B h^2)$ | $l_{e^{-0.5}}$ | $l_{\rm m}$ | $l_{e^{-0.5}}$ | $l_{e^{-0.5}}$ | $l_{\rm m}$ | $l_{e^{-0.5}}$ |
| (1) | (2) | (3) | (4) | (5) | (6) | (7) | (8) |
| O1 | (0.1, 0.75, 0.0125) | 7 | 38 | 74 | 14 | 55 | 102 |
| O2 | (0.2, 0.65, 0.0175) | 14 | 43 | 79 | 22 | 63 | 108 |
| O3 | (0.2, 0.70, 0.0125) | 13 | 42 | 78 | 21 | 61 | 107 |
| O4 | (0.2, 0.75, 0.0075) | 13 | 41 | 77 | 20 | 59 | 104 |
| O5 | (0.3, 0.60, 0.0175) | 17 | 47 | 81 | 27 | 66 | 111 |
| O6 | (0.3, 0.65, 0.0125) | 16 | 46 | 80 | 25 | 64 | 108 |
| O7 | (0.3, 0.70, 0.0075) | 16 | 44 | 78 | 24 | 61 | 106 |
| O8 | (0.4, 0.60, 0.0175) | 19 | 48 | 82 | 29 | 67 | 111 |
| O9 | (0.4, 0.65, 0.0125) | 18 | 47 | 80 | 27 | 64 | 109 |
| O10 | (0.4, 0.70, 0.0075) | 18 | 45 | 79 | 25 | 62 | 106 |
| O11 | (0.5, 0.55, 0.0175) | 21 | 50 | 83 | 31 | 68 | 113 |
| O12 | (0.5, 0.60, 0.0125) | 20 | 48 | 81 | 29 | 66 | 110 |
| O13 | (0.5, 0.65, 0.0075) | 19 | 46 | 79 | 27 | 63 | 107 |
| O14 | (1.0, 0.50, 0.0125) | 22 | 49 | 83 | 31 | 68 | 115 |
| Λ1 | (0.1, 0.90, 0.0125) | 28 | 57 | 90 | 40 | 76 | 121 |
| Λ2 | (0.2, 0.80, 0.0075) | 24 | 53 | 87 | 35 | 73 | 120 |
| Λ3 | (0.2, 0.75, 0.0125) | 26 | 55 | 88 | 37 | 75 | 121 |
| Λ4 | (0.2, 0.70, 0.0175) | 27 | 56 | 89 | 39 | 76 | 122 |
| Λ5 | (0.3, 0.70, 0.0075) | 23 | 52 | 86 | 34 | 72 | 119 |
| Λ6 | (0.3, 0.65, 0.0125) | 25 | 54 | 88 | 36 | 74 | 120 |
| Λ7 | (0.3, 0.60, 0.0175) | 27 | 56 | 89 | 38 | 75 | 121 |
| Λ8 | (0.4, 0.65, 0.0075) | 23 | 51 | 85 | 32 | 70 | 118 |
| Λ9 | (0.4, 0.60, 0.0125) | 24 | 53 | 87 | 35 | 73 | 119 |
| Λ10 | (0.4, 0.55, 0.0175) | 26 | 55 | 88 | 37 | 75 | 121 |
| Λ11 | (0.5, 0.60, 0.0125) | 23 | 51 | 86 | 33 | 71 | 118 |
| Λ12 | (1.0, 0.50, 0.0125) | 22 | 49 | 83 | 31 | 68 | 115 |
| Flat | ... | 16 | 41 | 73 | 21 | 54 | 96 |
| $W_l$ | ... | 32[a] | 51.6[b] | 89[a] | 42[a] | 70.4[b] | 118[a] |

[a] $l_{e^{-0.5}}$ for the window, computed according to the prescription above eq. (4).

[b] $l_e$ for the window.

Table 3: Numerical Values for Rms Temperature Anisotropies[a]

| Channel: | Ka1 | Ka2 | Ka3 | Ka4 | Q1 | Q2 | Q3 |
|---|---|---|---|---|---|---|---|
| "Sky"[b] | 39 | 18 | 36 | 51 | 33 | 47 | 42 |
| FBP, Nom. $W_l$[c] | 51 | 17 | 37 | 33 | 34 | 49 | 52 |

[a] $\delta T_{\rm rms}$ in $\mu$K.

[b] Estimated from the data of Fig. 4, as discussed in §2.

[c] Converted to rms (using eq. [7]) from the results of the likelihood analysis for the flat bandpower (FBP) angular spectrum (eq. [3]), and ignoring beamwidth and calibration uncertainties.

Table 4: Numerical Values for Bandtemperature[a] from Likelihood Analyses Assuming a Flat Bandpower Spectrum and Ignoring Beamwidth and Calibration Uncertainties

| Data Set | Analysis | $-1\sigma$ | Peak | $+1\sigma$ | Ave. Abs. Err.[e] | Ave. Fra. Err.[f] |
|---|---|---|---|---|---|---|
| Ka | G95[b] | 24 | 29 | 43 | ±10 | ±33% |
| | Here[c] | 25 | 30 | 46 | ±11 | ±35% |
| | Here[d] | 22 | 30 | 42 | ±10 | ±33% |
| Q | G95[b] | 33 | 40 | 55 | ±11 | ±28% |
| | Here[c] | 35 | 41 | 59 | ±12 | ±29% |
| | Here[d] | 32 | 41 | 55 | ±12 | ±28% |
| Ka + Q | G95[b] | 30 | 36 | 50 | ±10 | ±27% |
| | Here[c] | 33 | 39 | 56 | ±11 | ±29% |
| | Here[d] | 30 | 39 | 52 | ±11 | ±27% |

[a] $\delta T_l$ (eq. [7]) in $\mu$K.

[b] ET (16.0% & 84.0%) prescription.

[c] ET (15.87% & 84.13%) prescription.

[d] HPD (68.27%) prescription.

[e] Average absolute error in $\mu$K.

[f] Average fractional error, as a fraction of the central value.

Table 5: Numerical Values for $\beta$, from the "Sky" Rms,[a] and from a Theoretical Flat Bandpower CMB Angular Spectrum[b]

| Channel | Ka1 | Ka2 | Ka3 | Ka4 | Q1 | Q2 | Q3 |
|---|---|---|---|---|---|---|---|
| Ka1 | ... | 0.55 | 0.53 | 0.52 | 0.48 | 0.47 | 0.46 |
| Ka2 | −8.7 | ... | 0.51 | 0.50 | 0.46 | 0.45 | 0.44 |
| Ka3 | −0.47 | 8.6 | ... | 0.48 | 0.44 | 0.43 | 0.42 |
| Ka4 | 1.1 | 6.7 | 4.6 | ... | 0.42 | 0.41 | 0.40 |
| Q1 | −0.49 | 2.1 | −0.51 | −3.7 | ... | 0.40 | 0.39 |
| Q2 | 0.43 | 2.9 | 1.0 | −0.48 | 6.3 | ... | 0.38 |
| Q3 | 0.15 | 2.2 | 0.49 | −0.85 | 2.2 | −2.1 | ... |

[a]Lower left-hand triangle. Estimated from the ratio of the signal in the channel of the first row, to the signal in the channel of the first column (eq. [21]).

[b]Upper right-hand triangle. Estimated from the ratio of the signal in the channel of the first column, to the signal in the channel of the first row (eq. [22]).

Table 6: Numerical Values for Bandtemperature[a] from Likelihood Analyses Assuming a Flat Bandpower Spectrum, for the Individual-Channel Data Sets, and Ignoring Beamwidth and Calibration Uncertainties

| Data Set | Prescript. | $-1\sigma$ | Peak | $+1\sigma$ | Ave. Abs. Err.[b] | Ave. Fra. Err.[c] |
|---|---|---|---|---|---|---|
| Ka1 | ET | 40 | 51 | 79 | ±20 | ±39% |
| | HPD | 35 | 51 | 73 | ±19 | ±37% |
| Ka2 | ET | 7.7 | 16 | 30 | ±11 | ±69% |
| | HPD | 5.4 | 16 | 28 | ±11 | ±67% |
| Ka3 | ET | 27 | 34 | 53 | ±13 | ±40% |
| | HPD | 23 | 34 | 49 | ±13 | ±38% |
| Ka4 | ET | 23 | 29 | 45 | ±11 | ±39% |
| | HPD | 20 | 29 | 42 | ±11 | ±37% |
| Q1 | ET | 19 | 28 | 45 | ±13 | ±47% |
| | HPD | 17 | 28 | 42 | ±13 | ±46% |
| Q2 | ET | 33 | 39 | 58 | ±12 | ±32% |
| | HPD | 30 | 39 | 53 | ±12 | ±30% |
| Q3 | ET | 35 | 42 | 60 | ±13 | ±31% |
| | HPD | 31 | 42 | 56 | ±12 | ±29% |

[a] $\delta T_l$ (eq. [7]) in $\mu$K. ET limits are 15.87% & 84.13% and HPD limits are 68.27%.

[b] Average absolute error in $\mu$K.

[c] Average fractional error, as a fraction of the central value.

Table 7: Numerical Values for Bandtemperature[a] from Likelihood Analyses Assuming a Flat Bandpower Spectrum and Accounting for Beamwidth and Calibration Uncertainties

| Data Set | Prescript. | $-1\sigma$ | Peak | $+1\sigma$ | Ave. Abs. Err.[d] | Ave. Fra. Err.[e] |
|---|---|---|---|---|---|---|
| Ka | JOG[b] | 22 | 29 | 44 | ±11 | ±37% |
| | ET[c] | 25 | 31 | 48 | ±12 | ±38% |
| | HPD[c] | 22 | 31 | 43 | ±11 | ±36% |
| Q | JOG[b] | 31 | 40 | 57 | ±13 | ±32% |
| | ET[c] | 35 | 42 | 61 | ±14 | ±32% |
| | HPD[c] | 31 | 42 | 57 | ±13 | ±31% |
| Ka + Q | JOG[b] | 28 | 36 | 51 | ±11 | ±31% |
| | ET[c] | 33 | 39 | 58 | ±13 | ±32% |
| | HPD[c] | 30 | 39 | 53 | ±12 | ±30% |

[a] $\delta T_l$ (eq. [7]) in $\mu$K.

[b] JOG prescription to account for calibration and beamwidth uncertainties by adding 15% of the central value in quadrature to the G95 ET (16.0% & 84.0%) prescription error bars.

[c] ET (15.87% & 84.13%) and HPD (68.27%) limits from the beamwidth- and calibration-uncertainty corrected probability density distribution functions.

[d] Average absolute error in $\mu$K.

[e] Average fractional error, as a fraction of the central value.

Table 8: Numerical Values for $Q_{\mathrm{rms-PS}}$ (in $\mu$K) from the Ka2 and Q2 Data[a]

| | Data Set: | Ka2 | | | Q2 | | |
|---|---|---|---|---|---|---|---|
| Model | $(\Omega_0,\ h,\ \Omega_B h^2)$ | Bare[b] | Beam.[c] | B.+C.[d] | Bare[b] | Beam.[c] | B.+C.[d] |
| O1 | (0.1, 0.75, 0.0125) | $17^{42}_{\dots}$ | $17^{43}_{\dots}$ | $17^{44}_{\dots}$ | $35^{48}_{27}$ | $36^{48}_{27}$ | $36^{49}_{26}$ |
| O2 | (0.2, 0.65, 0.0175) | $16^{41}_{\dots}$ | $16^{41}_{\dots}$ | $16^{43}_{\dots}$ | $35^{47}_{26}$ | $35^{47}_{26}$ | $35^{48}_{26}$ |
| O3 | (0.2, 0.70, 0.0125) | $17^{42}_{\dots}$ | $17^{42}_{\dots}$ | $17^{44}_{\dots}$ | $36^{48}_{27}$ | $36^{48}_{27}$ | $36^{49}_{26}$ |
| O4 | (0.2, 0.75, 0.0075) | $17^{43}_{\dots}$ | $17^{43}_{\dots}$ | $17^{45}_{\dots}$ | $37^{49}_{27}$ | $37^{49}_{27}$ | $37^{51}_{27}$ |
| O5 | (0.3, 0.60, 0.0175) | $14^{35}_{\dots}$ | $14^{35}_{\dots}$ | $14^{37}_{\dots}$ | $30^{40}_{23}$ | $30^{41}_{23}$ | $30^{42}_{22}$ |
| O6 | (0.3, 0.65, 0.0125) | $14^{36}_{\dots}$ | $14^{36}_{\dots}$ | $14^{38}_{\dots}$ | $31^{41}_{23}$ | $31^{42}_{23}$ | $31^{43}_{23}$ |
| O7 | (0.3, 0.70, 0.0075) | $15^{37}_{\dots}$ | $15^{38}_{\dots}$ | $15^{39}_{\dots}$ | $32^{43}_{24}$ | $32^{43}_{24}$ | $32^{44}_{24}$ |
| O8 | (0.4, 0.60, 0.0175) | $12^{29}_{\dots}$ | $12^{29}_{\dots}$ | $12^{30}_{\dots}$ | $25^{33}_{19}$ | $25^{34}_{19}$ | $25^{35}_{19}$ |
| O9 | (0.4, 0.65, 0.0125) | $12^{30}_{\dots}$ | $12^{30}_{\dots}$ | $12^{31}_{\dots}$ | $26^{35}_{19}$ | $26^{35}_{19}$ | $26^{36}_{19}$ |
| O10 | (0.4, 0.70, 0.0075) | $12^{31}_{\dots}$ | $12^{31}_{\dots}$ | $12^{33}_{\dots}$ | $27^{36}_{20}$ | $27^{36}_{20}$ | $27^{37}_{20}$ |
| O11 | (0.5, 0.55, 0.0175) | $9.2^{23}_{\dots}$ | $9.2^{23}_{\dots}$ | $9.3^{24}_{\dots}$ | $20^{27}_{15}$ | $20^{27}_{15}$ | $20^{28}_{15}$ |
| O12 | (0.5, 0.60, 0.0125) | $9.6^{24}_{\dots}$ | $9.6^{24}_{\dots}$ | $9.7^{25}_{\dots}$ | $21^{28}_{16}$ | $21^{28}_{16}$ | $21^{29}_{16}$ |
| O13 | (0.5, 0.65, 0.0075) | $10^{25}_{\dots}$ | $10^{25}_{\dots}$ | $10^{27}_{\dots}$ | $22^{29}_{16}$ | $22^{29}_{16}$ | $22^{30}_{16}$ |
| O14 | (1.0, 0.50, 0.0125) | $8.3^{21}_{\dots}$ | $8.3^{21}_{\dots}$ | $8.4^{22}_{\dots}$ | $18^{24}_{14}$ | $18^{24}_{14}$ | $18^{25}_{13}$ |
| Λ1 | (0.1, 0.90, 0.0125) | $9.8^{24}_{\dots}$ | $9.8^{24}_{\dots}$ | $9.9^{25}_{\dots}$ | $22^{29}_{16}$ | $22^{29}_{16}$ | $22^{30}_{16}$ |
| Λ2 | (0.2, 0.80, 0.0075) | $9.6^{24}_{\dots}$ | $9.6^{24}_{\dots}$ | $9.7^{25}_{\dots}$ | $21^{28}_{16}$ | $21^{28}_{16}$ | $21^{29}_{16}$ |
| Λ3 | (0.2, 0.75, 0.0125) | $9.0^{22}_{\dots}$ | $9.0^{22}_{\dots}$ | $9.1^{23}_{\dots}$ | $20^{26}_{15}$ | $20^{27}_{15}$ | $20^{27}_{15}$ |
| Λ4 | (0.2, 0.70, 0.0175) | $8.6^{21}_{\dots}$ | $8.6^{21}_{\dots}$ | $8.7^{22}_{\dots}$ | $19^{25}_{14}$ | $19^{25}_{14}$ | $19^{26}_{14}$ |
| Λ5 | (0.3, 0.70, 0.0075) | $8.9^{22}_{\dots}$ | $8.9^{22}_{\dots}$ | $9.1^{23}_{\dots}$ | $20^{26}_{15}$ | $20^{26}_{15}$ | $20^{27}_{15}$ |
| Λ6 | (0.3, 0.65, 0.0125) | $8.4^{21}_{\dots}$ | $8.4^{21}_{\dots}$ | $8.5^{22}_{\dots}$ | $18^{25}_{14}$ | $19^{25}_{14}$ | $19^{25}_{14}$ |
| Λ7 | (0.3, 0.60, 0.0175) | $8.0^{20}_{\dots}$ | $8.0^{20}_{\dots}$ | $8.0^{21}_{\dots}$ | $18^{23}_{13}$ | $18^{24}_{13}$ | $18^{24}_{13}$ |
| Λ8 | (0.4, 0.65, 0.0075) | $8.6^{21}_{\dots}$ | $8.6^{22}_{\dots}$ | $8.7^{22}_{\dots}$ | $19^{25}_{14}$ | $19^{25}_{14}$ | $19^{26}_{14}$ |
| Λ9 | (0.4, 0.60, 0.0125) | $8.1^{20}_{\dots}$ | $8.1^{20}_{\dots}$ | $8.2^{21}_{\dots}$ | $18^{24}_{13}$ | $18^{24}_{13}$ | $18^{25}_{13}$ |
| Λ10 | (0.4, 0.55, 0.0175) | $7.7^{19}_{\dots}$ | $7.7^{19}_{\dots}$ | $7.7^{20}_{\dots}$ | $17^{22}_{13}$ | $17^{23}_{13}$ | $17^{23}_{13}$ |
| Λ11 | (0.5, 0.60, 0.0125) | $8.1^{20}_{\dots}$ | $8.1^{20}_{\dots}$ | $8.2^{21}_{\dots}$ | $18^{24}_{13}$ | $18^{24}_{13}$ | $18^{24}_{13}$ |
| Λ12 | (1.0, 0.50, 0.0125) | $8.3^{21}_{\dots}$ | $8.3^{21}_{\dots}$ | $8.4^{22}_{\dots}$ | $18^{24}_{14}$ | $18^{24}_{14}$ | $18^{25}_{13}$ |
| Flat | ... | $12^{30}_{\dots}$ | $12^{30}_{\dots}$ | $12^{32}_{\dots}$ | $25^{34}_{19}$ | $26^{35}_{19}$ | $26^{35}_{19}$ |

[a] The primary entry in each of the last six columns is the value at which the probability density distribution function peaks. Ellipses as the lower entry in a vertical pair denotes a non-detection; the corresponding upper entry is the $2\sigma$ (97.72% ET) upper limit. For detections, the vertical pair of numbers are the $\pm 1\sigma$ (68.27% HPD) upper and lower limits.

[b] Computed using eq. (8). Ignores beamwidth and calibration uncertainties.

[c] Accounts for beamwidth uncertainty using eq. (12). Ignores calibration uncertainty.

[d] Computed using eqs. (12) and (16). Accounts for beamwidth and calibration uncertainties.

Table 9: Numerical Values for $Q_{\rm rms-PS}$ (in $\mu$K) from the Ka and Ka + Q Data[a]

| | Data Set: | Ka | | | Ka + Q | | |
|---|---|---|---|---|---|---|---|
| Model | ($\Omega_0$, $h$, $\Omega_B h^2$) | Bare[b] | Beam.[c] | B.+C.[d] | Bare[b] | Beam.[c] | B.+C.[d] |
| O1 | (0.1, 0.75, 0.0125) | $27^{38}_{20}$ | $27^{38}_{20}$ | $28^{39}_{20}$ | $35^{46}_{27}$ | $35^{46}_{27}$ | $35^{47}_{26}$ |
| O2 | (0.2, 0.65, 0.0175) | $27^{37}_{20}$ | $27^{37}_{20}$ | $27^{38}_{19}$ | $34^{44}_{26}$ | $34^{44}_{26}$ | $34^{46}_{26}$ |
| O3 | (0.2, 0.70, 0.0125) | $28^{38}_{20}$ | $28^{38}_{20}$ | $28^{39}_{20}$ | $35^{45}_{27}$ | $35^{46}_{27}$ | $35^{47}_{26}$ |
| O4 | (0.2, 0.75, 0.0075) | $28^{39}_{21}$ | $28^{39}_{21}$ | $28^{40}_{20}$ | $36^{47}_{28}$ | $36^{47}_{28}$ | $36^{48}_{27}$ |
| O5 | (0.3, 0.60, 0.0175) | $23^{32}_{17}$ | $23^{32}_{17}$ | $23^{33}_{17}$ | $29^{38}_{23}$ | $29^{38}_{23}$ | $29^{39}_{22}$ |
| O6 | (0.3, 0.65, 0.0125) | $24^{33}_{18}$ | $24^{33}_{17}$ | $24^{34}_{17}$ | $30^{39}_{23}$ | $30^{39}_{23}$ | $30^{40}_{23}$ |
| O7 | (0.3, 0.70, 0.0075) | $25^{34}_{18}$ | $25^{34}_{18}$ | $25^{35}_{18}$ | $31^{40}_{24}$ | $31^{41}_{24}$ | $31^{42}_{24}$ |
| O8 | (0.4, 0.60, 0.0175) | $19^{26}_{14}$ | $19^{26}_{14}$ | $19^{27}_{14}$ | $24^{31}_{19}$ | $24^{31}_{19}$ | $24^{32}_{18}$ |
| O9 | (0.4, 0.65, 0.0125) | $20^{27}_{15}$ | $20^{27}_{15}$ | $20^{28}_{14}$ | $25^{32}_{19}$ | $25^{33}_{19}$ | $25^{34}_{19}$ |
| O10 | (0.4, 0.70, 0.0075) | $21^{28}_{15}$ | $21^{29}_{15}$ | $21^{29}_{15}$ | $26^{34}_{20}$ | $26^{34}_{20}$ | $26^{35}_{20}$ |
| O11 | (0.5, 0.55, 0.0175) | $15^{21}_{11}$ | $15^{21}_{11}$ | $16^{22}_{11}$ | $19^{25}_{15}$ | $19^{25}_{15}$ | $19^{26}_{15}$ |
| O12 | (0.5, 0.60, 0.0125) | $16^{22}_{12}$ | $16^{22}_{12}$ | $16^{23}_{12}$ | $20^{26}_{16}$ | $20^{26}_{16}$ | $20^{27}_{15}$ |
| O13 | (0.5, 0.65, 0.0075) | $17^{23}_{12}$ | $17^{23}_{12}$ | $17^{24}_{12}$ | $21^{27}_{16}$ | $21^{28}_{16}$ | $21^{28}_{16}$ |
| O14 | (1.0, 0.50, 0.0125) | $14^{19}_{10}$ | $14^{19}_{10}$ | $14^{20}_{10}$ | $17^{22}_{14}$ | $17^{23}_{14}$ | $17^{23}_{13}$ |
| Λ1 | (0.1, 0.90, 0.0125) | $17^{23}_{12}$ | $17^{23}_{12}$ | $17^{23}_{12}$ | $21^{27}_{16}$ | $21^{27}_{16}$ | $21^{28}_{16}$ |
| Λ2 | (0.2, 0.80, 0.0075) | $16^{22}_{12}$ | $16^{22}_{12}$ | $16^{23}_{12}$ | $20^{26}_{16}$ | $20^{26}_{16}$ | $20^{27}_{15}$ |
| Λ3 | (0.2, 0.75, 0.0125) | $15^{21}_{11}$ | $15^{21}_{11}$ | $15^{21}_{11}$ | $19^{24}_{15}$ | $19^{25}_{15}$ | $19^{25}_{15}$ |
| Λ4 | (0.2, 0.70, 0.0175) | $14^{20}_{11}$ | $14^{20}_{11}$ | $15^{20}_{11}$ | $18^{23}_{14}$ | $18^{23}_{14}$ | $18^{24}_{14}$ |
| Λ5 | (0.3, 0.70, 0.0075) | $15^{20}_{11}$ | $15^{21}_{11}$ | $15^{21}_{11}$ | $19^{24}_{15}$ | $19^{24}_{15}$ | $19^{25}_{14}$ |
| Λ6 | (0.3, 0.65, 0.0125) | $14^{19}_{10}$ | $14^{19}_{10}$ | $14^{20}_{10}$ | $18^{23}_{14}$ | $18^{23}_{14}$ | $18^{24}_{14}$ |
| Λ7 | (0.3, 0.60, 0.0175) | $13^{18}_{10}$ | $13^{18}_{10}$ | $14^{19}_{10}$ | $17^{21}_{13}$ | $17^{22}_{13}$ | $17^{22}_{13}$ |
| Λ8 | (0.4, 0.65, 0.0075) | $14^{20}_{11}$ | $14^{20}_{11}$ | $15^{20}_{10}$ | $18^{23}_{14}$ | $18^{23}_{14}$ | $18^{24}_{14}$ |
| Λ9 | (0.4, 0.60, 0.0125) | $14^{19}_{10}$ | $14^{19}_{10}$ | $14^{19}_{10}$ | $17^{22}_{13}$ | $17^{22}_{13}$ | $17^{23}_{13}$ |
| Λ10 | (0.4, 0.55, 0.0175) | $13^{18}_{10}$ | $13^{18}_{10}$ | $13^{18}_{9}$ | $16^{21}_{13}$ | $16^{21}_{13}$ | $16^{21}_{12}$ |
| Λ11 | (0.5, 0.60, 0.0125) | $14^{19}_{10}$ | $14^{19}_{10}$ | $14^{19}_{10}$ | $17^{22}_{13}$ | $17^{22}_{13}$ | $17^{23}_{13}$ |
| Λ12 | (1.0, 0.50, 0.0125) | $14^{19}_{10}$ | $14^{19}_{10}$ | $14^{20}_{10}$ | $17^{22}_{14}$ | $17^{23}_{14}$ | $17^{23}_{13}$ |
| Flat | ... | $20^{27}_{14}$ | $20^{27}_{14}$ | $20^{28}_{14}$ | $25^{33}_{20}$ | $25^{34}_{19}$ | $25^{34}_{19}$ |

[a]Conventions are the same as for Table 8.

[b]Computed using eq. (8). Ignores beamwidth and calibration uncertainties.

[c]Accounts for beamwidth uncertainty using eq. (12). Ignores calibration uncertainty.

[d]Computed using eqs. (12) and (16). Accounts for beamwidth and calibration uncertainties.

Table 10: Renormalized Maximum Values of the Probability Density Distribution Functions[a]

| Data: | Ka2 | | | Q2 | | | Ka | | | Ka + Q | | |
|---|---|---|---|---|---|---|---|---|---|---|---|---|
| Model | Bare[b] | B.[c] | B.+C.[d] | Bare[b] | B.[c] | B.+C.[d] | Bare[b] | B.[c] | B.+C.[d] | Bare[b] | B.[c] | B.+C.[d] |
| O1 | 1.0 | 1.0 | 1.0 | 1.0 | 1.0 | 1.0 | 1.0 | 1.0 | 1.0 | 1.0 | 1.0 | 1.0 |
| O2 | 0.99 | 1.0 | 1.0 | 0.75 | 0.75 | 0.75 | 0.81 | 0.81 | 0.81 | 0.89 | 0.89 | 0.89 |
| O3 | 0.99 | 0.99 | 0.99 | 0.79 | 0.79 | 0.79 | 0.84 | 0.84 | 0.85 | 0.91 | 0.91 | 0.91 |
| O4 | 0.99 | 0.99 | 0.99 | 0.84 | 0.84 | 0.85 | 0.88 | 0.88 | 0.88 | 0.92 | 0.92 | 0.92 |
| O5 | 0.98 | 0.98 | 0.98 | 0.63 | 0.62 | 0.62 | 0.71 | 0.71 | 0.71 | 0.82 | 0.82 | 0.81 |
| O6 | 0.98 | 0.98 | 0.98 | 0.67 | 0.68 | 0.67 | 0.75 | 0.75 | 0.75 | 0.84 | 0.84 | 0.84 |
| O7 | 0.98 | 0.98 | 0.98 | 0.74 | 0.73 | 0.74 | 0.80 | 0.80 | 0.80 | 0.86 | 0.86 | 0.86 |
| O8 | 0.97 | 0.97 | 0.97 | 0.57 | 0.57 | 0.57 | 0.66 | 0.66 | 0.66 | 0.78 | 0.78 | 0.77 |
| O9 | 0.97 | 0.97 | 0.97 | 0.62 | 0.62 | 0.62 | 0.71 | 0.71 | 0.71 | 0.80 | 0.80 | 0.80 |
| O10 | 0.96 | 0.97 | 0.97 | 0.69 | 0.69 | 0.69 | 0.76 | 0.76 | 0.76 | 0.83 | 0.83 | 0.83 |
| O11 | 0.97 | 0.97 | 0.97 | 0.51 | 0.51 | 0.51 | 0.61 | 0.61 | 0.61 | 0.74 | 0.74 | 0.73 |
| O12 | 0.96 | 0.97 | 0.97 | 0.57 | 0.57 | 0.57 | 0.66 | 0.66 | 0.66 | 0.77 | 0.77 | 0.77 |
| O13 | 0.96 | 0.96 | 0.96 | 0.64 | 0.64 | 0.64 | 0.71 | 0.71 | 0.71 | 0.80 | 0.80 | 0.80 |
| O14 | 0.97 | 0.97 | 0.97 | 0.49 | 0.49 | 0.49 | 0.59 | 0.58 | 0.58 | 0.73 | 0.73 | 0.72 |
| Λ1 | 0.96 | 0.96 | 0.96 | 0.31 | 0.31 | 0.31 | 0.41 | 0.41 | 0.41 | 0.57 | 0.57 | 0.56 |
| Λ2 | 0.97 | 0.97 | 0.97 | 0.39 | 0.39 | 0.39 | 0.50 | 0.49 | 0.49 | 0.65 | 0.65 | 0.65 |
| Λ3 | 0.97 | 0.97 | 0.97 | 0.35 | 0.35 | 0.35 | 0.45 | 0.45 | 0.45 | 0.61 | 0.61 | 0.60 |
| Λ4 | 0.96 | 0.96 | 0.96 | 0.32 | 0.32 | 0.32 | 0.42 | 0.42 | 0.42 | 0.58 | 0.58 | 0.57 |
| Λ5 | 0.97 | 0.97 | 0.97 | 0.42 | 0.42 | 0.42 | 0.52 | 0.52 | 0.52 | 0.68 | 0.67 | 0.67 |
| Λ6 | 0.97 | 0.97 | 0.97 | 0.37 | 0.37 | 0.37 | 0.47 | 0.47 | 0.47 | 0.63 | 0.63 | 0.62 |
| Λ7 | 0.96 | 0.96 | 0.96 | 0.33 | 0.33 | 0.33 | 0.43 | 0.43 | 0.43 | 0.59 | 0.59 | 0.59 |
| Λ8 | 0.97 | 0.97 | 0.97 | 0.45 | 0.44 | 0.44 | 0.54 | 0.54 | 0.54 | 0.70 | 0.70 | 0.69 |
| Λ9 | 0.97 | 0.97 | 0.97 | 0.39 | 0.39 | 0.39 | 0.49 | 0.49 | 0.49 | 0.65 | 0.65 | 0.64 |
| Λ10 | 0.96 | 0.96 | 0.96 | 0.35 | 0.35 | 0.35 | 0.45 | 0.45 | 0.45 | 0.61 | 0.61 | 0.60 |
| Λ11 | 0.97 | 0.97 | 0.97 | 0.43 | 0.42 | 0.42 | 0.53 | 0.53 | 0.52 | 0.68 | 0.68 | 0.67 |
| Λ12 | 0.97 | 0.97 | 0.97 | 0.49 | 0.49 | 0.49 | 0.59 | 0.58 | 0.58 | 0.73 | 0.73 | 0.72 |
| Flat | 0.94 | 0.95 | 0.95 | 0.93 | 0.94 | 0.94 | 0.92 | 0.92 | 0.93 | 0.84 | 0.84 | 0.84 |

[a]Renormalized such that it is unity for the model with the highest maximum value of the probability density distribution function, for the data set and the appropriate beamwidth- and calibration-uncertainty treatment. With the normalization set such that $L(Q_{\rm rms-PS} = 0) = 1$, these highest maximum values of the likelihoods are roughly 2, $5 \cdot 10^6$, $2 \cdot 10^6$, and $5 \cdot 10^{19}$ for Ka2, Q2, Ka and Ka+Q, respectively.

[b]Computed using eq. (8). Ignores beamwidth and calibration uncertainties.

[c]Accounts for beamwidth uncertainty using eq. (12). Ignores calibration uncertainty.

[d]Computed using eqs. (12) and (16). Accounts for beamwidth and calibration uncertainties.

Table 11: Renormalized Marginal Values of the Probability Density Distribution Functions[a]

| Data: | Ka2 | | | Q2 | | | Ka | | | Ka + Q | | |
|---|---|---|---|---|---|---|---|---|---|---|---|---|
| Model | Bare[b] | B.[c] | B.+C.[d] | Bare[b] | B.[c] | B.+C.[d] | Bare[b] | B.[c] | B.+C.[d] | Bare[b] | B.[c] | B.+C.[d] |
| O1 | 0.99 | 0.99 | 0.99 | 1.0 | 1.0 | 1.0 | 1.0 | 1.0 | 1.0 | 1.0 | 1.0 | 1.0 |
| O2 | 0.96 | 0.96 | 0.96 | 0.73 | 0.73 | 0.73 | 0.79 | 0.79 | 0.79 | 0.85 | 0.85 | 0.85 |
| O3 | 0.98 | 0.98 | 0.98 | 0.79 | 0.79 | 0.80 | 0.84 | 0.84 | 0.84 | 0.89 | 0.89 | 0.89 |
| O4 | 1.0 | 1.0 | 1.0 | 0.86 | 0.86 | 0.87 | 0.90 | 0.90 | 0.90 | 0.93 | 0.94 | 0.94 |
| O5 | 0.81 | 0.81 | 0.82 | 0.52 | 0.52 | 0.53 | 0.59 | 0.59 | 0.59 | 0.66 | 0.67 | 0.67 |
| O6 | 0.84 | 0.84 | 0.84 | 0.58 | 0.58 | 0.59 | 0.64 | 0.64 | 0.64 | 0.71 | 0.71 | 0.71 |
| O7 | 0.86 | 0.86 | 0.86 | 0.66 | 0.66 | 0.66 | 0.71 | 0.71 | 0.71 | 0.75 | 0.76 | 0.76 |
| O8 | 0.67 | 0.67 | 0.67 | 0.39 | 0.40 | 0.40 | 0.45 | 0.45 | 0.45 | 0.52 | 0.52 | 0.52 |
| O9 | 0.69 | 0.69 | 0.69 | 0.45 | 0.45 | 0.45 | 0.50 | 0.50 | 0.50 | 0.56 | 0.56 | 0.56 |
| O10 | 0.72 | 0.71 | 0.71 | 0.52 | 0.52 | 0.52 | 0.56 | 0.56 | 0.56 | 0.61 | 0.61 | 0.61 |
| O11 | 0.53 | 0.53 | 0.53 | 0.29 | 0.29 | 0.29 | 0.34 | 0.34 | 0.34 | 0.39 | 0.40 | 0.40 |
| O12 | 0.56 | 0.55 | 0.55 | 0.33 | 0.33 | 0.33 | 0.38 | 0.38 | 0.38 | 0.43 | 0.43 | 0.44 |
| O13 | 0.58 | 0.58 | 0.58 | 0.39 | 0.39 | 0.39 | 0.43 | 0.43 | 0.43 | 0.48 | 0.48 | 0.48 |
| O14 | 0.48 | 0.48 | 0.48 | 0.25 | 0.25 | 0.25 | 0.29 | 0.29 | 0.29 | 0.35 | 0.35 | 0.35 |
| Λ1 | 0.56 | 0.56 | 0.56 | 0.19 | 0.19 | 0.19 | 0.24 | 0.24 | 0.24 | 0.32 | 0.32 | 0.32 |
| Λ2 | 0.55 | 0.55 | 0.55 | 0.23 | 0.23 | 0.23 | 0.28 | 0.28 | 0.28 | 0.35 | 0.36 | 0.36 |
| Λ3 | 0.52 | 0.52 | 0.52 | 0.19 | 0.19 | 0.19 | 0.24 | 0.24 | 0.24 | 0.31 | 0.31 | 0.31 |
| Λ4 | 0.49 | 0.49 | 0.49 | 0.17 | 0.17 | 0.17 | 0.21 | 0.21 | 0.21 | 0.28 | 0.28 | 0.28 |
| Λ5 | 0.52 | 0.51 | 0.52 | 0.23 | 0.23 | 0.23 | 0.27 | 0.28 | 0.28 | 0.34 | 0.34 | 0.35 |
| Λ6 | 0.48 | 0.48 | 0.48 | 0.19 | 0.19 | 0.19 | 0.24 | 0.24 | 0.24 | 0.30 | 0.30 | 0.30 |
| Λ7 | 0.46 | 0.46 | 0.45 | 0.16 | 0.16 | 0.16 | 0.20 | 0.21 | 0.21 | 0.27 | 0.27 | 0.27 |
| Λ8 | 0.50 | 0.50 | 0.50 | 0.23 | 0.23 | 0.23 | 0.28 | 0.28 | 0.28 | 0.34 | 0.34 | 0.35 |
| Λ9 | 0.47 | 0.47 | 0.47 | 0.19 | 0.19 | 0.19 | 0.24 | 0.24 | 0.24 | 0.30 | 0.30 | 0.30 |
| Λ10 | 0.44 | 0.44 | 0.44 | 0.16 | 0.16 | 0.16 | 0.20 | 0.20 | 0.21 | 0.26 | 0.26 | 0.27 |
| Λ11 | 0.47 | 0.47 | 0.47 | 0.21 | 0.21 | 0.21 | 0.25 | 0.25 | 0.25 | 0.31 | 0.31 | 0.32 |
| Λ12 | 0.48 | 0.48 | 0.48 | 0.25 | 0.25 | 0.25 | 0.29 | 0.29 | 0.29 | 0.35 | 0.35 | 0.35 |
| Flat | 0.67 | 0.67 | 0.67 | 0.68 | 0.68 | 0.68 | 0.67 | 0.67 | 0.67 | 0.62 | 0.62 | 0.62 |

[a]Renormalized such that it is unity for the model with the highest marginal probability density distribution function value, for the data set and the appropriate beamwidth- and calibration-uncertainty treatment. When the likelihood normalization is set such that is it unity for $Q_{\mathrm{rms-PS}} = 0$, these highest marginal values are roughly 50, $1 \cdot 10^8$, $6 \cdot 10^7$ and $1 \cdot 10^{21}$ for Ka2, Q2, Ka and Ka+Q respectively.

[b]Computed using eq. (8). Ignores beamwidth and calibration uncertainties.

[c]Accounts for beamwidth uncertainty using eq. (12). Ignores calibration uncertainty.

[d]Computed using eqs. (12) and (16). Accounts for beamwidth and calibration uncertainties.

Table 12: Numerical Values for $Q_{\mathrm{rms-PS}}$ (in $\mu$K), Accounting for Beamwidth and Calibration Uncertainties[a]

| Model | $(\Omega_0,\, h,\, \Omega_B h^2)$ | DMR[b] | Ka1 | Ka2 | Ka3 | Ka4 | Ka |
|---|---|---|---|---|---|---|---|
| O1 | (0.1, 0.75, 0.0125) | $23.1^{25.6}_{20.5}$ | $46^{66}_{32}$ | $17^{44}_{\dots}$ | $31^{45}_{21}$ | $26^{38}_{18}$ | $28^{39}_{20}$ |
| O2 | (0.2, 0.65, 0.0175) | $26.5^{29.4}_{23.6}$ | $45^{64}_{31}$ | $16^{43}_{\dots}$ | $30^{44}_{21}$ | $26^{38}_{18}$ | $27^{38}_{19}$ |
| O3 | (0.2, 0.70, 0.0125) | $26.5^{29.4}_{23.6}$ | $46^{65}_{32}$ | $17^{44}_{\dots}$ | $31^{45}_{21}$ | $27^{39}_{18}$ | $28^{39}_{20}$ |
| O4 | (0.2, 0.75, 0.0075) | $26.5^{29.4}_{23.6}$ | $47^{67}_{33}$ | $17^{45}_{\dots}$ | $32^{46}_{22}$ | $27^{40}_{18}$ | $28^{40}_{20}$ |
| O5 | (0.3, 0.60, 0.0175) | $25.8^{28.6}_{23.0}$ | $38^{55}_{26}$ | $14^{37}_{\dots}$ | $26^{38}_{18}$ | $23^{33}_{15}$ | $23^{33}_{17}$ |
| O6 | (0.3, 0.65, 0.0125) | $25.8^{28.6}_{23.0}$ | $39^{57}_{27}$ | $14^{38}_{\dots}$ | $27^{39}_{18}$ | $23^{33}_{16}$ | $24^{34}_{17}$ |
| O7 | (0.3, 0.70, 0.0075) | $25.8^{28.6}_{23.0}$ | $41^{59}_{28}$ | $15^{39}_{\dots}$ | $28^{40}_{19}$ | $24^{34}_{16}$ | $25^{35}_{18}$ |
| O8 | (0.4, 0.60, 0.0175) | $23.4^{25.9}_{20.8}$ | $32^{46}_{22}$ | $12^{30}_{\dots}$ | $22^{31}_{15}$ | $19^{27}_{13}$ | $19^{27}_{14}$ |
| O9 | (0.4, 0.65, 0.0125) | $23.4^{25.9}_{20.8}$ | $33^{47}_{22}$ | $12^{31}_{\dots}$ | $22^{32}_{15}$ | $19^{28}_{13}$ | $20^{28}_{14}$ |
| O10 | (0.4, 0.70, 0.0075) | $23.4^{25.9}_{20.8}$ | $34^{49}_{23}$ | $12^{33}_{\dots}$ | $23^{34}_{16}$ | $20^{29}_{13}$ | $21^{29}_{15}$ |
| O11 | (0.5, 0.55, 0.0175) | $20.5^{22.8}_{18.3}$ | $25^{37}_{17}$ | $9.3^{24}_{\dots}$ | $17^{25}_{12}$ | $15^{22}_{10}$ | $16^{22}_{11}$ |
| O12 | (0.5, 0.60, 0.0125) | $20.5^{22.8}_{18.3}$ | $26^{38}_{18}$ | $9.7^{25}_{\dots}$ | $18^{26}_{12}$ | $16^{23}_{11}$ | $16^{23}_{12}$ |
| O13 | (0.5, 0.65, 0.0075) | $20.5^{22.8}_{18.3}$ | $28^{40}_{19}$ | $10^{27}_{\dots}$ | $19^{27}_{13}$ | $16^{24}_{11}$ | $17^{24}_{12}$ |
| O14 | (1.0, 0.50, 0.0125) | $19.8^{22.0}_{17.7}$ | $23^{33}_{16}$ | $8.4^{22}_{\dots}$ | $16^{23}_{11}$ | $14^{20}_{9.1}$ | $14^{20}_{10}$ |
| Λ1 | (0.1, 0.90, 0.0125) | $25.6^{28.4}_{22.8}$ | $27^{39}_{19}$ | $9.9^{25}_{\dots}$ | $19^{27}_{13}$ | $16^{23}_{11}$ | $17^{23}_{12}$ |
| Λ2 | (0.2, 0.80, 0.0075) | $23.7^{26.3}_{21.1}$ | $26^{38}_{18}$ | $9.7^{25}_{\dots}$ | $18^{26}_{12}$ | $16^{23}_{11}$ | $16^{23}_{12}$ |
| Λ3 | (0.2, 0.75, 0.0125) | $23.7^{26.3}_{21.1}$ | $25^{35}_{17}$ | $9.1^{23}_{\dots}$ | $17^{25}_{12}$ | $15^{21}_{9.9}$ | $15^{21}_{11}$ |
| Λ4 | (0.2, 0.70, 0.0175) | $23.7^{26.3}_{21.1}$ | $23^{34}_{16}$ | $8.7^{22}_{\dots}$ | $16^{23}_{11}$ | $14^{20}_{9.4}$ | $15^{20}_{11}$ |
| Λ5 | (0.3, 0.70, 0.0075) | $22.2^{24.7}_{19.8}$ | $24^{35}_{17}$ | $9.1^{23}_{\dots}$ | $17^{24}_{12}$ | $15^{21}_{9.8}$ | $15^{21}_{11}$ |
| Λ6 | (0.3, 0.65, 0.0125) | $22.2^{24.7}_{19.8}$ | $23^{33}_{16}$ | $8.5^{22}_{\dots}$ | $16^{23}_{11}$ | $14^{20}_{9.2}$ | $14^{20}_{10}$ |
| Λ7 | (0.3, 0.60, 0.0175) | $22.2^{24.7}_{19.8}$ | $22^{31}_{15}$ | $8.0^{21}_{\dots}$ | $15^{22}_{10}$ | $13^{19}_{8.8}$ | $14^{19}_{9.8}$ |
| Λ8 | (0.4, 0.65, 0.0075) | $21.3^{23.7}_{19.0}$ | $24^{34}_{16}$ | $8.7^{22}_{\dots}$ | $16^{24}_{11}$ | $14^{20}_{9.5}$ | $15^{20}_{10}$ |
| Λ9 | (0.4, 0.60, 0.0125) | $21.3^{23.7}_{19.0}$ | $22^{32}_{15}$ | $8.2^{21}_{\dots}$ | $15^{22}_{10}$ | $13^{19}_{8.9}$ | $14^{19}_{9.9}$ |
| Λ10 | (0.4, 0.55, 0.0175) | $21.3^{23.7}_{19.0}$ | $21^{30}_{14}$ | $7.7^{20}_{\dots}$ | $14^{21}_{9.9}$ | $13^{18}_{8.4}$ | $13^{18}_{9.4}$ |
| Λ11 | (0.5, 0.60, 0.0125) | $20.8^{23.1}_{18.5}$ | $22^{32}_{15}$ | $8.2^{21}_{\dots}$ | $15^{22}_{10}$ | $13^{19}_{8.9}$ | $14^{19}_{9.9}$ |
| Λ12 | (1.0, 0.50, 0.0125) | $20.4^{22.7}_{18.2}$ | $23^{33}_{16}$ | $8.4^{22}_{\dots}$ | $16^{23}_{11}$ | $14^{20}_{9.1}$ | $14^{20}_{10}$ |
| Flat | ... | $20.4^{22.6}_{18.2}$ | $33^{48}_{22}$ | $12^{32}_{\dots}$ | $22^{32}_{15}$ | $19^{27}_{13}$ | $20^{28}_{14}$ |

[a]For each model, the first of the three entries in each of the last six columns is the value at which the probability density distribution function peaks. Ellipses as the lower entry in a vertical pair denotes a non-detection; the corresponding upper entry is the $2\sigma$ (97.72% ET) upper limit. For detections, the vertical pair of numbers are the $\pm 1\sigma$ (68.27% HPD) upper and lower limits.

[b]DMR values for open models is from two-year DMR (galactic-frame, quadrupole-excluded) normalization of Górski et al. (1995) (also see RSBG), for flat-$\Lambda$ models from two-year DMR (ecliptic-frame, quadrupole-excluded) normalization of Bunn & Sugiyama (1995) (also see RSBG), and for flat bandpower from two-year DMR (galactic-frame, quadrupole-excluded) normalization of Górski et al. (1994). DMR $1\sigma$ range accounts for DMR statistical and systematic uncertainties following Stompor et al. (1995) (also see RSBG).

Table 13: Numerical Values for $Q_{\mathrm{rms-PS}}$ (in $\mu$K), Accounting for Beamwidth and Calibration Uncertainties[a]

| Model | $(\Omega_0,\, h,\, \Omega_B h^2)$ | DMR | Q1 | Q2 | Q3 | Q | Ka + Q |
|---|---|---|---|---|---|---|---|
| O1 | (0.1, 0.75, 0.0125) | $23.1^{25.6}_{20.5}$ | $25^{39}_{15}$ | $36^{49}_{26}$ | $38^{51}_{28}$ | $37^{50}_{28}$ | $35^{47}_{26}$ |
| O2 | (0.2, 0.65, 0.0175) | $26.5^{29.4}_{23.6}$ | $25^{38}_{14}$ | $35^{48}_{26}$ | $37^{50}_{27}$ | $36^{49}_{27}$ | $34^{46}_{26}$ |
| O3 | (0.2, 0.70, 0.0125) | $26.5^{29.4}_{23.6}$ | $25^{39}_{15}$ | $36^{49}_{26}$ | $37^{51}_{28}$ | $37^{50}_{28}$ | $35^{47}_{26}$ |
| O4 | (0.2, 0.75, 0.0075) | $26.5^{29.4}_{23.6}$ | $26^{40}_{15}$ | $37^{51}_{27}$ | $38^{52}_{28}$ | $38^{52}_{29}$ | $36^{48}_{27}$ |
| O5 | (0.3, 0.60, 0.0175) | $25.8^{28.6}_{23.0}$ | $21^{33}_{12}$ | $30^{42}_{22}$ | $31^{42}_{23}$ | $31^{42}_{23}$ | $29^{39}_{22}$ |
| O6 | (0.3, 0.65, 0.0125) | $25.8^{28.6}_{23.0}$ | $22^{34}_{12}$ | $31^{43}_{23}$ | $32^{44}_{24}$ | $32^{43}_{24}$ | $30^{40}_{23}$ |
| O7 | (0.3, 0.70, 0.0075) | $25.8^{28.6}_{23.0}$ | $22^{35}_{13}$ | $32^{44}_{24}$ | $33^{45}_{25}$ | $33^{45}_{25}$ | $31^{42}_{24}$ |
| O8 | (0.4, 0.60, 0.0175) | $23.4^{25.9}_{20.8}$ | $17^{27}_{9.8}$ | $25^{35}_{19}$ | $26^{35}_{19}$ | $26^{35}_{19}$ | $24^{32}_{18}$ |
| O9 | (0.4, 0.65, 0.0125) | $23.4^{25.9}_{20.8}$ | $18^{28}_{10}$ | $26^{36}_{19}$ | $27^{36}_{20}$ | $27^{36}_{20}$ | $25^{34}_{19}$ |
| O10 | (0.4, 0.70, 0.0075) | $23.4^{25.9}_{20.8}$ | $19^{29}_{11}$ | $27^{37}_{20}$ | $28^{38}_{21}$ | $28^{38}_{21}$ | $26^{35}_{20}$ |
| O11 | (0.5, 0.55, 0.0175) | $20.5^{22.8}_{18.3}$ | $14^{22}_{7.8}$ | $20^{28}_{15}$ | $21^{28}_{15}$ | $21^{28}_{16}$ | $19^{26}_{15}$ |
| O12 | (0.5, 0.60, 0.0125) | $20.5^{22.8}_{18.3}$ | $15^{23}_{8.2}$ | $21^{29}_{16}$ | $22^{29}_{16}$ | $22^{29}_{16}$ | $20^{27}_{15}$ |
| O13 | (0.5, 0.65, 0.0075) | $20.5^{22.8}_{18.3}$ | $15^{24}_{8.7}$ | $22^{30}_{16}$ | $23^{31}_{17}$ | $23^{30}_{17}$ | $21^{28}_{16}$ |
| O14 | (1.0, 0.50, 0.0125) | $19.8^{22.0}_{17.7}$ | $13^{19}_{7.0}$ | $18^{25}_{13}$ | $19^{25}_{14}$ | $19^{25}_{14}$ | $17^{23}_{13}$ |
| Λ1 | (0.1, 0.90, 0.0125) | $25.6^{28.4}_{22.8}$ | $16^{36}_{\dots}$ | $22^{30}_{16}$ | $22^{30}_{16}$ | $22^{30}_{17}$ | $21^{28}_{16}$ |
| Λ2 | (0.2, 0.80, 0.0075) | $23.7^{26.3}_{21.1}$ | $14^{22}_{7.9}$ | $21^{29}_{16}$ | $21^{29}_{16}$ | $22^{29}_{16}$ | $20^{27}_{15}$ |
| Λ3 | (0.2, 0.75, 0.0125) | $23.7^{26.3}_{21.1}$ | $15^{33}_{\dots}$ | $20^{27}_{15}$ | $20^{27}_{15}$ | $20^{27}_{15}$ | $19^{25}_{15}$ |
| Λ4 | (0.2, 0.70, 0.0175) | $23.7^{26.3}_{21.1}$ | $14^{31}_{\dots}$ | $19^{26}_{14}$ | $19^{26}_{14}$ | $19^{26}_{15}$ | $18^{24}_{14}$ |
| Λ5 | (0.3, 0.70, 0.0075) | $22.2^{24.7}_{19.8}$ | $13^{21}_{7.4}$ | $20^{27}_{15}$ | $20^{27}_{15}$ | $20^{27}_{15}$ | $19^{25}_{14}$ |
| Λ6 | (0.3, 0.65, 0.0125) | $22.2^{24.7}_{19.8}$ | $14^{31}_{\dots}$ | $19^{25}_{14}$ | $19^{25}_{14}$ | $19^{25}_{14}$ | $18^{24}_{14}$ |
| Λ7 | (0.3, 0.60, 0.0175) | $22.2^{24.7}_{19.8}$ | $13^{29}_{\dots}$ | $18^{24}_{13}$ | $18^{24}_{13}$ | $18^{24}_{14}$ | $17^{22}_{13}$ |
| Λ8 | (0.4, 0.65, 0.0075) | $21.3^{23.7}_{19.0}$ | $13^{20}_{7.2}$ | $19^{26}_{14}$ | $19^{26}_{14}$ | $19^{26}_{15}$ | $18^{24}_{14}$ |
| Λ9 | (0.4, 0.60, 0.0125) | $21.3^{23.7}_{19.0}$ | $12^{19}_{6.7}$ | $18^{25}_{13}$ | $18^{24}_{13}$ | $18^{24}_{14}$ | $17^{23}_{13}$ |
| Λ10 | (0.4, 0.55, 0.0175) | $21.3^{23.7}_{19.0}$ | $12^{28}_{\dots}$ | $17^{23}_{13}$ | $17^{23}_{13}$ | $17^{23}_{13}$ | $16^{21}_{12}$ |
| Λ11 | (0.5, 0.60, 0.0125) | $20.8^{23.1}_{18.5}$ | $12^{19}_{6.7}$ | $18^{24}_{13}$ | $18^{24}_{13}$ | $18^{24}_{14}$ | $17^{23}_{13}$ |
| Λ12 | (1.0, 0.50, 0.0125) | $20.4^{22.7}_{18.2}$ | $13^{19}_{7.0}$ | $18^{25}_{13}$ | $19^{25}_{14}$ | $19^{25}_{14}$ | $17^{23}_{13}$ |
| Flat | ... | $20.4^{22.6}_{18.2}$ | $18^{28}_{10}$ | $26^{35}_{19}$ | $27^{37}_{20}$ | $27^{37}_{20}$ | $25^{34}_{19}$ |

[a]Conventions are the same as for Table 12.

Table 14: Numerical Values for $\delta T_l$ (in $\mu$K), Accounting for Beamwidth and Calibration Uncertainties[a]

| Model | ($\Omega_0$, $h$, $\Omega_B h^2$) | Ka1 | Ka2 | Ka3 | Ka4 | Q1 | Q2 | Q3 |
|---|---|---|---|---|---|---|---|---|
| O1 | (0.1, 0.75, 0.0125) | $51^{73}_{35}$ | $19^{49}_{\dots}$ | $35^{50}_{23}$ | $29^{43}_{20}$ | $28^{44}_{16}$ | $40^{55}_{29}$ | $42^{57}_{31}$ |
| O2 | (0.2, 0.65, 0.0175) | $50^{71}_{34}$ | $18^{48}_{\dots}$ | $34^{50}_{23}$ | $30^{43}_{20}$ | $28^{43}_{16}$ | $40^{55}_{30}$ | $42^{57}_{31}$ |
| O3 | (0.2, 0.70, 0.0125) | $50^{71}_{35}$ | $18^{48}_{\dots}$ | $34^{50}_{23}$ | $29^{43}_{20}$ | $28^{44}_{16}$ | $40^{55}_{30}$ | $42^{57}_{31}$ |
| O4 | (0.2, 0.75, 0.0075) | $50^{72}_{35}$ | $18^{48}_{\dots}$ | $34^{50}_{23}$ | $29^{43}_{20}$ | $28^{43}_{16}$ | $40^{55}_{30}$ | $42^{57}_{31}$ |
| O5 | (0.3, 0.60, 0.0175) | $49^{70}_{34}$ | $18^{47}_{\dots}$ | $34^{49}_{23}$ | $30^{43}_{20}$ | $28^{43}_{16}$ | $41^{56}_{30}$ | $42^{57}_{31}$ |
| O6 | (0.3, 0.65, 0.0125) | $49^{71}_{34}$ | $18^{48}_{\dots}$ | $34^{49}_{23}$ | $30^{43}_{20}$ | $28^{43}_{16}$ | $40^{56}_{30}$ | $42^{57}_{31}$ |
| O7 | (0.3, 0.70, 0.0075) | $50^{71}_{34}$ | $18^{48}_{\dots}$ | $34^{50}_{23}$ | $29^{43}_{20}$ | $28^{43}_{16}$ | $40^{55}_{30}$ | $42^{57}_{31}$ |
| O8 | (0.4, 0.60, 0.0175) | $48^{70}_{33}$ | $18^{47}_{\dots}$ | $34^{49}_{23}$ | $30^{43}_{20}$ | $28^{43}_{16}$ | $41^{56}_{30}$ | $42^{57}_{31}$ |
| O9 | (0.4, 0.65, 0.0125) | $49^{71}_{33}$ | $18^{47}_{\dots}$ | $34^{49}_{23}$ | $29^{43}_{20}$ | $28^{43}_{16}$ | $41^{56}_{30}$ | $42^{57}_{31}$ |
| O10 | (0.4, 0.70, 0.0075) | $49^{71}_{34}$ | $18^{48}_{\dots}$ | $34^{49}_{23}$ | $29^{43}_{20}$ | $28^{43}_{16}$ | $40^{55}_{30}$ | $42^{57}_{31}$ |
| O11 | (0.5, 0.55, 0.0175) | $48^{70}_{33}$ | $18^{47}_{\dots}$ | $34^{49}_{23}$ | $30^{43}_{20}$ | $28^{43}_{16}$ | $41^{56}_{30}$ | $42^{57}_{31}$ |
| O12 | (0.5, 0.60, 0.0125) | $48^{70}_{33}$ | $18^{47}_{\dots}$ | $34^{49}_{23}$ | $30^{43}_{20}$ | $28^{43}_{16}$ | $41^{56}_{30}$ | $42^{57}_{31}$ |
| O13 | (0.5, 0.65, 0.0075) | $49^{71}_{33}$ | $18^{47}_{\dots}$ | $34^{49}_{23}$ | $29^{43}_{20}$ | $28^{43}_{16}$ | $41^{56}_{30}$ | $42^{57}_{31}$ |
| O14 | (1.0, 0.50, 0.0125) | $48^{69}_{33}$ | $18^{46}_{\dots}$ | $34^{49}_{23}$ | $30^{43}_{20}$ | $28^{43}_{16}$ | $41^{56}_{30}$ | $42^{57}_{31}$ |
| Λ1 | (0.1, 0.90, 0.0125) | $47^{67}_{32}$ | $18^{45}_{\dots}$ | $34^{49}_{23}$ | $30^{43}_{20}$ | $30^{68}_{\dots}$ | $42^{58}_{31}$ | $43^{58}_{32}$ |
| Λ2 | (0.2, 0.80, 0.0075) | $47^{68}_{33}$ | $18^{46}_{\dots}$ | $34^{49}_{23}$ | $30^{43}_{20}$ | $28^{43}_{15}$ | $42^{57}_{31}$ | $42^{57}_{31}$ |
| Λ3 | (0.2, 0.75, 0.0125) | $47^{67}_{32}$ | $18^{46}_{\dots}$ | $34^{49}_{23}$ | $30^{43}_{20}$ | $30^{68}_{\dots}$ | $42^{57}_{31}$ | $42^{57}_{32}$ |
| Λ4 | (0.2, 0.70, 0.0175) | $47^{67}_{32}$ | $18^{45}_{\dots}$ | $34^{49}_{23}$ | $30^{43}_{20}$ | $30^{68}_{\dots}$ | $42^{58}_{31}$ | $43^{58}_{32}$ |
| Λ5 | (0.3, 0.70, 0.0075) | $47^{68}_{33}$ | $18^{46}_{\dots}$ | $34^{49}_{23}$ | $30^{43}_{20}$ | $28^{43}_{15}$ | $42^{57}_{31}$ | $42^{57}_{31}$ |
| Λ6 | (0.3, 0.65, 0.0125) | $47^{68}_{33}$ | $18^{46}_{\dots}$ | $34^{49}_{23}$ | $30^{43}_{20}$ | $30^{68}_{\dots}$ | $42^{57}_{31}$ | $42^{57}_{32}$ |
| Λ7 | (0.3, 0.60, 0.0175) | $47^{67}_{32}$ | $18^{45}_{\dots}$ | $34^{49}_{23}$ | $30^{43}_{20}$ | $30^{68}_{\dots}$ | $42^{58}_{31}$ | $42^{57}_{32}$ |
| Λ8 | (0.4, 0.65, 0.0075) | $48^{68}_{33}$ | $18^{46}_{\dots}$ | $34^{49}_{23}$ | $30^{43}_{20}$ | $28^{43}_{16}$ | $41^{57}_{30}$ | $42^{57}_{31}$ |
| Λ9 | (0.4, 0.60, 0.0125) | $47^{68}_{33}$ | $18^{46}_{\dots}$ | $34^{49}_{23}$ | $30^{43}_{20}$ | $28^{43}_{15}$ | $42^{57}_{31}$ | $42^{57}_{31}$ |
| Λ10 | (0.4, 0.55, 0.0175) | $47^{67}_{32}$ | $18^{45}_{\dots}$ | $34^{49}_{23}$ | $30^{43}_{20}$ | $30^{68}_{\dots}$ | $42^{57}_{31}$ | $42^{57}_{32}$ |
| Λ11 | (0.5, 0.60, 0.0125) | $47^{68}_{33}$ | $18^{46}_{\dots}$ | $34^{49}_{23}$ | $30^{43}_{20}$ | $28^{43}_{15}$ | $41^{57}_{31}$ | $42^{57}_{31}$ |
| Λ12 | (1.0, 0.50, 0.0125) | $48^{69}_{33}$ | $18^{46}_{\dots}$ | $34^{49}_{23}$ | $30^{43}_{20}$ | $28^{43}_{16}$ | $41^{56}_{30}$ | $42^{57}_{31}$ |
| Flat | ... | $50^{74}_{34}$ | $18^{49}_{\dots}$ | $34^{50}_{23}$ | $29^{43}_{20}$ | $28^{43}_{16}$ | $40^{55}_{29}$ | $42^{58}_{31}$ |

[a]For each model, the first of the three entries in each of the last seven columns is the value at which the probability density distribution function peaks. Ellipses as the lower entry in a vertical pair denotes a non-detection; the corresponding upper entry is the $2\sigma$ (97.72% ET) upper limit. For detections, the vertical pair of numbers are the $\pm 1\sigma$ (68.27% HPD) upper and lower limits.

Table 15: Renormalized Maximum Values of the Probability Density Distribution Functions[a]

| Model | Ka1 | Ka2 | Ka3 | Ka4 | Q1 | Q2 | Q3 | Ka | Q | Ka + Q |
|---|---|---|---|---|---|---|---|---|---|---|
| O1 | 1.0 | 1.0 | 1.0 | 1.0 | 1.0 | 1.0 | 1.0 | 1.0 | 1.0 | 1.0 |
| O2 | 1.0 | 1.0 | 0.88 | 0.80 | 0.87 | 0.75 | 0.82 | 0.81 | 0.81 | 0.89 |
| O3 | 0.99 | 0.99 | 0.90 | 0.84 | 0.90 | 0.79 | 0.85 | 0.85 | 0.84 | 0.91 |
| O4 | 0.99 | 0.99 | 0.92 | 0.88 | 0.93 | 0.85 | 0.89 | 0.88 | 0.88 | 0.92 |
| O5 | 0.99 | 0.98 | 0.82 | 0.70 | 0.80 | 0.62 | 0.72 | 0.71 | 0.70 | 0.81 |
| O6 | 0.98 | 0.98 | 0.84 | 0.74 | 0.83 | 0.67 | 0.76 | 0.75 | 0.74 | 0.84 |
| O7 | 0.97 | 0.98 | 0.87 | 0.78 | 0.87 | 0.74 | 0.81 | 0.80 | 0.79 | 0.86 |
| O8 | 0.98 | 0.97 | 0.78 | 0.65 | 0.77 | 0.57 | 0.68 | 0.66 | 0.65 | 0.77 |
| O9 | 0.97 | 0.97 | 0.81 | 0.69 | 0.80 | 0.62 | 0.72 | 0.71 | 0.70 | 0.80 |
| O10 | 0.96 | 0.97 | 0.84 | 0.74 | 0.84 | 0.69 | 0.78 | 0.76 | 0.75 | 0.83 |
| O11 | 0.97 | 0.97 | 0.75 | 0.59 | 0.73 | 0.51 | 0.63 | 0.61 | 0.60 | 0.73 |
| O12 | 0.97 | 0.97 | 0.78 | 0.64 | 0.77 | 0.57 | 0.68 | 0.66 | 0.65 | 0.77 |
| O13 | 0.96 | 0.96 | 0.82 | 0.70 | 0.81 | 0.64 | 0.74 | 0.71 | 0.71 | 0.80 |
| O14 | 0.98 | 0.97 | 0.74 | 0.57 | 0.72 | 0.49 | 0.61 | 0.58 | 0.58 | 0.72 |
| Λ1 | 0.95 | 0.96 | 0.60 | 0.41 | 0.58 | 0.31 | 0.44 | 0.41 | 0.40 | 0.56 |
| Λ2 | 0.98 | 0.97 | 0.67 | 0.49 | 0.65 | 0.39 | 0.52 | 0.49 | 0.49 | 0.65 |
| Λ3 | 0.97 | 0.97 | 0.64 | 0.45 | 0.62 | 0.35 | 0.48 | 0.45 | 0.45 | 0.60 |
| Λ4 | 0.96 | 0.96 | 0.61 | 0.41 | 0.59 | 0.32 | 0.45 | 0.42 | 0.41 | 0.57 |
| Λ5 | 0.98 | 0.97 | 0.69 | 0.51 | 0.67 | 0.42 | 0.55 | 0.52 | 0.52 | 0.67 |
| Λ6 | 0.97 | 0.97 | 0.65 | 0.46 | 0.63 | 0.37 | 0.50 | 0.47 | 0.47 | 0.62 |
| Λ7 | 0.96 | 0.96 | 0.62 | 0.43 | 0.60 | 0.33 | 0.46 | 0.43 | 0.43 | 0.59 |
| Λ8 | 0.98 | 0.97 | 0.71 | 0.53 | 0.69 | 0.44 | 0.57 | 0.54 | 0.54 | 0.69 |
| Λ9 | 0.97 | 0.97 | 0.67 | 0.48 | 0.65 | 0.39 | 0.52 | 0.49 | 0.49 | 0.64 |
| Λ10 | 0.96 | 0.96 | 0.64 | 0.44 | 0.61 | 0.35 | 0.48 | 0.45 | 0.44 | 0.60 |
| Λ11 | 0.98 | 0.97 | 0.69 | 0.51 | 0.67 | 0.42 | 0.55 | 0.52 | 0.52 | 0.67 |
| Λ12 | 0.98 | 0.97 | 0.74 | 0.57 | 0.72 | 0.49 | 0.61 | 0.58 | 0.58 | 0.72 |
| Flat | 0.91 | 0.95 | 0.93 | 0.92 | 0.99 | 0.94 | 0.95 | 0.93 | 0.91 | 0.84 |
| | $6 \cdot 10^3$ | 2 | $1 \cdot 10^3$ | $3 \cdot 10^2$ | $1 \cdot 10^1$ | $5 \cdot 10^6$ | $1 \cdot 10^7$ | $2 \cdot 10^6$ | $6 \cdot 10^9$ | $5 \cdot 10^{19}$ |

[a]Accounts for beamwidth and calibration uncertainties. Renormalized such that it is unity for the model with the highest maximum value of the probability density distribution function for the data set. The last line of the table gives this highest maximum likelihood value when the normalization is set such that $L(Q_{\mathrm{rms-PS}} = 0\ \mu\mathrm{K}) = 1$.

Table 16: Renormalized Marginal Values of the Probability Density Distribution Functions[a]

| Model | Ka1 | Ka2 | Ka3 | Ka4 | Q1 | Q2 | Q3 | Ka | Q | Ka + Q |
|---|---|---|---|---|---|---|---|---|---|---|
| O1 | 1.0 | 0.99 | 1.0 | 1.0 | 1.0 | 1.0 | 1.0 | 1.0 | 1.0 | 1.0 |
| O2 | 0.97 | 0.96 | 0.86 | 0.79 | 0.86 | 0.73 | 0.79 | 0.79 | 0.77 | 0.85 |
| O3 | 0.99 | 0.98 | 0.90 | 0.84 | 0.90 | 0.80 | 0.85 | 0.84 | 0.83 | 0.89 |
| O4 | 1.0 | 1.0 | 0.95 | 0.90 | 0.95 | 0.87 | 0.91 | 0.90 | 0.90 | 0.94 |
| O5 | 0.83 | 0.82 | 0.68 | 0.59 | 0.68 | 0.53 | 0.59 | 0.59 | 0.58 | 0.67 |
| O6 | 0.85 | 0.84 | 0.73 | 0.64 | 0.72 | 0.59 | 0.65 | 0.64 | 0.63 | 0.71 |
| O7 | 0.87 | 0.86 | 0.77 | 0.71 | 0.78 | 0.66 | 0.72 | 0.71 | 0.70 | 0.76 |
| O8 | 0.68 | 0.67 | 0.54 | 0.45 | 0.54 | 0.40 | 0.46 | 0.45 | 0.44 | 0.52 |
| O9 | 0.71 | 0.69 | 0.58 | 0.50 | 0.58 | 0.45 | 0.51 | 0.50 | 0.49 | 0.56 |
| O10 | 0.73 | 0.71 | 0.63 | 0.56 | 0.63 | 0.52 | 0.58 | 0.56 | 0.55 | 0.61 |
| O11 | 0.54 | 0.53 | 0.42 | 0.34 | 0.41 | 0.29 | 0.34 | 0.34 | 0.33 | 0.40 |
| O12 | 0.57 | 0.55 | 0.45 | 0.38 | 0.45 | 0.33 | 0.39 | 0.38 | 0.37 | 0.44 |
| O13 | 0.59 | 0.58 | 0.49 | 0.43 | 0.50 | 0.39 | 0.44 | 0.43 | 0.42 | 0.48 |
| O14 | 0.49 | 0.48 | 0.37 | 0.29 | 0.37 | 0.25 | 0.30 | 0.29 | 0.29 | 0.35 |
| Λ1 | 0.55 | 0.56 | 0.36 | 0.25 | 0.36 | 0.19 | 0.25 | 0.24 | 0.23 | 0.32 |
| Λ2 | 0.55 | 0.55 | 0.38 | 0.29 | 0.38 | 0.23 | 0.29 | 0.28 | 0.27 | 0.36 |
| Λ3 | 0.51 | 0.52 | 0.34 | 0.25 | 0.35 | 0.19 | 0.25 | 0.24 | 0.24 | 0.31 |
| Λ4 | 0.48 | 0.49 | 0.31 | 0.22 | 0.32 | 0.17 | 0.22 | 0.21 | 0.21 | 0.28 |
| Λ5 | 0.52 | 0.52 | 0.37 | 0.28 | 0.37 | 0.23 | 0.28 | 0.28 | 0.27 | 0.35 |
| Λ6 | 0.48 | 0.48 | 0.33 | 0.24 | 0.33 | 0.19 | 0.24 | 0.24 | 0.23 | 0.30 |
| Λ7 | 0.45 | 0.45 | 0.30 | 0.21 | 0.30 | 0.16 | 0.21 | 0.21 | 0.20 | 0.27 |
| Λ8 | 0.50 | 0.50 | 0.36 | 0.28 | 0.36 | 0.23 | 0.29 | 0.28 | 0.27 | 0.35 |
| Λ9 | 0.47 | 0.47 | 0.32 | 0.24 | 0.32 | 0.19 | 0.24 | 0.24 | 0.23 | 0.30 |
| Λ10 | 0.44 | 0.44 | 0.29 | 0.21 | 0.29 | 0.16 | 0.21 | 0.21 | 0.20 | 0.27 |
| Λ11 | 0.47 | 0.47 | 0.33 | 0.26 | 0.33 | 0.21 | 0.26 | 0.25 | 0.25 | 0.32 |
| Λ12 | 0.49 | 0.48 | 0.37 | 0.29 | 0.37 | 0.25 | 0.30 | 0.29 | 0.29 | 0.35 |
| Flat | 0.68 | 0.67 | 0.67 | 0.66 | 0.71 | 0.68 | 0.71 | 0.67 | 0.67 | 0.62 |
| | $3 \cdot 10^5$ | 50 | $3 \cdot 10^4$ | $8 \cdot 10^3$ | $3 \cdot 10^2$ | $1 \cdot 10^8$ | $3 \cdot 10^8$ | $6 \cdot 10^7$ | $2 \cdot 10^{11}$ | $1 \cdot 10^{21}$ |

[a]Accounts for beamwidth and calibration uncertainties. Renormalized such that it is unity for the model with the highest marginal probability density distribution function value for the data set. The last line of the table gives the marginal value for the model with highest marginal probability density distribution function value when the likelihoods are normalized such that $L(Q_{\rm rms-PS} = 0\ \mu{\rm K}) = 1$.

## FIGURE CAPTIONS

Fig. 1.— CMB anisotropy multipole moments $l(l+1)C_l/(2\pi) \times 10^{10}$ (broken lines, scale on left axis) as a function of multipole $l$, to $l = 300$, for selected models O1, O11, O14, $\Lambda$2, $\Lambda$10, and Flat, normalized to the two-year DMR maps. O1, O11, O14, and Flat are normalized to the galactic-coordinates quadrupole-excluded maps, while $\Lambda$2 and $\Lambda$10 are normalized to the ecliptic-coordinates quadrupole-excluded ones. See Table 13 for model-parameter values. Also shown are two SP94 individual-channel, zero-lag, nominal beamwidth, window functions $W_l$ (solid lines, scale on right axis), sensitive to the largest (Ka1) and smallest (Q3) angular scales. See Table 1 for window function parameter values.

Fig. 2.— Individual-channel, nominal beamwidth, zero-lag SP94 window functions $W_l$ (solid lines), as a function of $l$, to $l = 300$. Also shown (hatched regions) is the uncertainty in the windows sensitive to the largest (Ka1) and smallest (Q3) angular scales, due to the one standard deviation uncertainty in the beamwidths. See Table 1 for window function parameter values. The $W_l$s are computed using eq. (1) of G95. The $W_l$ used here assume that the SP94 beam is gaussian and do not explicitly account for the small ellipticity.

Fig. 3.— $(\delta T_{\rm rms}{}^2)_l$ (eq. [6]) as a function of $l$, to $l = 250$, for (upper panel) the larger beamwidth (upper $1\sigma$) Ka1 channel window, and for (lower panel) the smaller beamwidth (lower $1\sigma$) Q3 channel window, for the selected models shown in Fig. 1, normalized to the two-year DMR data. See Tables 2, 12, and 13 for numerical values. Note that the peak sensitivity of an SP94 individual-channel window function corresponds to a different angular scale in each of the models.

Fig. 4.— SP94 data of G95. (a) Individual-channel Ka data. (b) Individual-channel Q data.

Fig. 5.— Likelihood functions as a function of $Q_{\rm rms-PS}$ (to $Q_{\rm rms-PS} = 60\mu$K) derived from analyses: ignoring beamwidth and calibration uncertainties (solid lines); accounting for beamwidth uncertainty but ignoring calibration uncertainty (dotted lines); and accounting for beamwidth and calibration uncertainties (dashed lines). In (a) are those for the Ka2 data (which does not have a detection), in (b) those for the Q2 data (which has one of the best individual-channel detections), in (c) those for the Ka data, and in (d) those for the Ka + Q data. The 6 individual panels shown for each of these SP94 data sets correspond to the 6 selected models of Fig. 1. See Table 2 for model-parameter values. In each panel the likelihood functions have been renormalized such that the peak value is unity for the one with the highest peak value. See Tables 8 – 16 for numerical values derived from the corresponding probability density distribution functions.

Fig. 6.— Likelihood functions, as a function of $Q_{\rm rms-PS}$ for the 6 selected models (O1, O11, O14, Λ2, Λ10, and Flat) of Fig. 1 (line styles are identical to those used in Figs. 1 and 3), from the beamwidth- and calibration-uncertainty corrected analyses, for the Ka + Q, Ka, Ka2, and Q2 data sets. See Table 2 for model-parameter values, and Tables 8 – 16 for numerical values derived from the corresponding probability density distribution functions.

Fig. 7.— CMB anisotropy bandtemperature predictions and observational results. Hatched regions are what would be seen by a series of ideal, Kronecker-delta window-function, experiments, for the model normalized to the $1\sigma$ range from the two-year DMR data (hatched region with smaller vertical extent), and for the model normalized to the $1\sigma$ range from the SP94 combined Q (upper panel), Ka (central panel), and full Ka + Q (lower panel) data subsets (hatched region with larger vertical extent). These are evaluated using the appropriate model spectrum in eq. (7) for a Kronecker-delta $W_l$. Both normalizations account for all major known uncertainties. Also shown are the corresponding SP94 individual-channel observational results. Solid squares correspond to detections, and are placed at the peak of the probability density distribution function. The vertical error bars on the points corresponding to detections are $\pm 1\sigma$ HPD error bars. Solid inverted triangles are $2\sigma$ ET upper limits corresponding to non-detections. From left to right, the observational data points correspond to the Ka1 to Q3 data results, and they are placed at $l_{\rm m}$ determined from $(\delta T_{\rm rms}{}^2)_l$ (eq. [6]). The horizontal lines on the SP94 individual-channel observational results represent the $l$-space width of the corresponding $W_l$, and terminate at $l_{e^{-0.5}}$ determined from $(\delta T_{\rm rms}{}^2)_l$ using the $+1\sigma$ beamwidth $W_l$ for the lower value of $l_{e^{-0.5}}$ and the $-1\sigma$ beamwidth $W_l$ for the higher value. The 6 sets of 3 panels each [(a) to (f)] correspond to the 6 selected models of Fig. 1. See Table 2 for model-parameter values, Tables 12 and 13 for the SP94 combined data model normalizations and the two-year DMR model normalizations, and Table 14 for the SP94 individual-channel results.

Figure 1

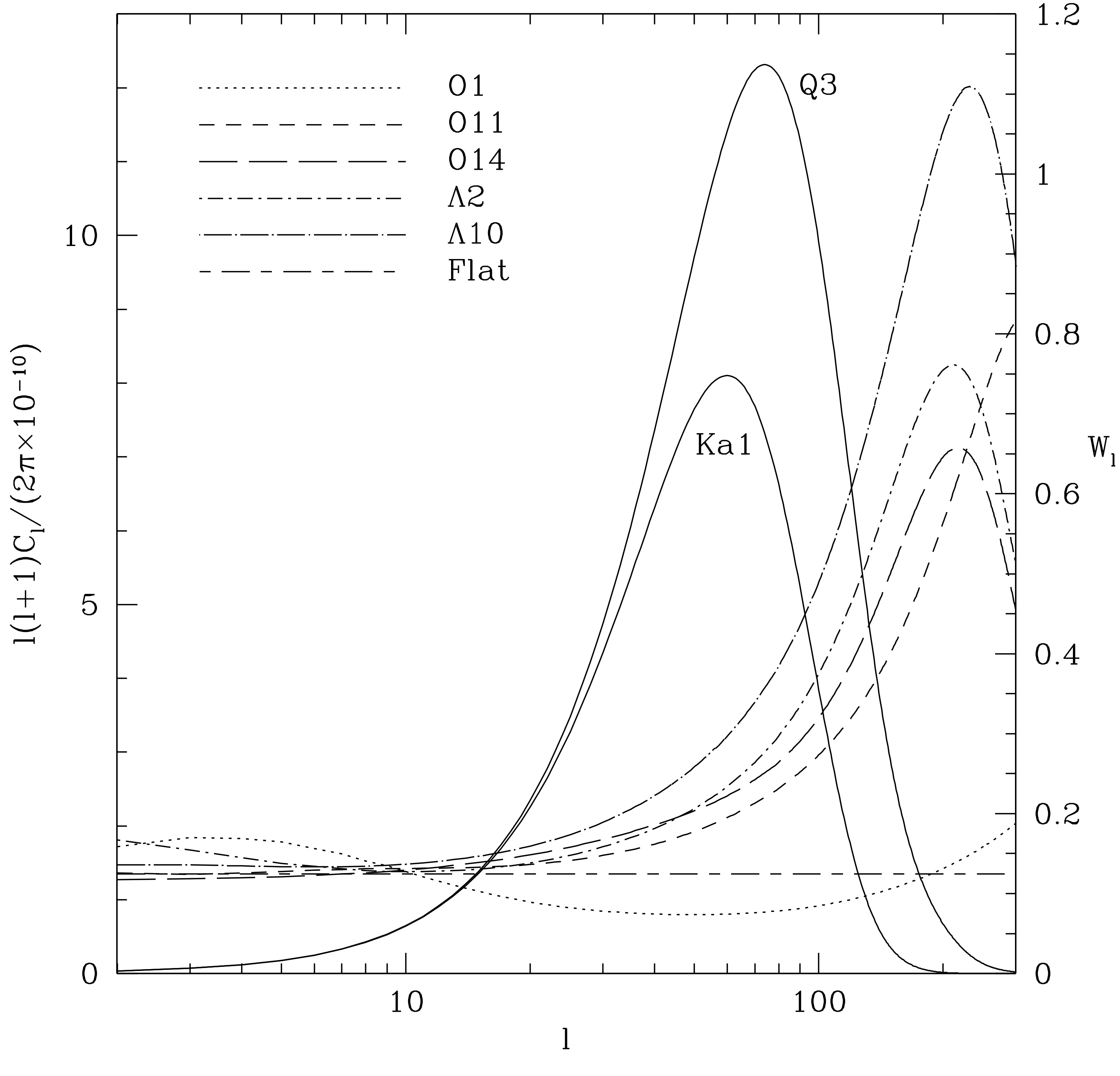

Figure 2

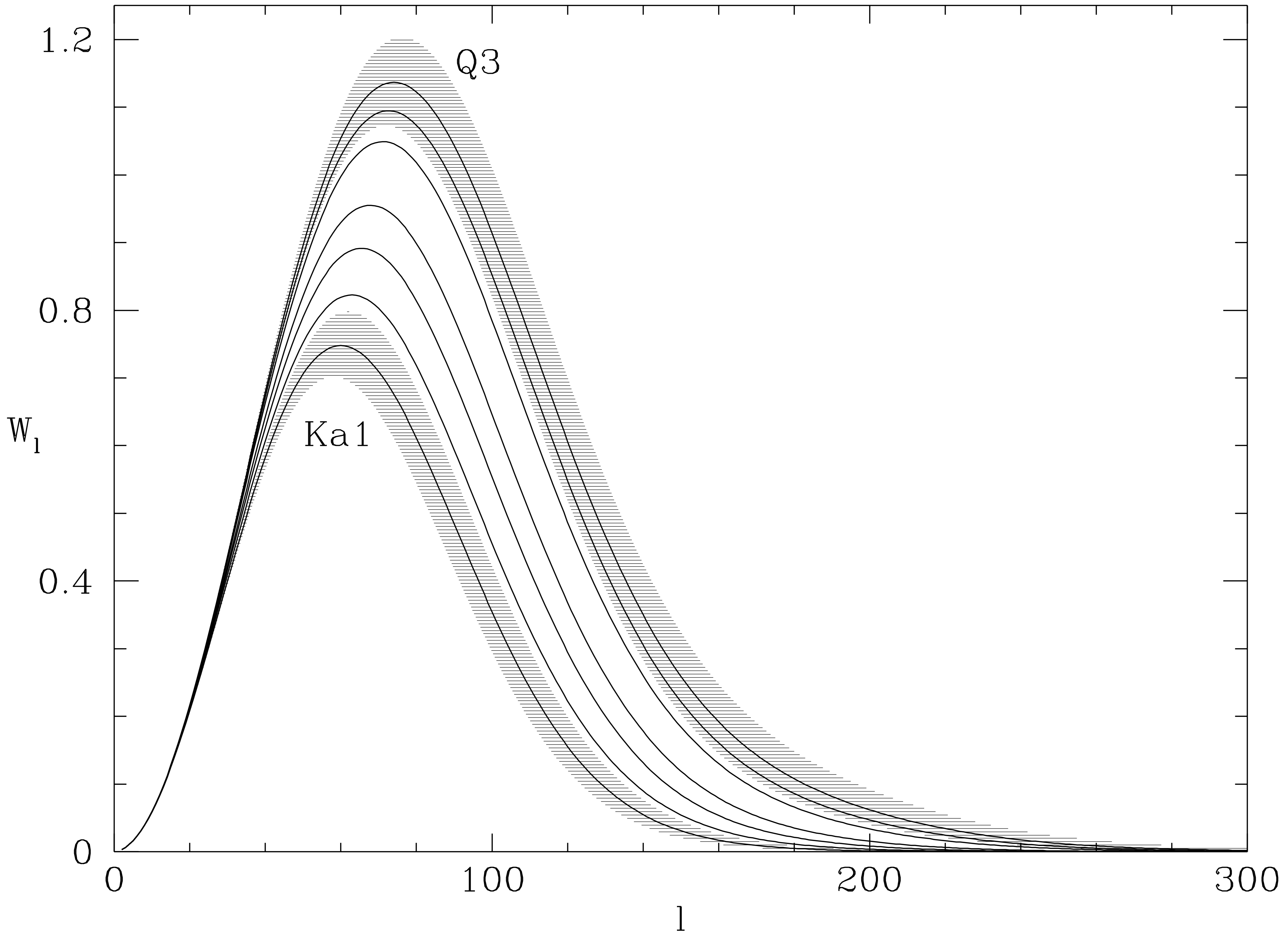

Figure 3

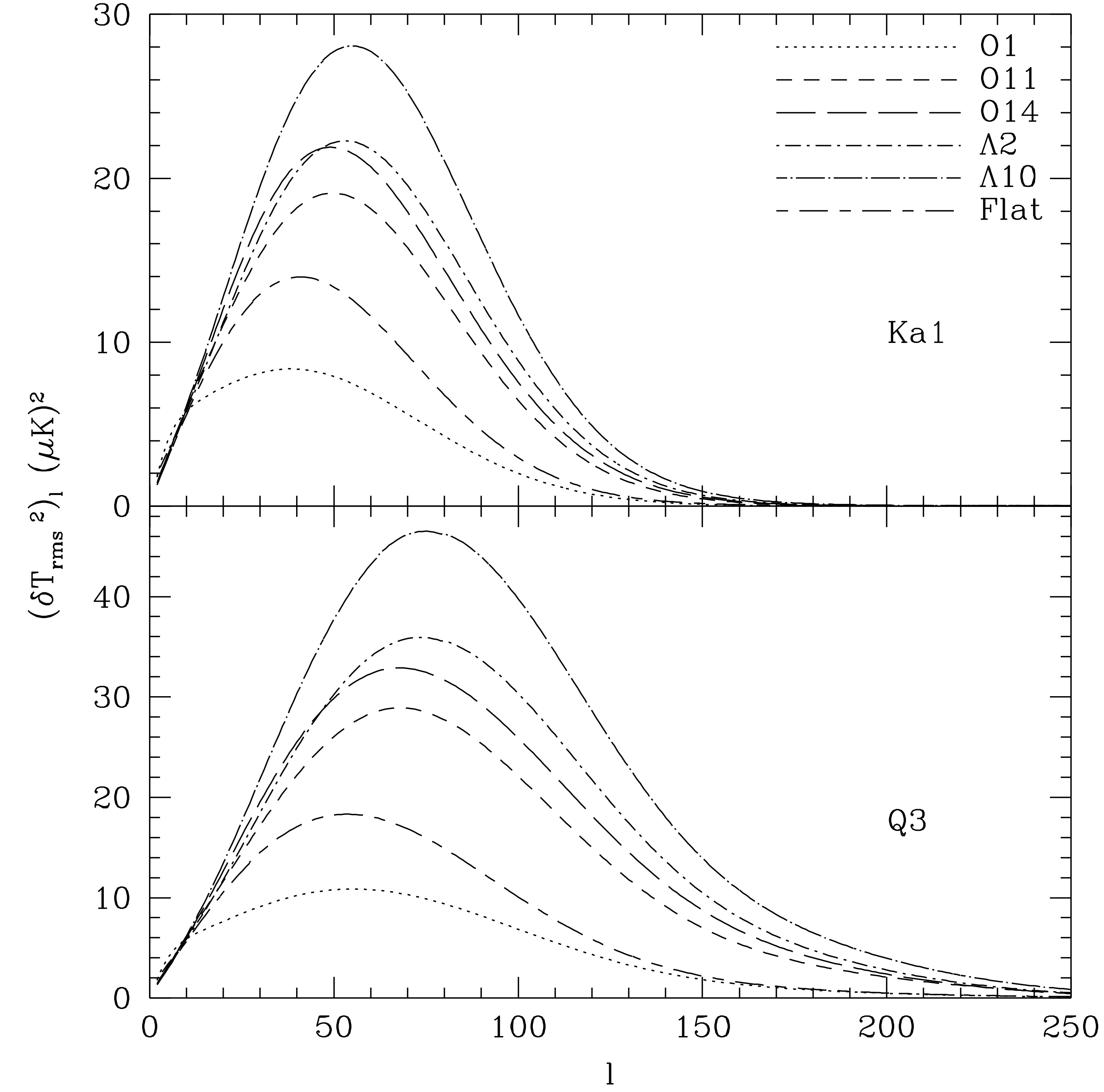

Figure 4(a)

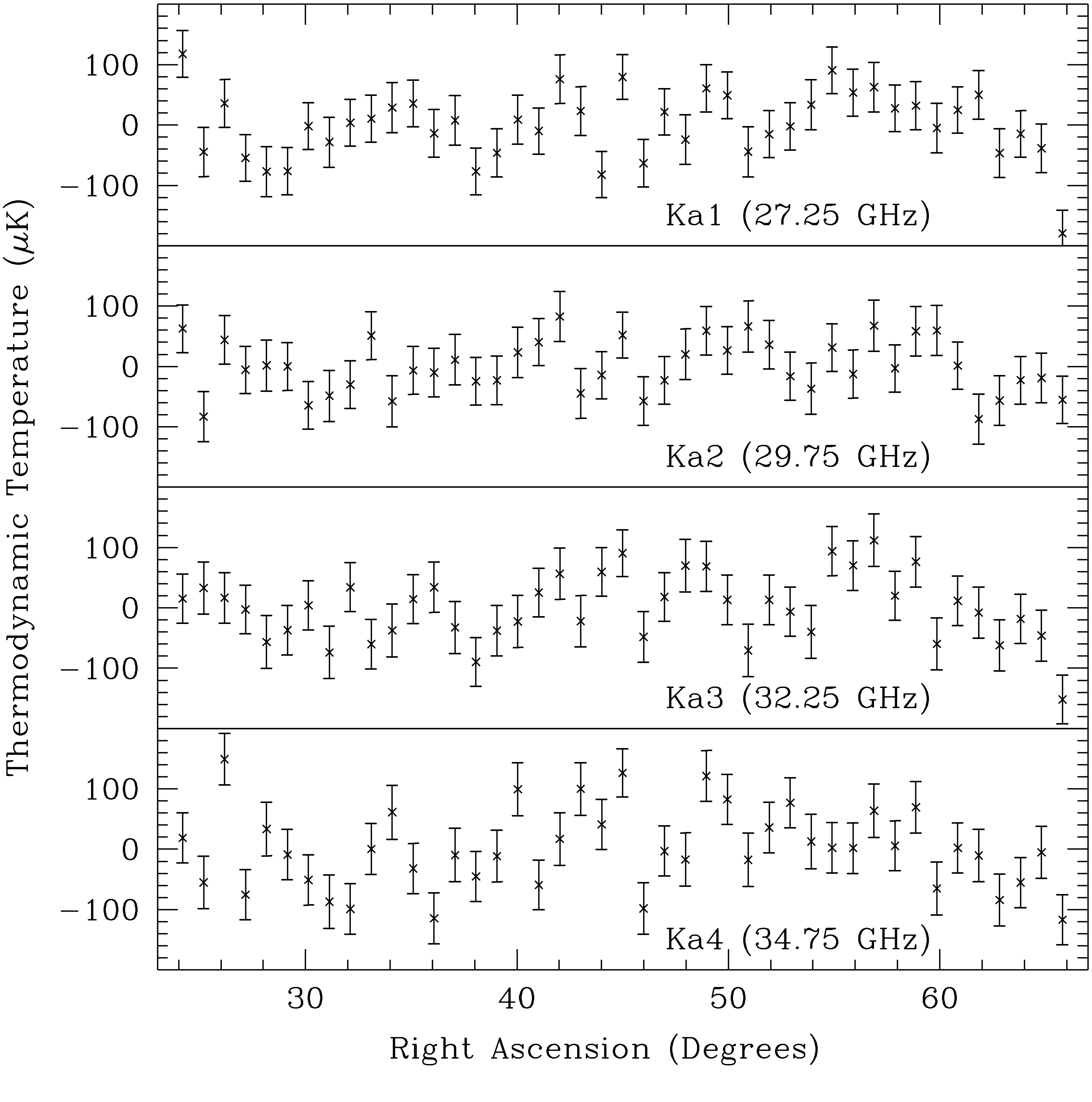

Figure 4(b)

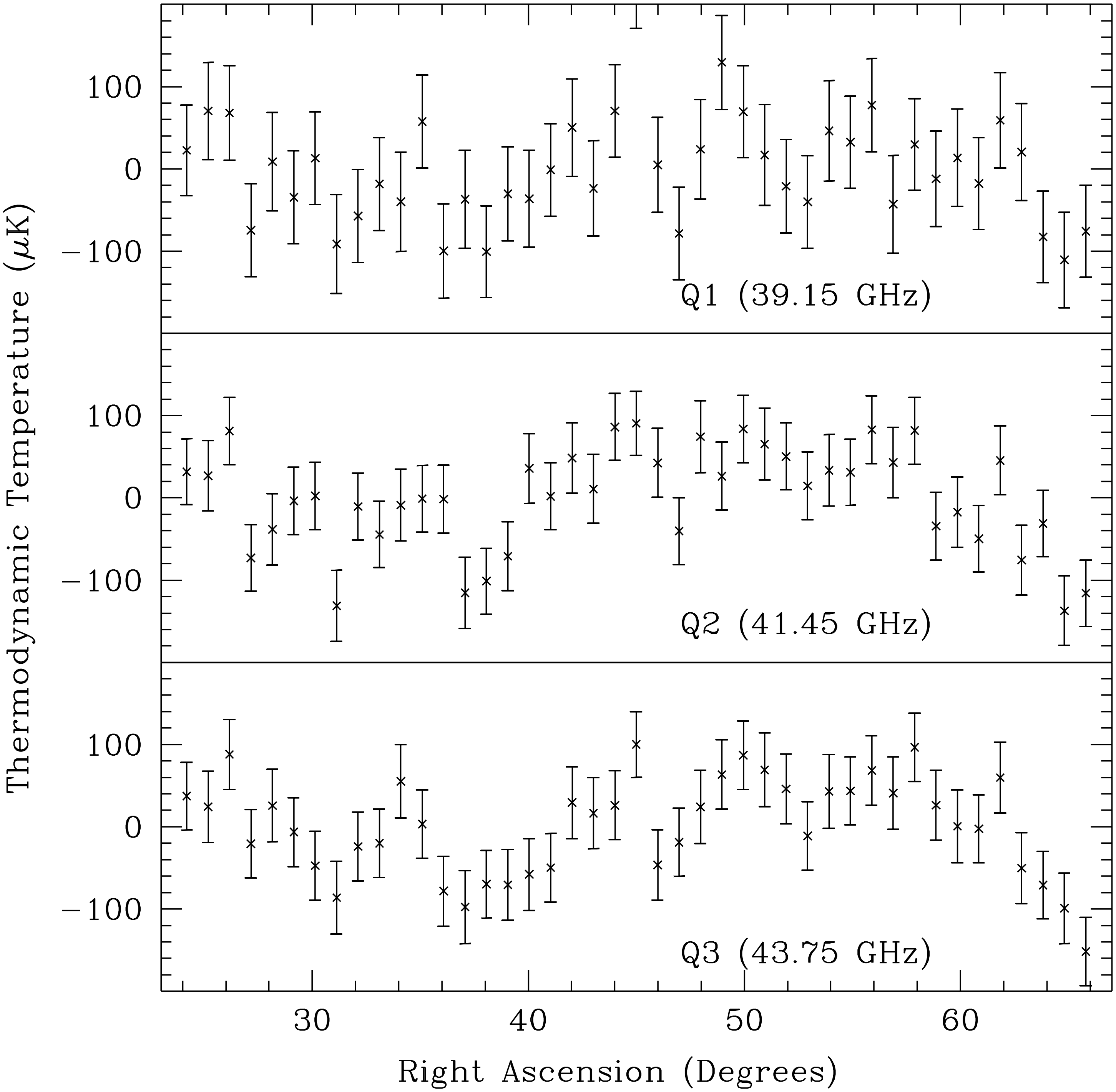

Figure 5(a)

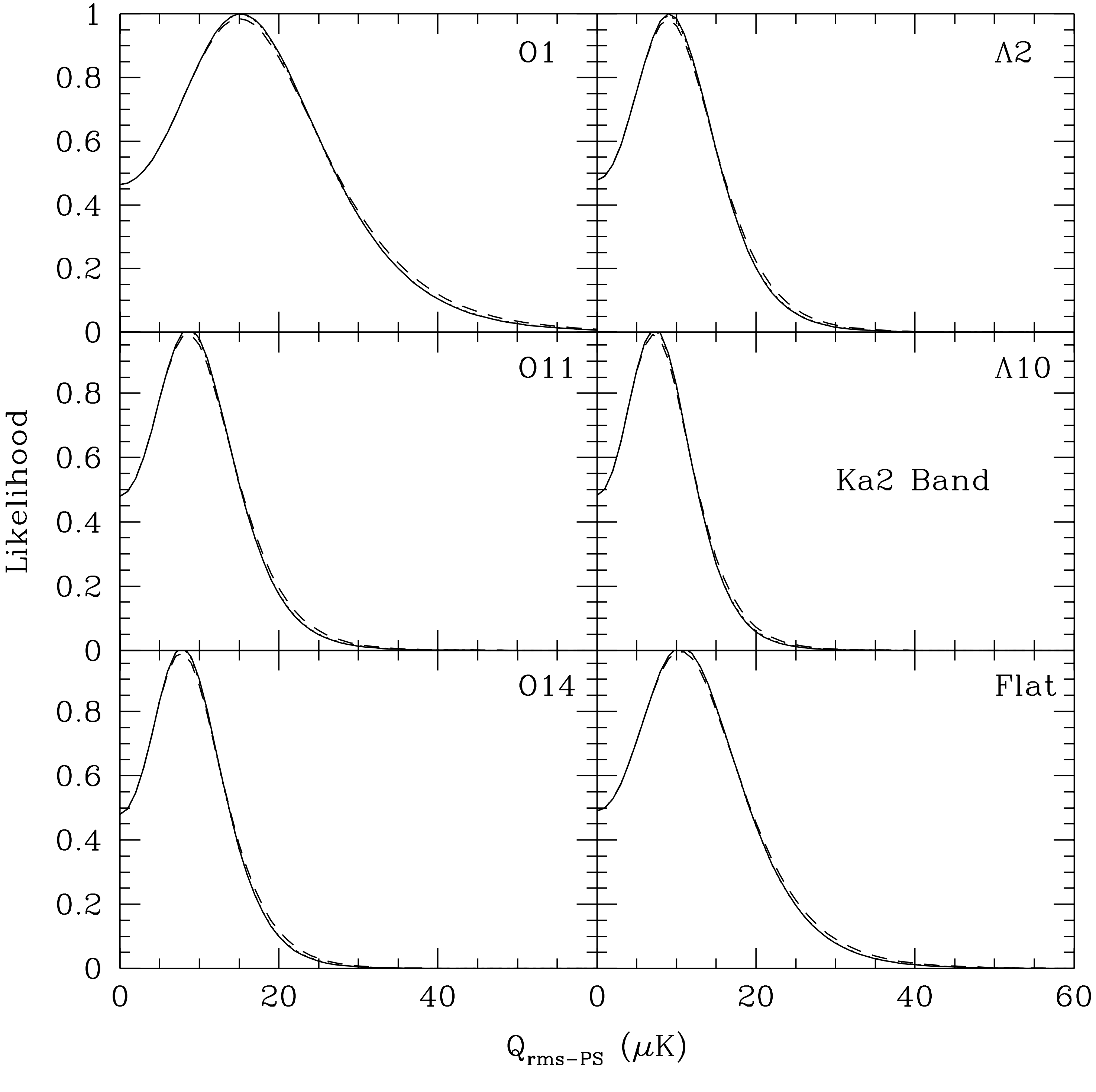

Figure 5(b)

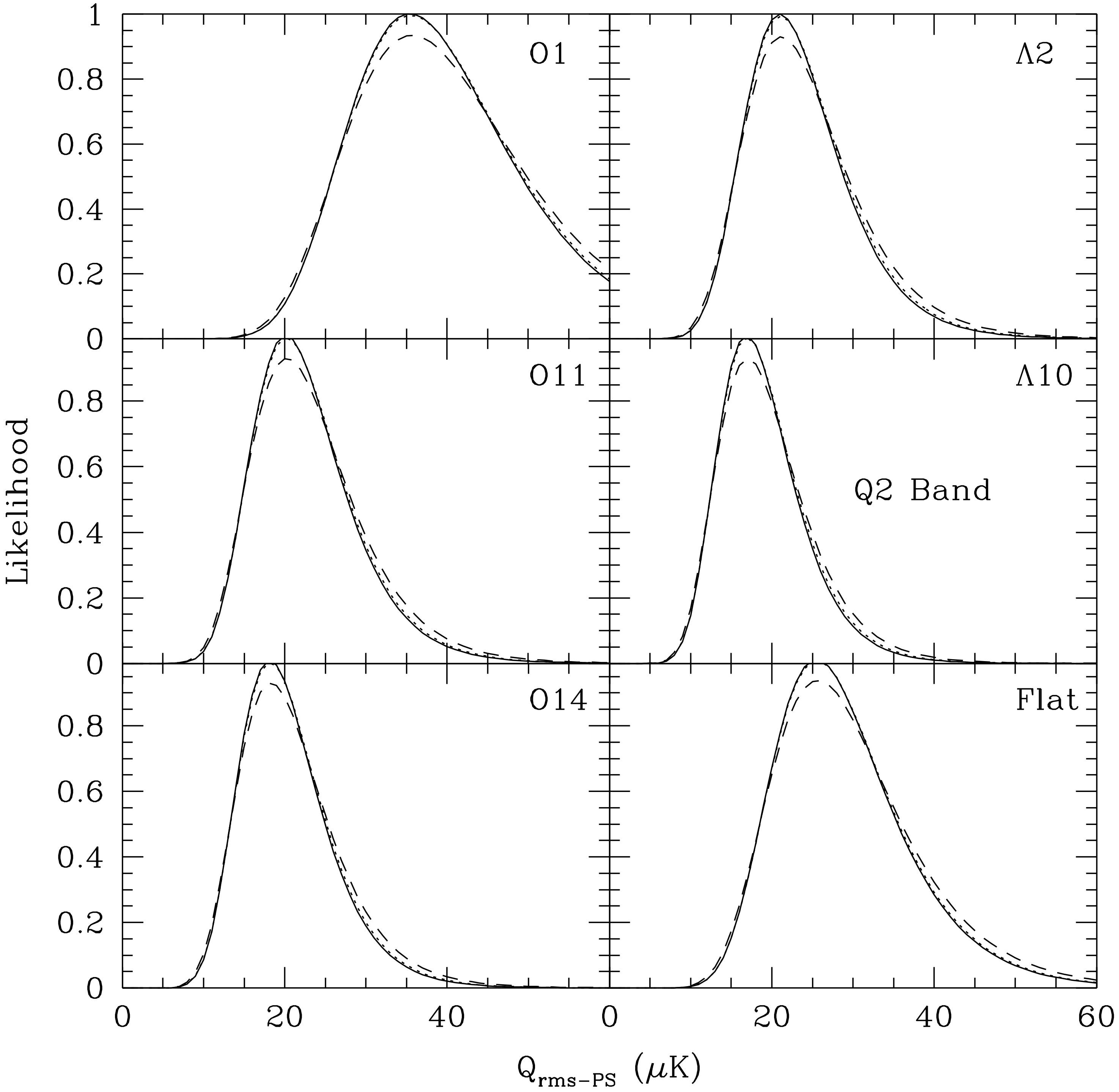

Figure 5(c)

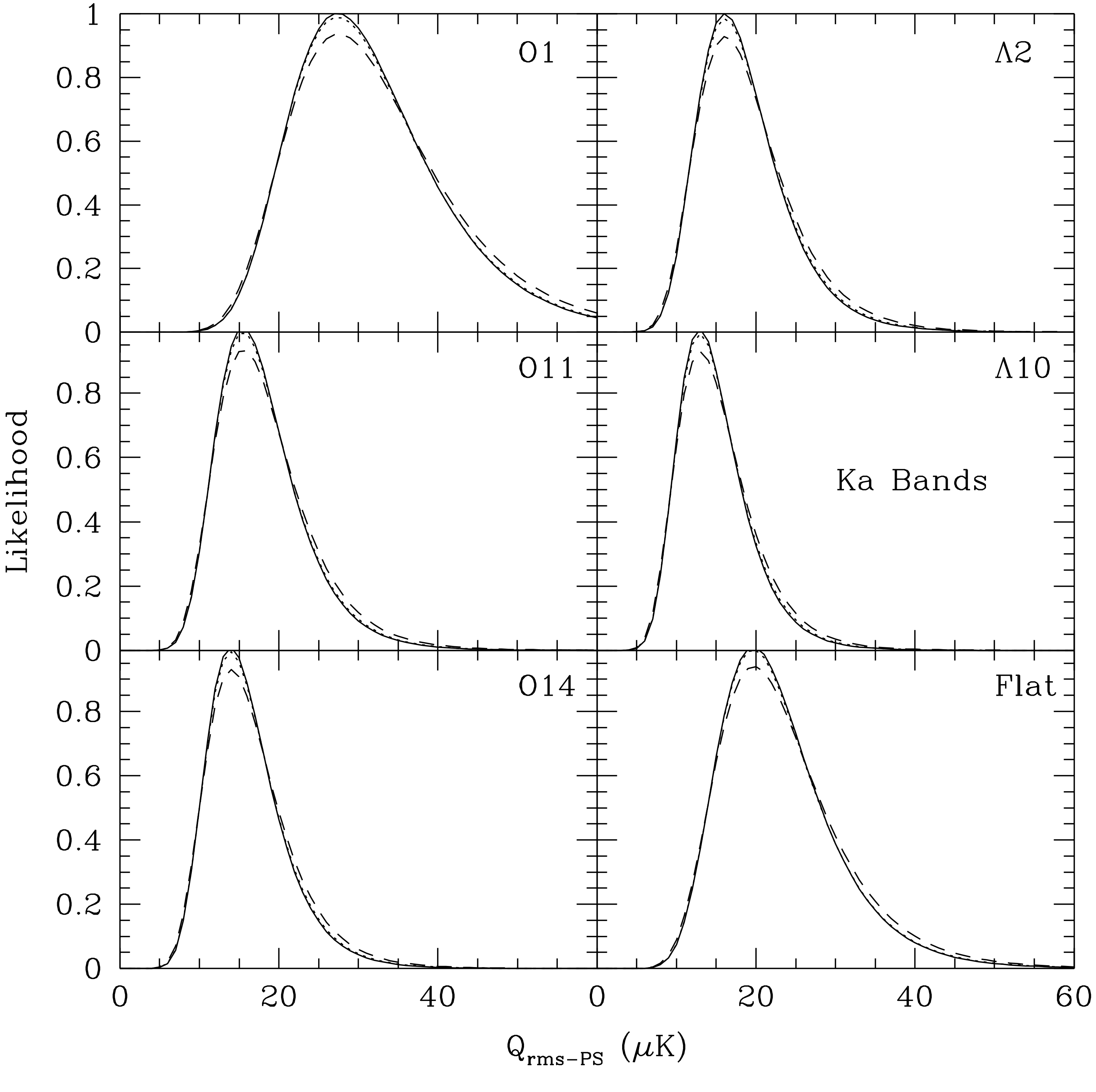

Figure 5(d)

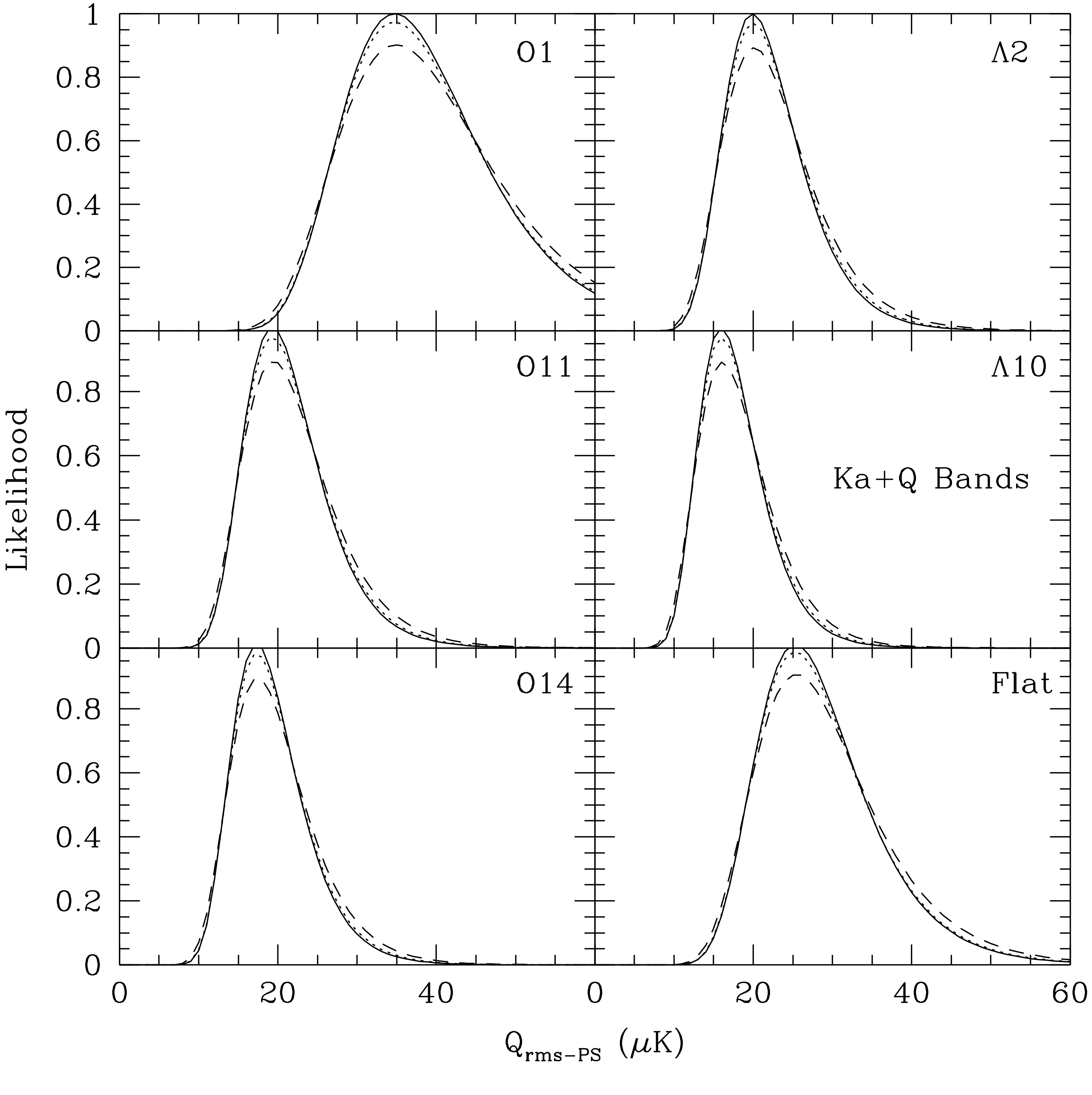

Figure 6

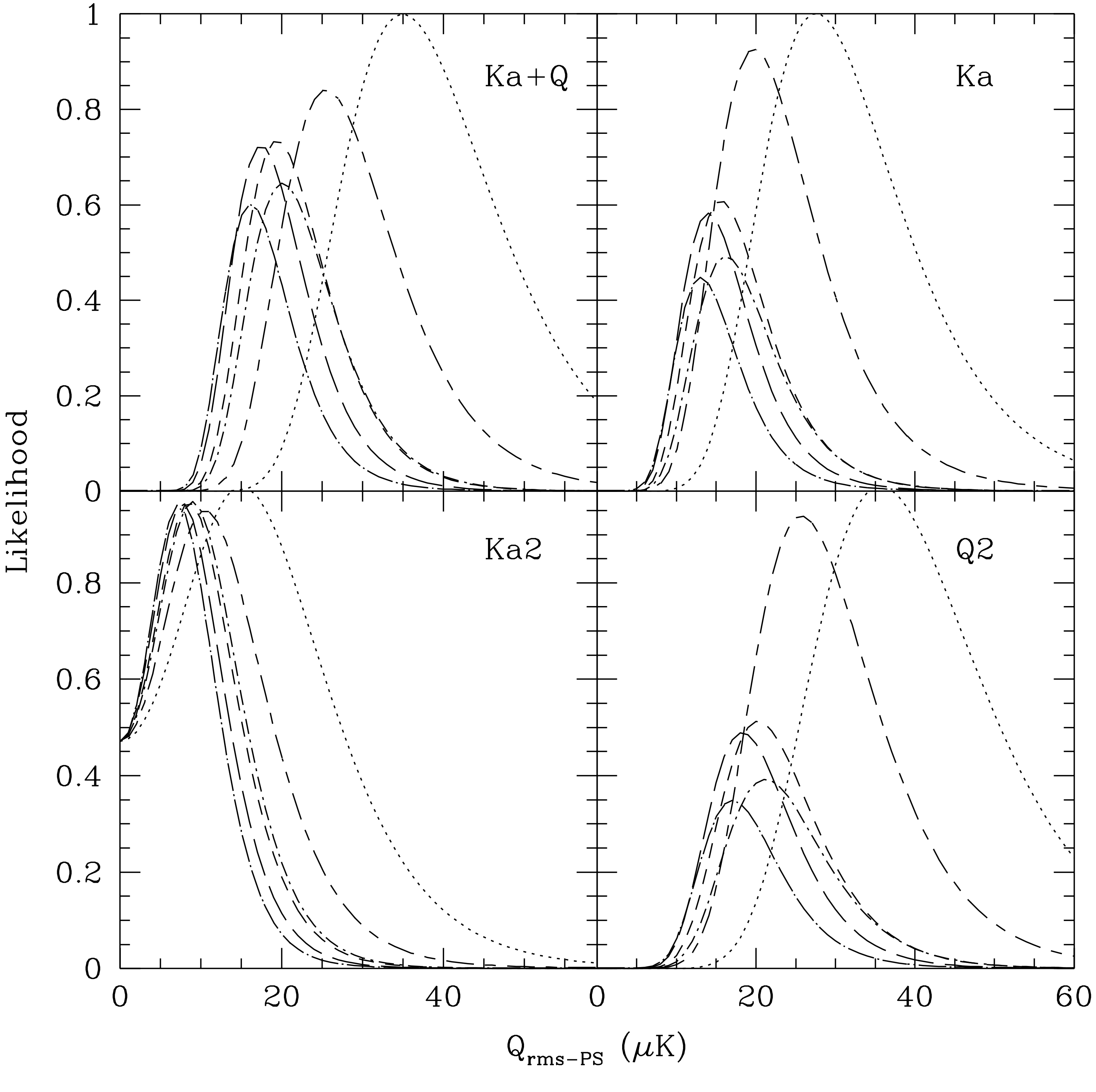

Figure 7(a)

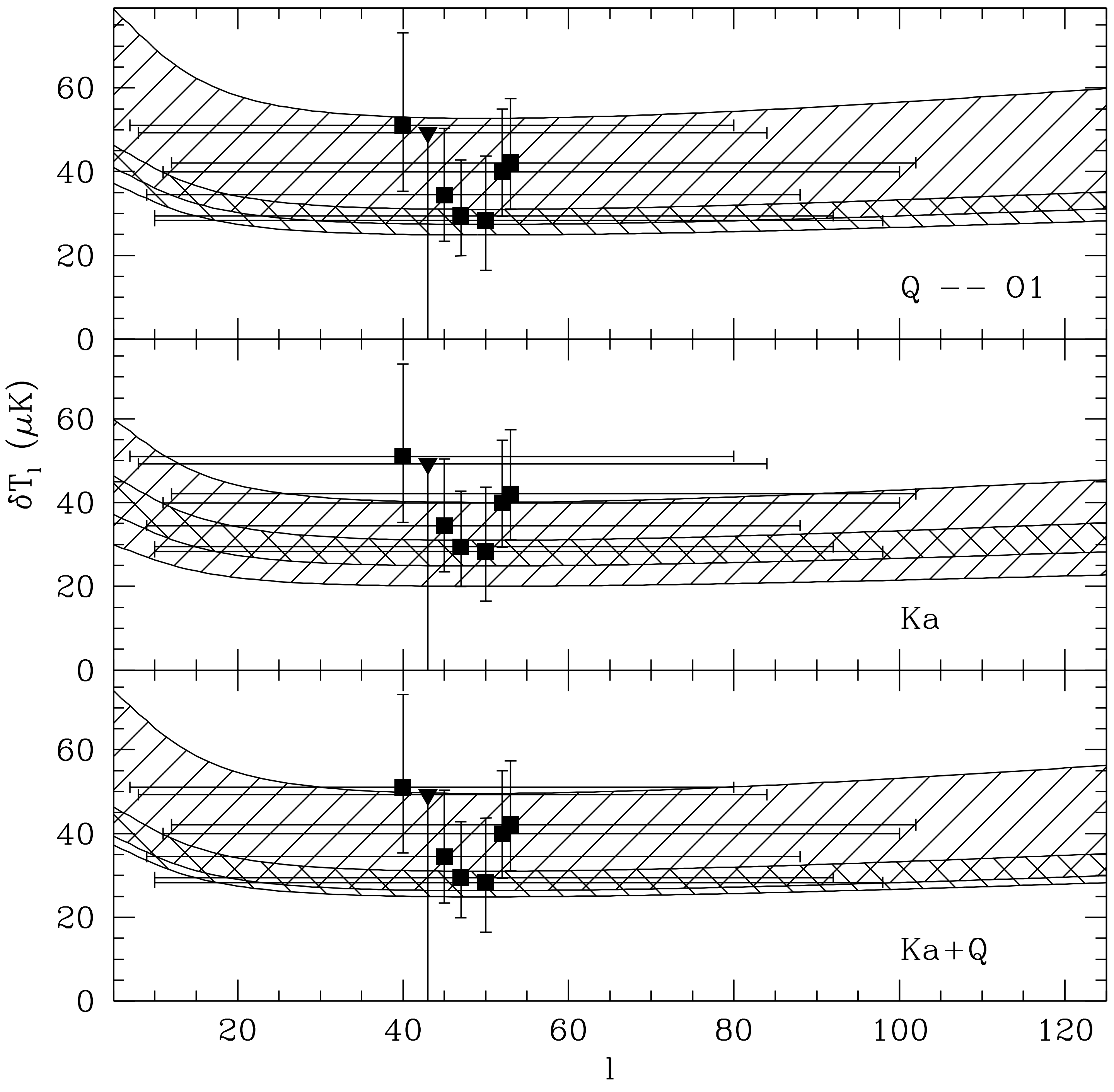

Figure 7(b)

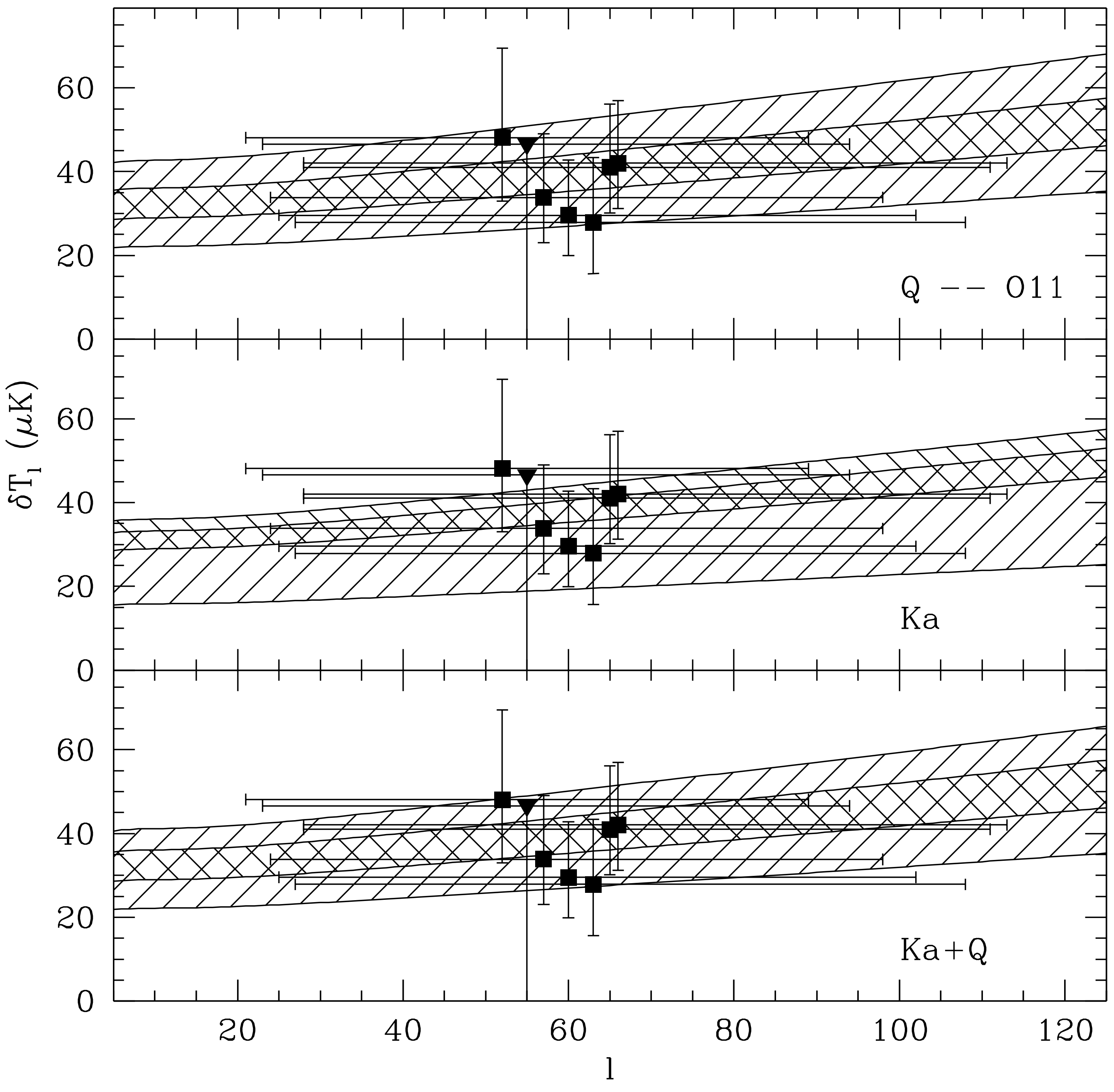

Figure 7(c)

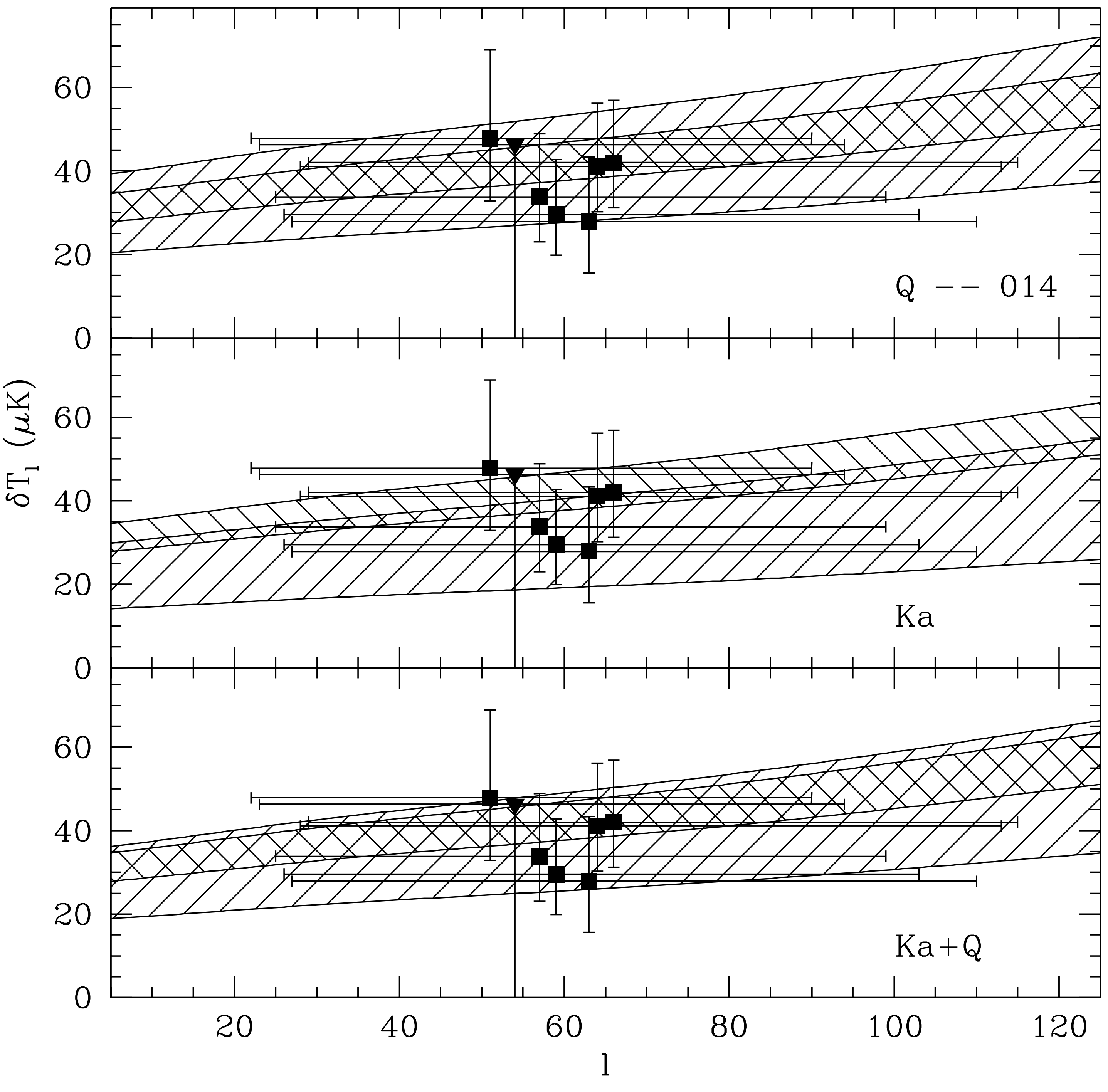

Figure 7(d)

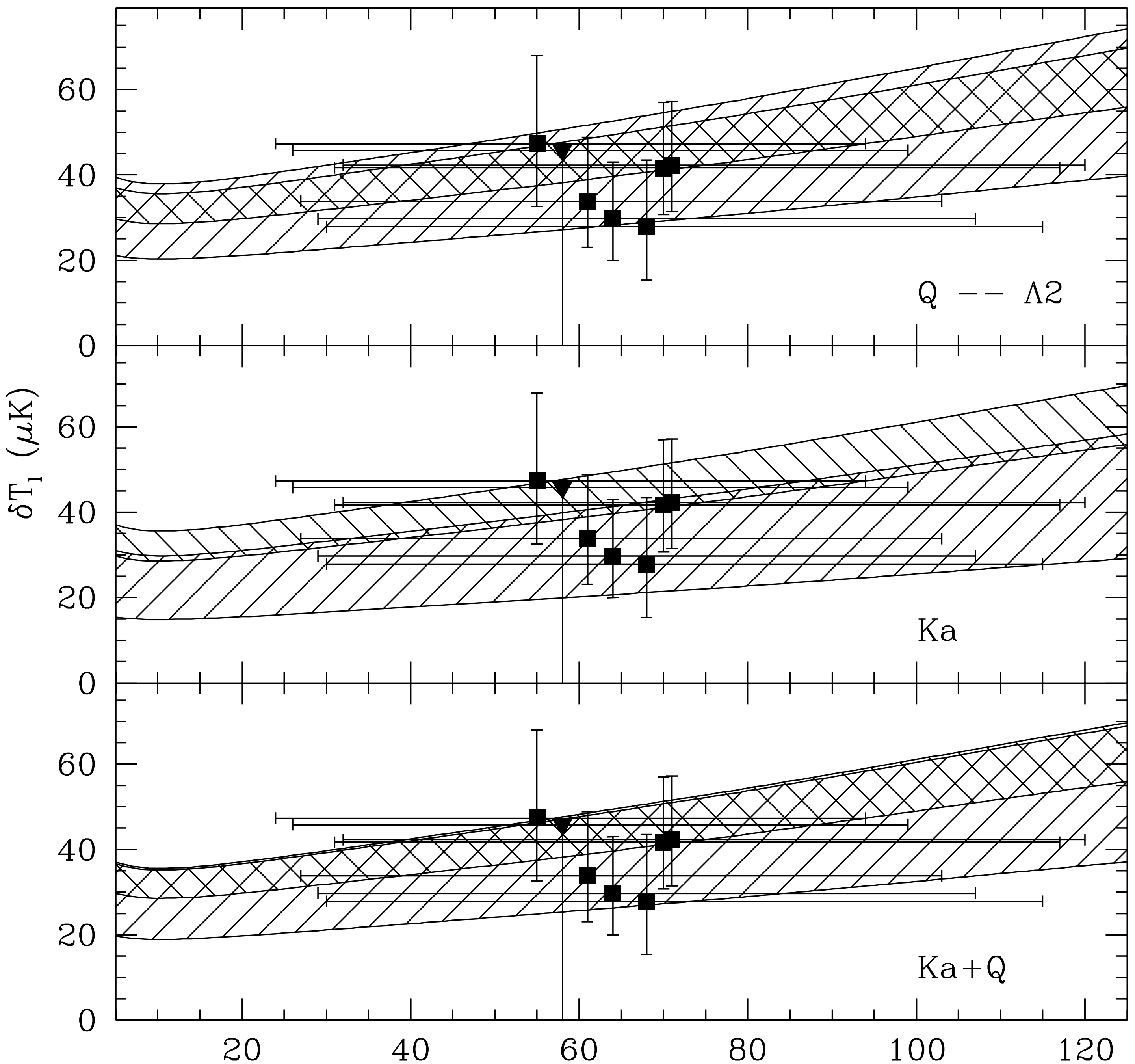

Figure 7(e)

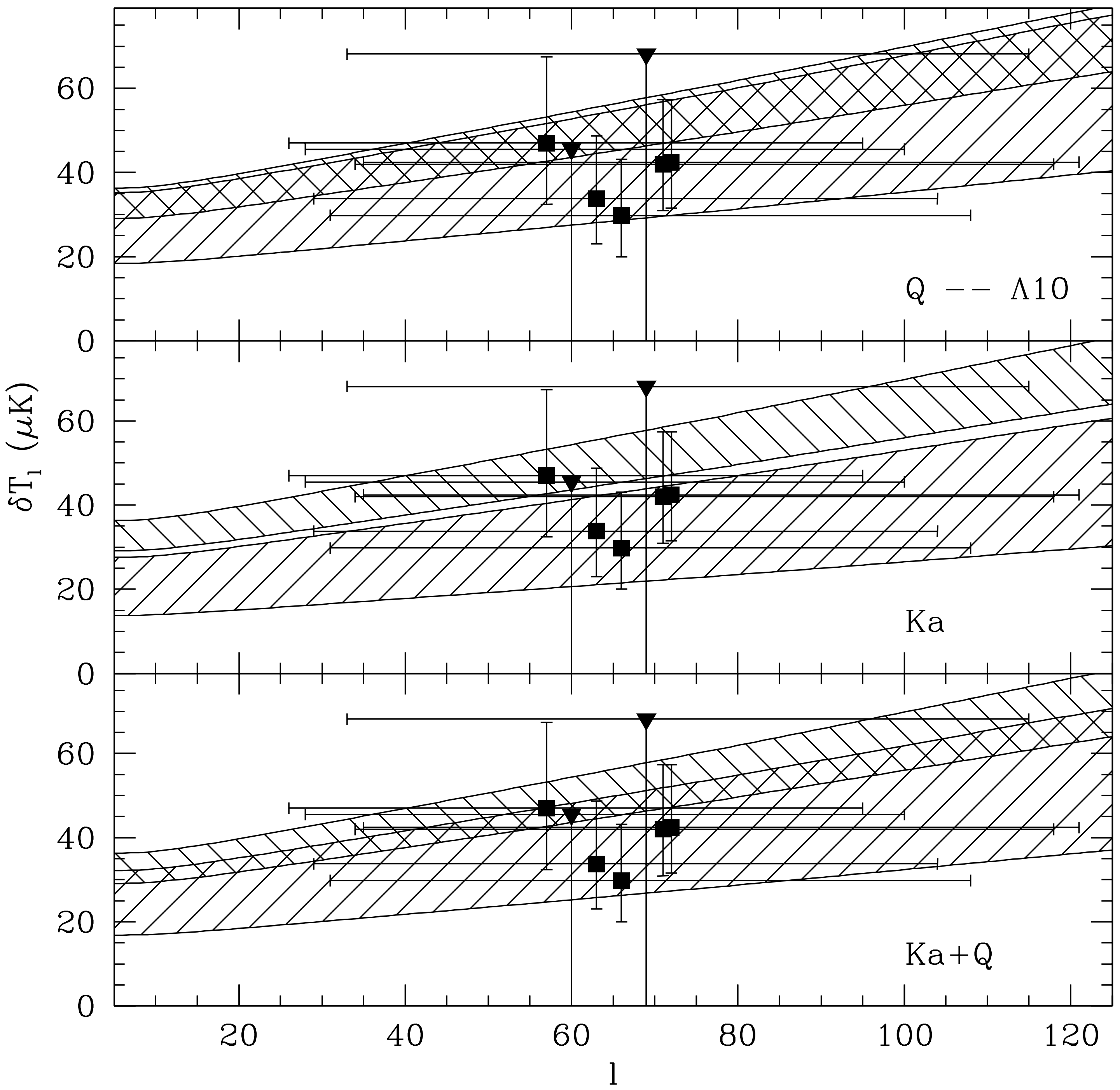

Figure 7(f)

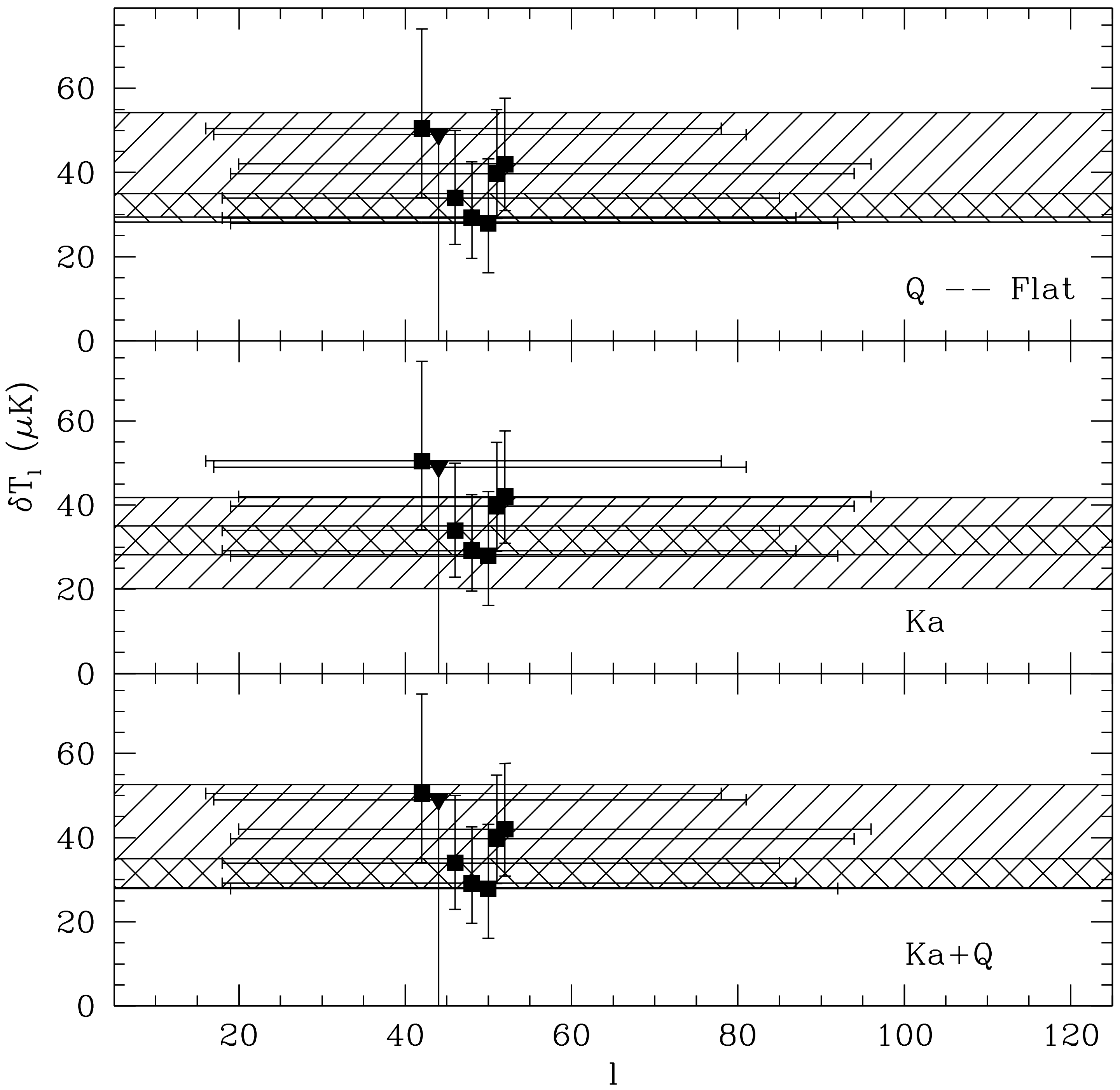